\documentclass[a4paper,11pt]{article}
\pdfoutput=1
\usepackage{jcappub}
\usepackage{amsthm,graphicx}
\usepackage{epsfig}
\usepackage{latexsym, amssymb} 
\usepackage{amsmath}
\usepackage[normalem]{ulem}
\usepackage{xcolor}
\usepackage{verbatim}
\usepackage{wrapfig,lipsum,booktabs}

\usepackage{rotating}
\usepackage{amsmath}
\usepackage{epsfig}
\usepackage{txfonts}
\usepackage{xspace}

\def\lb{\textit{LiteBIRD}\xspace}
\def\Planck{\textit{Planck}\xspace}
\newcommand{\nB}{{n_{\rm B}}}
\newcommand{\PB}{{P_{\rm B}}}
\newcommand{\AB}{{A_{\rm B}}}

\newcommand{\Mpc}{{\rm Mpc}}

\def\ba{\begin{eqnarray}}
\def\ea{\end{eqnarray}}
\def\be{\begin{equation}}
\def\ee{\end{equation}}
\usepackage{hyperref}
\usepackage{grffile}
\usepackage{graphics}
\usepackage{booktabs}
\usepackage{lineno}
\begin{document}
\include{aas_macros.sty}
\title{LiteBIRD Science Goals and Forecasts: Primordial Magnetic Fields}
\author[1,2]{D.\,Paoletti,}
\author[3,4]{J.\,Rubino-Martin,}
\author[5]{M.\,Shiraishi,}
\author[6,1]{D.\,Molinari,}
\author[7]{J.\,Chluba,}
\author[1,2]{F.\,Finelli,}
\author[8,9,10]{C.\,Baccigalupi,}
\author[11]{J.\,Errard,}
\author[1,2]{A.\,Gruppuso,}
\author[12]{A.\,I.\,Lonappan,}
% ...
\affiliation[1]{INAF - OAS Bologna, via Piero Gobetti, 93/3, 40129 Bologna, Italy}
\affiliation[2]{INFN Sezione di Bologna, Viale C. Berti Pichat, 6/2 – 40127 Bologna Italy}
\affiliation[3]{Instituto de Astrofísica de Canarias, E-38200 La Laguna, Tenerife, Canary Islands, Spain}
\affiliation[4]{Departamento de Astrofísica, Universidad de La Laguna (ULL), E-38206, La Laguna, Tenerife, Spain}
\affiliation[5]{Suwa University of Science, Chino, Nagano 391-0292, Japan}
\affiliation[6]{High Performance Computing Department, CINECA, via Magnanelli 2/3, Casalecchio di Reno, 40033, Italy}
\affiliation[7]{Jodrell Bank Centre for Astrophysics, Alan Turing Building, Department of Physics and Astronomy, School of Natural Sciences, The University of Manchester, Oxford Road, Manchester M13 9PL, UK}
\affiliation[8]{International School for Advanced Studies (SISSA), Via Bonomea 265, 34136, Trieste, Italy}
\affiliation[9]{INFN Sezione di Trieste, via Valerio 2, 34127 Trieste, Italy}
\affiliation[10]{IFPU, Via Beirut, 2, 34151 Grignano, Trieste, Italy}
\affiliation[11]{Université de Paris, CNRS, Astroparticule et Cosmologie, F-75013 Paris, France}
\affiliation[12]{Dipartimento di Fisica, Università di Roma Tor Vergata, Via della Ricerca Scientifica, 1, 00133, Roma, Italy}
% ...

\defcitealias{LiteBIRD:2022cnt}{PTEP}

\emailAdd{daniela.paoletti@inaf.it}

\abstract
{We present detailed forecasts for the constraints on the characteristics of primordial magnetic fields (PMFs) generated prior to recombination that will be obtained with the \lb satellite. The constraints are driven by some of the main physical effects of PMFs on the CMB anisotropies: the gravitational effects of magnetically-induced perturbations; the effects on the thermal and ionization history of the Universe; the Faraday rotation imprint on the CMB polarization spectra; and the non-Gaussianities induced in polarization anisotropies. \lb represents a sensitive probe for PMFs.
We explore different levels of complexity, for \lb data and PMF configurations, accounting for possible degeneracies with primordial gravitational waves from inflation. By exploiting all the physical effects, \lb will be able to improve the current limit on PMFs at intermediate and large scales coming from \Planck. In particular, thanks to its accurate $B$-mode polarization measurement, \lb will improve the constraints on infrared configurations for the gravitational effect, giving $B_{\rm 1\,Mpc}^{\nB =-2.9} < 0.8$\,nG at 95\% C.L., potentially opening the possibility to detect nanogauss fields with high significance. We also observe a significant improvement in the limits when marginalized over the spectral index, $B_{1\,{\rm Mpc}}^{\rm marg}< 2.2$\,nG at 95\,\% C.L. From the thermal history effect, which relies mainly on $E$-mode polarization data, we obtain a significant improvement for all PMF configurations, with the marginalized case, $\sqrt{\langle B^2\rangle}^{\rm marg}<0.50$\,nG at 95\,\% C.L. Faraday rotation constraints will take advantage of the wide frequency coverage of \lb and the high sensitivity in $B$ modes, improving the limits by orders of magnitude with respect to current results, $B_{1\,{\rm Mpc}}^{\nB =-2.9} < 3.2$\,nG at 95\,\% C.L. Finally, non-Gaussianities of the $B$-mode polarization can probe PMFs at the level of 1\,nG, again significantly improving the current bounds from \Planck. Altogether our forecasts represent a broad collection of complementary probes based on widely tested methodologies, providing conservative limits on PMF characteristics that will be achieved with the \lb satellite.}

\maketitle
\section{Introduction}
Magnetic fields are ubiquitous in the Universe. They exist at all scales, from the smallest scales in stars and planets, representing a necessary condition for the development of life, up to the largest scales observable, filling the entire Universe in both structures and voids. 
Though the origin of some of these magnetic fields, especially those on the smallest scales, is known, the origin of magnetism on cosmological scales is an open issue \citep{Grasso:2000wj, Widrow:2002ud, Kandus:2010nw, Widrow:2011hs, Durrer:2013pga,Subramanian:2015lua}. This investigation requires putting together different pieces of information, in the form of different cosmological probes. The cosmic microwave background (CMB) and its observations with \lb~\cite{LiteBIRD:2022cnt} play a key role in solving this puzzle. 

The cosmic magnetism extends to the smallest scales of galaxies, clusters of galaxies and voids, and up to the largest scales observable in filaments connecting the large scale structures.  With the improvement of cosmological observations we can now question whether this cosmic magnetism is more a property of the entire Universe rather than of single objects. If future observations confirm that magnetization is a universal property its origin would at least partially lie in magnetic fields generated in the early Universe, the so-called primordial magnetic fields (PMFs). 

Magnetic fields have been observed in galaxies, and in particular in the Milky Way since 1949 when two independent observations of polarized optical light \citep{1949Sci...109..165H,1949Sci...109..166H} were later interpreted as the result of dust grain alignment due to a diffuse magnetic field in the Galaxy \citep{1951ApJ...114..206D}. Afterwards, Zeeman line splitting and Faraday rotation measurements confirmed the presence of a Galactic magnetic field, which is now mapped to a high degree of accuracy  \cite{Planck:2016gdp,2022A&A...657A..43H}.
With the improvement of observational techniques, magnetic fields have been determined to be a fundamental component of all galaxies with morphologies and characteristics that depend on the host, hinting at a co-evolution of the magnetic and matter components \citep{Beck:1995zs,1997FCPh...19....1V,Battaner:2000ef}. Astrophysical mechanisms, such as stellar dynamos, can produce magnetic fields in a galactic environment, but galaxy magnetic fields are observed out to high redshifts \cite{Kronberg:1993vk,Bernet:2008qp} setting strong constraints on the capabilities of astrophysical processes to fully generate these fields in the available cosmic time. Moreover, dynamo processes, which are responsible for the fields we observe on planetary and stellar scales, would have had very little time to form magnetic fields on galactic scales. And indeed, most models of dynamos require initial magnetic field seeds \cite{Subramanian:2018xlb,shukurov_subramanian_2021}.

Zooming out to larger scales,  magnetic fields on scales as large as Megaparsec (Mpc) are observed in galaxy clusters \cite{Carilli:2001hj,Govoni:2004as,Bonafede:2010wg,Botteon:2022umz} with amplitudes of the order of a few microgauss ($\mu$G). In galaxy clusters, astrophysical mechanisms capable of generating such coherent fields are more complex and involve mostly feedback from active galactic nuclei and galaxy winds \citep{Dolag:1999wvi,Donnert:2008sn,Vazza:2017qge,Mtchedlidze:2022ewp}. Although such mechanisms contribute to the overall magnetic fields observed in cosmic structures, to reproduce the current observations with only these astrophysical fields is difficult and often requires the presence of initial seed magnetic fields. Future observations with the Low Frequency Array and the Square Kilometre Array will help to understand the nature of such large-scale magnetic fields in more detail \citep{Vazza:2015rqa,Vazza:2017qge}.

However, cosmic magnetism goes beyond galaxies and clusters; in fact, in the past decade the presence of magnetic fields has been suggested on even larger cosmological structures, in voids and filaments. 
The presence of intergalactic magnetic fields (IGMFs) in voids can leave imprints in $\gamma$-ray observations. Electron-positron pairs generated by the interaction of blazar TeV emission with extragalactic background light are deflected by the IGMF \citep{Fermi-LAT:2018jdy}. Therefore, the secondary photon cascade at GeV energies due to the inverse-Compton interaction with CMB photons is spread in a low luminosity halo undetectable by current experiments (see Ref.\citep{Broderick:2011av} for an alternative explanation with plasma turbulence). 
The lack of detection of secondary GeV photons from some blazars in Fermi satellite data is compatible with the presence of magnetic fields in voids and this kind of measurement has led to \textit{lower} limits on the amplitude of the magnetic fields (contrary to the usual upper limits from the CMB)\citep{Neronov:2010gir,Tavecchio:2010mk,Taylor:2011bn,Vovk:2011aa,HESS:2023zwb}. Recently also $\gamma$-ray bursts have been proposed for such analyses, and through the same mechanism can again be compatible with lower limits on the field amplitude \cite{Wang:2020vyu,Vovk:2023qfk}.
The future data from the Cherenkov Telescope Array will have enough resolution to identify the low luminosity halos and finally confirm the hypothesis of IGMFs in voids (alternative explanations cannot justify the extended halo), improve the lower bounds \cite{Barai:2018msb,Batista:2021rgm} and possibly  help identify a helical structure of the IGMF \citep{Long:2015bda}. 
Voids are crucial for two reasons. On the one hand to generate and maintain magnetic fields on Mpc scales with astrophysical sources in voids is very difficult. On the other hand the void environment is such that magnetic fields have an almost completely passive evolution, making them the best candidate to understand the properties of a possible primordial seed magnetic field.

Large-scale magnetic fields have also been observed in bridges connecting galaxy clusters \citep{Govoni981,10.1093/mnrasl/slaa142} and some stacking analyses seem to indicate magnetic fields in the filaments connecting large-scale structures \citep{Vernstrom:2021hru,Carretti:2022fqk,Carretti:2022tbj}. In the future, Square Kilometre Array will allow to perform deeper stacking and provide information on the characteristics of such fields. This may finally resolve the origin of cosmic magnetism. Magnetic fields in filaments, which follow the filamentary structure and cannot be produced by astrophysical mechanisms \citep{Vazza:2021vwy}, would be the smoking gun of PMFs.

The idea of PMFs dates back many years and was meant as a purely theoretical hypothesis \cite{Thorne:1967zz}, but as the cosmic magnetism keeps unveiling it is becoming increasingly interesting for modern cosmology  \cite{Turner:1987bw,Enqvist:1998fw,Olinto:1997sj,Widrow:2002ud,Durrer:2006pc,Subramanian:2009fu,Widrow:2011hs,Ryu:2011hu,Durrer:2013pga,Subramanian:2015lua,Kahniashvili:2018mzl,Subramanian:2018xlb,Subramanian:2019jyd,shukurov_subramanian_2021}. We are now at a stage where we cannot ignore 
the effects of the potential presence of PMFs on the history of the Universe.

To explain cosmic magnetism is not the end of the story for PMFs. The role of PMFs in cosmology is twofold: on the one hand they may represent the seeds that generated the cosmic magnetism; and on the other hand their generation in the early Universe requires unique conditions for the physics of the early Universe. 
Therefore, PMFs represent a new window on the fundamental physics in the early Universe, providing an insight on aspects which will be difficult to investigate otherwise.

PMF generation mechanisms can be classified depending on the time at which they take place. The so-called \textit{causal mechanisms} are the generation processes that take place after inflation and that are bounded by causality. This bound is their greatest weakness, as such mechanisms limit the coherence length of the generated PMFs to the causal horizon at the generation time. In order for PMFs to seed cosmic magnetism and to be maintained against dissipation,  large coherence lengths are required. Therefore, causal fields require an inverse cascade process to increase the coherence length \citep{Brandenburg:1996fc,Banerjee:2004df}, which in turn gives a helical component of PMFs.
The main mechanisms of this class are related to first-order phase transitions, with both electroweak and quantum chromodynamics as plausible candidates, and rely on the instabilities at the interface of the transitioning regions, which can create currents that generate magnetic fields \citep{Quashnock:1988vs,Vachaspati:1991nm,Baym:1995fk,
Sigl:1996dm,Hindmarsh:1997tj,Grasso:1997nx,Ahonen:1997wh,Boyanovsky:2005ut,Caprini:2009yp,Tevzadze:2012kk,Zhang:2019vsb,Ellis:2019tjf}. However, the current standard scenario points towards simple cross-overs for the main phase transitions, instead of first-order transitions, meaning that finding evidence of a first-order phase transition would imply fundamental physics beyond the standard model. For example first order phase transitions have been associated with minimal supersymmetric standard model (MSSM), with the Higgs potential having a higher barrier and an additional degree of freedom through three-point self-coupling at tree level, or other extensions of the standard model with extra dimensions (e.g. Refs.\cite{Huang:2016cjm,Megias:2020vek}), with interesting perspectives also for direct detection of gravitational waves (GWs) \citep{Romero:2021kby}.

In the post inflationary Universe there is another mechanism unavoidably generating magnetic fields, namely second-order perturbations through the Harrison mechanism \citep{Harrison:1973zz}. The vorticity induced by second-order perturbations creates small currents that in turn generate magnetic fields \citep{Matarrese:2004kq,Fenu:2010kh,Fidler:2015kkt}. Although these PMFs are generated unavoidably, it has been demonstrated through simulations that they are too weak to provide by themselves alone the seeds of the cosmic magnetism \citep{Hutschenreuter:2018vkr}. During the reionization process, it is also possible
to generate weak magnetic fields with a Biermann battery effect (see for example Ref. \cite{Gnedin:2000ax}).
It has been shown how causally generated PMFs all share a common characteristic: a positive and even tilt in wavenumber space $\nB \ge 2$ \citep{2003JCAP...11..010D}, where $\nB$ is the spectral index of the power law describing PMF's scale dependence, defined in the following section. This characteristic represents a unique opportunity for constraining such fields. 

The other generation mechanisms are the inflationary ones. PMFs can be generated during inflation and, thanks to the nature of inflation, a large coherence length is not an issue. For inflationary PMFs, the main issue is the amplitude. As electromagnetism is conformally invariant, inflation dilutes everything, including the PMF amplitude. This implies that if the observed current cosmic magnetism is seeded by inflation, conformal invariance must be broken during inflation \citep{Turner:1987bw,Ratra:1991bn,Giovannini:2000dj,Tornkvist:2000js,Bamba:2003av,Ashoorioon:2004rs,Demozzi:2009fu,
Kanno:2009ei,Caldwell:2011ra,Jain:2012jy,Fujita:2015iga}, with also the possibility of further amplification during the preheating phase \citep{Finelli:2000sh,Bassett:2000aw,Byrnes:2011aa}. 
Inflationary mechanisms present additional challenges, such as the back reaction and strong coupling issues \citep{Sasaki:2022rat,Li:2022rqt,BazrafshanMoghaddam:2017zgx,Green:2015fss,Qian:2015mzl,Ferreira:2013sqa,Byrnes:2011aa,Demozzi:2009fu,Kanno:2009ei}.
PMFs from inflation can also be related to magnetic monopole generation and constraints \cite{Kobayashi:2022qpl}.
Inflationary PMFs have unbounded spectral indices, and $\nB$ is strongly related to the kind of inflationary mechanism at play in the generation of the fields. 
Thus for both inflationary and causal fields the value of $\nB$ represents a critical characteristic for inferring their nature and origin\footnote{Another characteristic spectral index is the Batchelor spectrum typical of turbulent processes \citep{10.2307/98382,1954RSPTA.247..163P,1976JAM....43..521M}.}.

How can we constrain these characteristics of PMFs and probe the physics of the early Universe and the origin of cosmic magnetism? Potentially crossing the entire history of the Universe, PMFs affect both early and late cosmological observables in direct and indirect ways; they contribute as a massless, relativistic component to the cosmological plasma, essentially affecting all the Universe's evolution at the background and perturbative levels. We will trace these effects up to the main observable for PMFs and our main interest in this work, CMB anisotropies, particularly in polarization.

The first stage of the Universe's history, where the contribution of PMFs can be indirectly observed, is the Big Bang nucleosynthesis (BBN). In this period, adding an effective extra radiation component modifies the interaction rates and the expansion rate, affecting the production of primordial elements. BBN was one of the first probes used to constrain PMFs \cite{1995APh.....3...95G,Kernan:1995bz,Cheng:1996yi,Giovannini:1997gp,Kahniashvili:2010wm,Luo:2018nth}. Although current constraints are around a fraction of a $\mu$G, two orders of magnitude weaker than the ones from other probes, BBN offers an interesting prospect for the future using GWs \citep{2022PhRvL.128v1301K}.

PMFs have a complex effect on cosmological perturbations, especially at small scales with a change in the formation of structures. We can account for effects on the matter power spectrum, together with effects on large-scale structure observables such as weak lensing and clustering, as well as the magnification bias that was studied more recently \cite{Kahniashvili:2012dy,Fedeli:2012rr,Pandey:2012ss,Kunze:2013hy,Camera:2013fva,Kunze:2021qxt,Kunze:2022mlr,ralegankar2024primordial}. 
 Always at the smallest scales, one effect that has recently gained interest (although it was already formulated a long time ago) is small-scale baryon inhomogeneities \citep{Jedamzik:2018itu,Ralegankar:2023pyx}. The presence of PMFs in the plasma can affect the evolution of baryons on the smallest scales, with the possibility of creating additional inhomogeneities, which in turn may affect recombination and CMB anisotropies. This effect has been associated with a partial relief of the Hubble parameter tension \citep{Jedamzik:2020krr,Jedamzik:2023csc,jedamzik2023cosmic}, and has been used to provide strong constraints on the PMF amplitude by using small-scale data \citep{Galli:2021mxk}. 
The effects on large-scale structure formation are quite strong and can provide tight constraints, also with future experiments such as Euclid and Rubin \citep{EUCLID:2011zbd,LSSTScience:2009jmu}, but they also represent a theoretical challenge.
The estimation of these effects requires a fully nonlinear treatment of the PMF evolution and its impact on cosmological perturbations on the smallest scales. Such a treatment is currently available only through numerical simulations, which are limited by the time requested to run each set of initial conditions, whereas a full data likelihood analysis requires several thousands of different initial conditions. Possible solutions to this issue are still in the embryonic stage and for this reason, although promising, this avenue is still very model dependent. 

Connected to the impact on large-scale structure, but observed at an earlier stage when structures are not yet fully nonlinear, the 21-cm signal \citep{Schleicher:2008hc,Shiraishi:2014fka,Gluscevic:2016gns,Minoda:2018gxj,Kunze:2019qpe,Cruz:2023rmo} offers interesting perspectives for forthcoming radio observatories.
More indirectly, but still in the domain of large-scale structure observations, we have several probes of PMFs. PMFs can have a strong impact on the Sunyaev-Zeldovich effect \citep{Sunyaev:1972eq} and in particular this can affect both clusters \citep{Tashiro:2009hx} and the intergalactic medium (IGM) \citep{Minoda:2018hiu}, with imprints on the smallest observable scales of the CMB. PMFs also affect the star-formation history \cite{Schleicher:2009zb}, with consequences for dwarf galaxy abundances \cite{Safarzadeh:2019kyq}. Another recent astrophysical probe uses the rotation measure from fast radio bursts to provide upper bounds on the possible PMF contribution to the IGM \cite{Hackstein:2019abb}. Some of the most recent probes provide a completely new avenue with direct GW detectors \cite{Saga:2018ont} or pulsar-timing arrays \citep{RoperPol:2022iel}. 
 All these new avenues to investigate the characteristics of PMFs offer interesting prospects for the future and will be crucial as complementary probes in the case of a scenario with a clear detection of signals compatible with PMFs.

One of the best probes of PMFs in the next decade is the CMB, with PMFs affecting it in several different ways and leaving imprints both on the CMB anisotropies and its absolute spectrum.
\lb will be revolutionary for CMB observations \citep{LiteBIRD:2022cnt}, providing incomparable precision measurements for not only the CMB in the standard model, but also the CMB in the presence of PMFs especially through the CMB polarization anisotropies. Our focus in this paper will be to investigate how much \lb measurements can improve our knowledge of PMFs.

Before going into details of the different imprints on CMB anisotropies, we should consider the possible kinds of intrinsic model of PMFs. The simplest form would be a homogeneous magnetic field across the Universe. Such a model---even if simple in terms of the physics of the magnetic field---is actually the most complex to generate and deal with. Indeed a homogeneous field would not be supported in the Friedmann-Robertson-Walker metric and could live only in a Bianchi-like Universe \citep{Collins:1973lda,Barrow:1985tda}. This property is one of the reasons for the strong constraints on such fields already provided by COBE and BBN \citep{Barrow:1997mj,Grasso:1996kk}, although it has been shown in Ref. \citep{Adamek:2011pr} that neutrinos can relax some of these constraints. A homogeneous field has a further effect on CMB statistics, generating correlations among different multipoles \citep{Durrer:1998ya,Naselsky:2004gm}.
The act of creating such a field is rather complex and usually the generation cannot rely on local processes, making it hard to produce without any collateral consequence on the background cosmology.
In light of this complexity, the standard model used for PMFs is usually a stochastic background, which is fully supported on the standard background cosmology and can be generated by local processes, in agreement with many PMF generation mechanisms. We will not consider a homogeneous field in this paper.

The first imprint on CMB anisotropies that we consider in this work is the cosmological perturbations of PMFs; PMFs contribute to the total energy-momentum tensor of the cosmological plasma in a unique way. They are a fully relativistic massless component, but do not contribute at the background level. This causes the generation of independent magnetically-induced modes in the scalar, vector and tensor sectors. These modes contribute to the CMB anisotropy angular power spectra in temperature and polarization through the so-called \textit{gravitational effect}. The scalar contribution is sourced by the energy density and the anisotropic pressure of PMFs and is affected by the Lorentz force on the baryons \citep{Giovannini:2004aw,Kahniashvili:2006hy,2007AIPC..957..449Y,2008PhRvD..77d3005Y,Finelli2008,2008PhRvD..77f1301G,
2008PhRvD..77l3001G,2008PhRvD..77f3003G,2010JCAP...05..022B,2011PhRvD..83b3006K}. The vector contribution is sourced by the anisotropic pressure of PMFs. The same goes for the magnetically-induced GWs \citep{1998PhRvL..81.3575S,2000PhRvD..61d3001D,Mack2002,Caprini:2001nb,2002MNRAS.335L..57S,2003MNRAS.344L..31S,2004PhRvD..70d3011L,2006AN....327..422C,Paoletti2009,Shaw:2009nf}. 
The gravitational effect provides constraints using current data of a few nanogauss (nG), with stronger constraints for specific values of $\nB$, reaching up to a few picogauss (pG) level for blue-tilted $\nB$ \cite{Paoletti:2010rx,Shaw:2010ea,Paoletti:2012bb,Planck2013params,Ade:2015cva,Zucca:2016iur,Paoletti:2019pdi}. As will be shown, the gravitational effect induces $B$-mode polarization, with contributions on large and intermediate scales, the main focus of \lb.

The second imprint on CMB anisotropy angular power spectra is dissipation of PMFs after recombination through ambipolar diffusion and magnetohydrodynamics (MHD) decaying turbulence. Dissipation injects energy and produces a magnetically-induced {\it heating} of the cosmological plasma, with major effects on CMB temperature and polarization \citep{Sethi2005,Sethi2009,2015JCAP...06..027K,2015MNRAS.451.2244C,Planck:2015zrl,Paoletti:2018uic,Paoletti:2022gsn}. \lb, with its cosmic-variance limited measurement of $E$-mode polarization on large and intermediate scales (where this effect is strong), represents one of the best opportunities for improving in this direction.
This energy injection also causes spectral distortions \citep{2000PhRvL..85..700J,Kunze2014, Wagstaff:2015jaa}. Although the signal is below the detectability level of current experiments, it is a good target for possible future spectrometers \citep{Delabrouille:2019thj,2021ExA....51.1515C}.

The third imprint is the \textit{non-Gaussian} contribution of PMFs on CMB an\-isotro\-pies and the non-negligible bispectrum generated in polarization. \lb will be pivotal, as this signal requires large sky fractions. The gravitational contribution of PMFs to cosmological perturbations is through the electromagnetic energy-momentum tensor, which is quadratic in the stochastic fields. The square of a random distribution is far from Gaussian, as is the PMF contribution to CMB anisotropies. PMFs excite all higher-order statistical moments, the bispectrum \cite{Brown:2006wv,Seshadri:2009sy,Caprini:2009vk,Shiraishi:2010yk,Trivedi:2010gi,Shiraishi:2011xvp,Shiraishi:2011fi,Shiraishi:2011dh,Shiraishi:2012sn,Shiraishi:2012rm,Shiraishi:2013wua,Ade:2015cva}, the trispectrum \cite{Trivedi:2011vt,Trivedi:2013wqa}, and so on, again with all different initial conditions contributing in different ways. 

The fourth imprint that we will discuss, again focusing on polarization on large and intermediate scales probed by \lb, is \textit{Faraday rotation}. Faraday rotation is an important effect involving light propagating through a magnetized medium, 
widely used in radio astronomy and one of the main probes of cosmic magnetism \citep{Aramburo-Garcia:2022ywn}. PMFs rotate the polarization plane of CMB photons and induce a secondary $B$-mode signal from the rotation of the $E$-modes slightly reducing the power of the $E$-modes \cite{Kosowsky:1996yc,Kosowsky:2004zh,Campanelli:2004pm,Pogosian:2013dya}. This creates a unique photon frequency-dependent effect on the CMB. This frequency dependence makes the signal brighter at the lowest frequencies and is one of the  key ingredients for distinguishing its contribution from other ones (e.g., cosmic birefringence \cite{Komatsu:2022nvu}). 

These four effects all modify the polarization pattern of CMB anisotropies, either generating additional new signals in $B$-mode polarization (the gravitational effect and the Faraday rotation), modifying the primary anisotropies  (the heating effect), or generating a non-negligible bispectrum in $B$-mode polarization. All affect polarization at large and intermediate angular scales. These are the focus and uniqueness of \lb; as only through satellite missions can we access the whole sky. \lb, as was the case for \Planck, will be capable of providing a number of different probes within the same experiments; however \lb will have the important advantage of sensitive polarization channels. For this reason, \lb represents the main future for PMFs studies with the CMB and will provide a huge improvement over current results as shown in this paper.

This work is part of a series of papers that present the science
achievable by the \lb space mission, expanding on the overview
published in Ref.~\cite{LiteBIRD:2022cnt} (hereafter \citetalias{LiteBIRD:2022cnt}). We investigate the capabilities of \lb to study PMFs through four of the main effects on CMB anisotropies.

The paper is structured as follows. Section~\ref{sec:formalism} presents the formalism used for PMFs throughout the paper. Section~\ref{sec:forecast} presents the settings of the forecasts, how data sets will be created and the different cases considered.  
In Section~\ref{sec:gravitational} we will present the gravitational effect on CMB anisotropy power spectra. We will then investigate the constraints that \lb can provide on PMFs using different assumptions on the mock data, considering both \lb alone and its combination with \Planck. We will show that \lb is capable of detecting PMFs at the level of current constraints. We will also show the interplay between PMF signals and GWs from inflation, and how \lb will be able to disentangle the two effects in most cases. In Section~\ref{sec:thermalhistory} we will present the effect of PMF dissipation on primary CMB anisotropies in both temperature and polarization. We will then  present the forecasts for \lb constraints, considering the effects of ambipolar diffusion and MHD decaying turbulence, both separately and in combination. We will also present the effects of dissipation on primordial $B$ modes from inflation and we will demonstrate that for \lb's sensitivity this is not an issue in terms of degeneracy between the two signals. Finally we will present a particular configuration of PMFs for which it is useful to combine the gravitational and dissipation effects in order to tighten the constraints on the PMF amplitude. In Section~\ref{sec:faraday} we will present the effects of Faraday rotation on the $B$-mode power spectrum and compare it with known uncertainties such as instrumental noise and foreground residuals. We will then show the dramatic improvement that \lb will provide on current constraints.
Section~\ref{sec:nongaussian} presents non-Gaussianity studies, and finally in Section~\ref{sec:conclusions} we draw our conclusions.

\section{Formalism}
\label{sec:formalism}

We model PMFs as a stochastic background, since this represents the most generic form that can be generated by local processes, such as the ones generally invoked for PMF generation.
We are interested in the PMF effects in the CMB observational window, which involves only linear scales. We neglect the possible nonlinear behaviour of the PMFs and assume the ideal MHD limit, in which the PMFs passively evolve with dilution by the Universe's expansion, and can be described by ${\vec B}^{(\mathrm{phys})}{({\vec x},\tau)}={\vec B}({\vec x})/a(\tau)^2$, where  ${\vec B}({\vec x})$ is the comoving field, $a(\tau)$ is the scale factor and $\tau$ the conformal time.\footnote{We choose the standard convention in which the scale factor is $a(\tau_0) = 1$ at the present conformal time $\tau_0$.}
These assumptions are justified in the cosmological environment \citep{Grasso:2000wj}, when the effects we will describe take place, and by the scales which we are interested in. However, non-idealities may lead to a different behaviour of PMFs on very small scales, where the evolution of the fields in time and spectral distribution under the back reaction of the fluid must be taken into account \citep{Saveliev:2012ea,Saveliev:2013uva,2014ApJ...794L..26Z,2015PhRvL.114g5001B,Brandenburg:2016odr}.

In Fourier space such a stochastic background of PMFs is described by the two-point correlation function for random fields,
\be
\Big\langle B_i({\vec k}) \, B_j^*({\vec k}')\Big\rangle=\frac{(2\pi)^3}{2} \, \delta^{(3)}({\vec k}-{\vec k}') \, \left(\delta_{ij}-\hat k_i\hat k_j\right) \, \PB (k)\,,
\label{PSpectrum}
\ee
where $\PB (k)= \AB \, k^{\nB}$ using the assumption of a power law power spectrum for the fields.\footnote{For the Fourier transform and its inverse, we use
\ba 
Y ({\vec k}{,}\tau) =  \int {\rm d}^3 x \, \mathrm{e}^{i {\vec k} \cdot {\vec x}} \,Y ({\vec x}{,}\tau)\,, \qquad Y ({\vec x}{,}\tau) = \int \frac{ {\rm d}^3 k}{(2 \pi)^3} \, \mathrm{e}^{-i {\vec k} \cdot {\vec x}} \,Y ({\vec k}{,}\tau) \,, \nonumber 
\label{Fourier}
\ea
where $Y$ is a generic function.} In this paper we do not consider a helical component in the PMFs, which we leave for future studies.
PMFs are characterized by two parameters, $A_\mathrm{B}$ and $\nB$.
The associated amplitude of the PMFs, which will be the actual variable of the treatment, can be expressed with different conventions. Two of the main parametrizations used are the comoving fields smoothed on a comoving scale $\lambda$ (where $\lambda =1$\,Mpc, for example):
\ba
B^2_\lambda &=& \int_0^{\infty} \frac{d{k \, k^2}}{2 \pi^2} \, \mathrm{e}^{-k^2 \lambda^2} \PB (k) = \frac{\AB}{4 \pi^2 \lambda^{\nB+3}} \, \Gamma \left( \frac{\nB +3}{2} \right)\,,
\label{gaussian}
\ea
and the amplitude of the stochastic PMFs expressed using their root mean square (rms):
\begin{equation}
\langle{B^2({k})}\rangle=\frac{\AB }{2\pi^2}\frac{k_{\mathrm D}^{\nB +3}}{\nB +3}\,,
\label{mean-squared}
\end{equation}
where $k_{\mathrm D}$ is the damping scale defined in \autoref{kd_def}.
The first parametrization is a common choice that aims to compare the amplitudes of the comoving fields with the ones observed in the large-scale structure such as clusters and voids. The second parametrization is a more generic choice in physical terms, but makes more difficult a direct comparison with the cosmic magnetism observations.
We will use both parametrizations depending on the analysis. Note that the constraints depend on the parametrization they are expressed with, therefore, comparison among different results should always consider the different  parametrizations used.

We need to model PMF dissipation on small scales. Magnetic effects survive the Silk damping and the magnetically induced perturbations provide strong effects precisely in the region where primary perturbations are suppressed. On smaller scales, a fraction of the Silk's one, PMFs and their effects are also damped. The modelling of this damping is one of the open points in PMF cosmology. It would technically require a simulation based approach to infer the decaying rate \citep{2014ApJ...794L..26Z,2015PhRvL.114g5001B}. However this approach does not allow for a parameter space exploration which provides constraints from CMB data.
When considering the 
damping of the magnetosonic and Alfven waves, the damping scale can be expressed as \citep{Jedamzik1998, Subramanian1998,Mack2002}:
\ba
k_\mathrm{D}&=&(5.5 \times 10^4)^{\frac{1}{\nB+5}} \left(\frac{B_{\lambda}}{\mathrm{nG}}\right)^{-\frac{2}{\nB +5}} 
\left(\frac{2\pi}{\lambda/{\mathrm{Mpc}}}\right)^{\frac{\nB +3}{\nB +5}} h^{\frac{1}{\nB+5}} \left(\frac{\Omega_{\mathrm b} h^2}{0.022}\right)^{\frac{1}{\nB +5}}\Big|_{\lambda=1\,{\mathrm Mpc}} \mathrm{Mpc}^{-1}\,.
\label{kd_def}
\ea
See Ref.\cite{2018JCAP...08..034B} for an alternative based on the rms of the field. We assume a sharp cut off in the power spectrum at this damping scale. This assumption, although approximate, is the only one allowing us to have analytical expressions for the contributions to the cosmological perturbations. Alternative models for the damping invoke, for example, a Gaussian damping \cite{Kunze2014,2015JCAP...06..027K} for which the energy-momentum tensor convolutions are not solvable analytically.

%
\section{Forecasts setting}
\label{sec:forecast}
The central objective of this work is to provide forecasts for the future capabilities of \lb to constrain PMFs. In order to pursue this objective and provide a complete set of forecasts, we will consider different types of mock data sets in our analyses, especially those more sensitive to data contamination issues. For these cases we will go from the simplest case, growing in complexity up to the most realistic case.

For the power spectrum based analyses including the treatments of gravitational and heating effects, we will use  an inverse Wishart likelihood \cite{knoxlike,Hamimeche:2008ai}. This kind of approach is typically used in idealized forecasts as in \cite{Core_inf} (something similar is also applied in \cite{Wolz:2023lzb}):
\be
\chi_\mathrm{eff}^2 = - 2 \ln {\cal L} = 
\sum_{\ell} (2\ell+1) f_\mathrm{sky} \left\{ \mathrm{Tr} [ \hat{{\bf C}}_\ell \bar{{\bf C}}_\ell^{-1}] +
\ln |\hat{{\bf C}}_\ell \bar{{\bf C}}_\ell^{-1}| - n \right\},
\label{like}
\end{equation}
with theoretical $\bar{{\bf C}}_\ell$ and observed $\hat{{\bf C}}_\ell$ covariance matrices. The covariance matrix depends on the power spectra 
$C_\ell^{XY}$, where $X$ and $Y$ 
can take the values $T, E$ (with $n=2$) or $T, E, B$ (with $n=3$).

We simulate the instrumental white noise with an inverse variance weighting of the \lb  instrumental characteristics \citepalias{LiteBIRD:2022cnt},
\be
N_\ell^{\mathrm XX, \rm inst} = \left[ \sum_\nu \frac{1}{w_{X \, \nu}^{-1} \exp \left[ \ell (\ell+1) \frac{\theta^2_{{\rm FWHM} \, \nu}}{8 \ln 2} \right]} \right]^{-1} \,,
\label{eq:noise}
\ee
where we convolve for each channel the sensitivity $w_{X \, \nu}$ with the beam resolution given at the full width half maximum $\theta_{\mathrm{FWHM}\nu}$.

We consider only the seven central frequency channels from 78 to 195\,GHz which are dominated by the CMB. We assume that the foreground dominated channels will be mostly used for component separation of the signal. This guarantees that we do not introduce biases in the frequency channel weighting of the white noise, which depends only on instrumental characteristics and is insensitive to the importance of the CMB signal at that frequency.\footnote{Typically higher frequency channels have the best angular resolution but the CMB is strongly subdominant, if ever present at all, so they are not useful for the CMB outside a pure component separation role.} 

We simulate all three components in temperature and polarization $TT$-$EE$-$BB$ and the correlation $TE$ up to $\ell_{\rm max}=1350$ when \lb alone is used. When in combination with a high multipole data set, we cut the maximum multipole at the crossing of the signal-to-noise ratio of the two experiments considered. 
For $TT$-$EE$-$TE$ we assume a 70\,\% sky fraction unless otherwise stated, whereas for $BB$ the sky fraction is reduced to 49.5\,\% due to the inclusion of the residuals after component separation that simulate a realistic data set \citepalias{LiteBIRD:2022cnt}.


We will consider three cases, which we describe in the following, independently of the analyses they are applied to.
\begin{itemize}
\item \textit{Ideal}: Only white instrumental noise and cosmic variance are implemented.
\item \textit{Baseline}:  This is the standard case common to all the analyses presented in the paper and reflects the methodology applied in Sect.~2 of Ref. \citetalias{LiteBIRD:2022cnt}. We will consider instrumental noise in $TT$-$EE$-$TE$, whereas for BB we will also include the statistical foreground residuals in the noise out to $\ell=191$ and use the post-component-separation noise in this region (as opposed to pure white noise for $\ell>191$). Considering only the contribution of statistical foreground residuals to the noise, we are implicitly assuming that we are perfectly cleaning the signal. This contribution to the noise is unavoidable once a component-separation pipeline is applied. Different pipelines can reduce this noise, but usually this is done at the cost of increasing the bias due to the foreground contamination. It is therefore necessary to find a trade off between the two and we decided to apply the same approach used in Ref. \citetalias{LiteBIRD:2022cnt}. The two separate treatments between multipoles below and above 192 of the noise in $BB$ mimic a hybrid likelihood approach where for large and intermediate scales we consider the component separated spectra whereas for smaller scales (where the Galactic contamination is subdominant) we can directly use cross-spectra, which should reduce the noise.
In the baseline approach the lensing $BB$ signal is also considered as an additional noise source and is set to zero in the theoretical angular power spectrum computed by the Einstein-Boltzmann code. This approach mimics a perfect modelling of the lensed $B$-mode power spectrum.
\item \textit{Realistic}: We will consider a setting of the typical data analysis pipeline, namely, we receive a sky signal after processing and component separation that contains the lensing signal, a residual bias from foreground cleaning and a systematic bias, together with a boosted noise from the component separation that includes statistical foreground residuals.
We fit these contamination signals with a template reproducing our knowledge of the actual signal with a nuisance amplitude that is used to marginalize the other cosmological parameters over the contamination. In the simplest case we consider only the lensing $BB$, which is fitted from the theoretical power spectrum derived from the cosmological model at every step of Markov Chain Monte Carlo (MCMC). For the foreground and systematic bias,  we instead directly consider the fiducial signal as a template and vary only the amplitudes, implicitly assuming that we will have a perfect knowledge of the power spectrum shape of these signals.
\end{itemize}
In this paper, we will not consider any other systematics but those presented in Ref. \citetalias{LiteBIRD:2022cnt}. We assume that temperature and $E$-mode polarization are free of systematics. For a satellite such as \lb, which is optimized for polarization and relies on a half-wave plate modulator (see \cite{LiteBIRD:2020khw} for technical details), a possible concern is the contamination of large scales by the $1/f$ noise for temperature anisotropy measurements. To address this specific concern, we have tested all the PMF effects that depend also on temperature anisotropies (namely the gravitational effects and the impact on the ionization history) against preliminary estimates of the $1/f$ effect based on \lb simulations. The results show a negligible effect on the constraints that we derive on PMFs.

\begin{figure}[t!]
    \centering
    \includegraphics[scale=0.50]{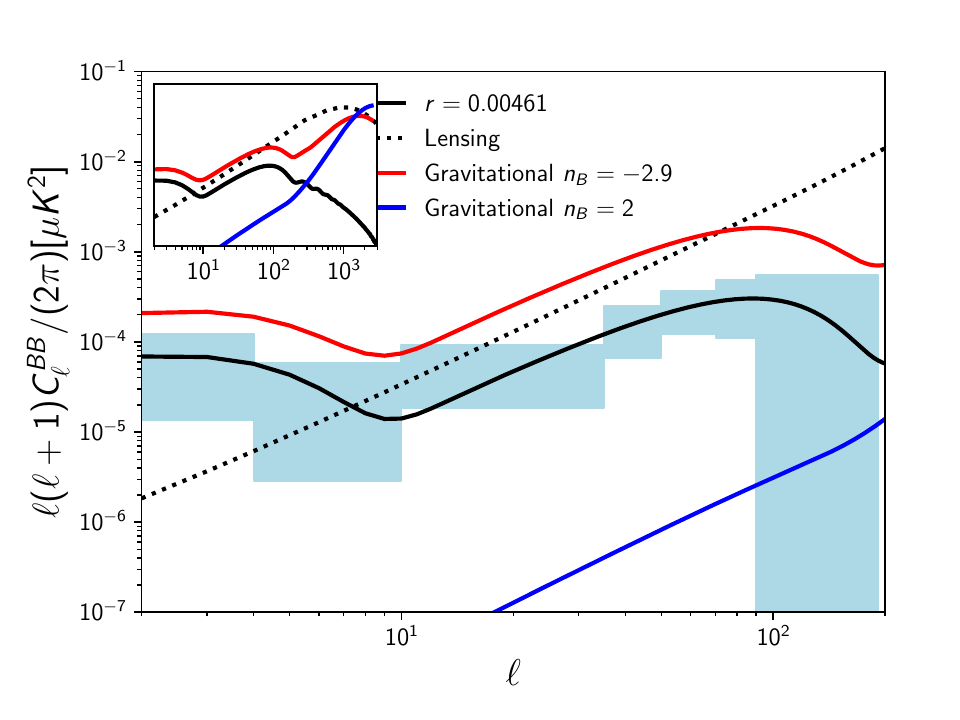}
    \includegraphics[scale=0.53]{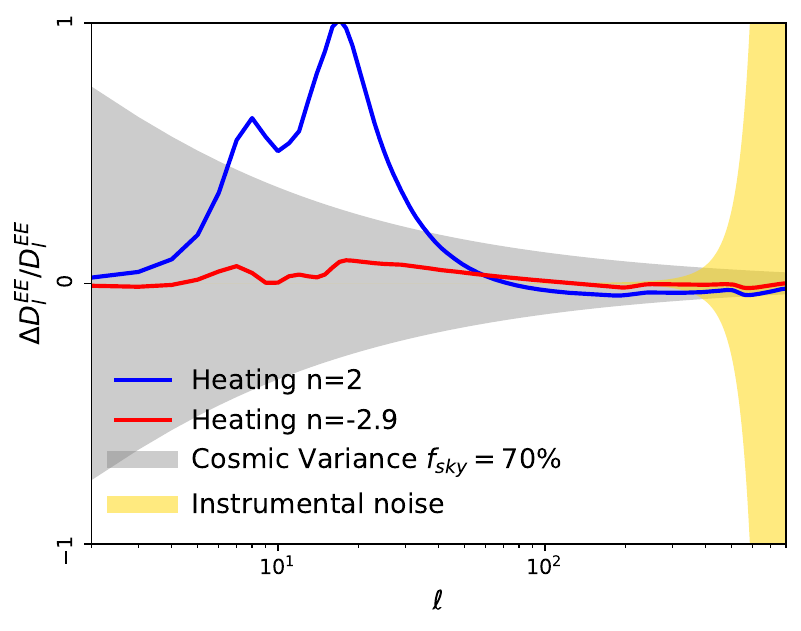}
    \caption{(Left) The comparison of the gravitational effect on $B$ modes created by PMFs for $\nB=2$ in blue and $\nB=-2.9$ in red, with the \lb error bars. In black solid and dotted lines, we show the primary CMB assuming a $R^2$-like model of inflation and the lensing contribution, respectively. In the inset we show the same curves but including also the small scales which are not observed by \lb. (Right) The heating effect on $E$ mode polarization, expressed in the relative differences to the $\Lambda$CDM model, compared with the instrument noise and cosmic variance. Again in red we show $\nB=-2.9$ and in blue $\nB=2$.}
    \label{CL_ptep}
\end{figure}
In \autoref{CL_ptep} we show an example of the signal in $BB$ coming from the gravitational effect compared with the error bars predicted for \lb \citepalias{LiteBIRD:2022cnt}. We show two representative cases that will be investigated in detail later, the minimum index for causally generated fields $\nB=2$ and the almost scale invariant case $\nB=-2.9$ that represents the most infrared spectral index that we can consider while keeping the field density finite unless we insert an infrared cut off. In the right panel we show the relative differences for the heating effect on the $E$-mode polarization. The details of these effects will be described in the following two sections.

For our fiducial models we will assume \Planck 2018 baseline marginalized results \citep{Planck:2018vyg}, as the non-magnetic underlying cosmological model. Our parameters are specifically baryon and cold dark matter densities $\Omega_b h^2=0.0224$, $\Omega_c h^2=0.1202$, angular scale of the sound horizon $\Theta=1.0409$, reionization optical depth $\tau=0.0544$, scalar spectral index $n_{\rm s}=0.9649$ and amplitude of scalar fluctuations $\log[10^{10}A_s]=3.045$. For both the gravitational and heating effects we will vary all the cosmological parameters together with the magnetic ones for which we use flat priors [0,10] for the amplitude ([0,1000] for the rms parametrization) and [-2.9,3] for $\nB$ in the gravitational effect, and [0,4] for the amplitude and [-2.9,2] for $\nB$ in the heating effect (due to the very powerful effect of positive $\nB$ which limits the numerical stability to maximum $\nB=2$). For the gravitational effect we will assume only massless neutrinos (contrary to the usual minimal mass of 0.06\,eV adopted in \Planck and elsewhere in the paper)  because we do not account for the large-scale modification to the magnetically induced modes due to neutrino mass. This is a subdominant effect, but we prefer to coherently treat magnetic and non-magnetic perturbations.

Due to the nature of some of the PMF effects it is useful to include data on the smaller scales that are inaccessible to \lb. We complement \lb with \Planck data in temperature and $E$-mode polarization in some of the analyses. In order to use the same simulated sky, and the same likelihood treatment for the two data sets, we do not use the real \Planck data likelihood, but rather we simulate a mock dataset up to $\ell=2700$ using \Planck instrumental characteristics, manipulating the simulated noise (with a series of boosting factors in different multipole ranges) in order to reproduce similar uncertainties in the cosmological parameters (as in Ref. \cite{Euclid:2021qvm}) from the \Planck CMB-only baseline \citep{Planck:2018vyg}. In order to avoid issues in the cross-correlation between the two experiments we avoid overlapping multipoles, cutting \lb at $\ell =800$ and starting the \Planck data set at $\ell  = 801$.

\section{Gravitational effect}
\label{sec:gravitational}
In this section we will study the gravitational effect of PMFs. This effect contributes to all the CMB angular power spectra in temperature and polarization and its main area of improvement with respect to current constraints is given by the accurate measurement of the  $B$-mode polarization. For this reason, \lb represents one of the best datasets to improve such constraints.

PMFs contribute as a massless fully relativistic component in the cosmological fluid and their energy-momentum tensor components are usually assumed to be first order on the same footing as cosmological perturbations:
\ba
\kappa^0_0&=&-\rho_B=-\frac{B^2({\vec x})}{8\pi a^4(\tau)}\,;\\
\kappa_i^0&=&0\,;\\
\kappa_j^i&=&\frac{1}{4\pi a^4(\tau)}
\left(\frac{B^2({\vec x})}{2}\,\delta_j^i-B_j({\vec x})\,B^i({\vec x})\right)\,.
\ea
The electromagnetic energy-momentum tensor sources magnetically induced perturbations:
\be
\delta G_{\mu\nu}=8\pi G \Big( \delta T_{\mu\nu}+\kappa_{\mu\nu} \Big)\,,
\ee
where we use natural units.
All the projections are excited by PMFs, namely scalar, vector and tensor perturbations. Scalar magnetic perturbations are sourced by the magnetic energy density and anisotropic pressure, with a contribution from the Lorentz force on baryons. Vector modes are fully sourced by the Lorentz force and anisotropic pressure; as primary vector modes are rapidly decaying in the standard model, PMFs represent the only source of this type of perturbations. Finally magnetically-induced tensor modes are sourced by the tensor projection of the anisotropic pressure.

The different modes are generated with different initial conditions, depending on whether they are compensated, passive, or inflationary. The compensated modes are the solution of the Einstein-Boltzmann equation system with their name coming from the compensation between the PMF energy-momentum tensor components and the fluid perturbations. This initial condition does not contribute to the total curvature perturbation at first order \cite{Giovannini:2004aw,Finelli2008,Paoletti2009}. The passive modes are relic modes in scalar and tensor projections, which result from the matching of initial conditions at neutrino decoupling. Before neutrino decoupling the additional magnetic anisotropic pressure sources logarithmic modes, which are then suppressed by the compensation from neutrino anisotropic pressure after decoupling; however, a residual mode in the form of an offset in the standard primary mode survives and the dependence on the magnetic anisotropic pressure power spectrum (which as in the compensated mode substitutes the primordial fluctuation power spectrum) imprints distinctive shapes \cite{Lewis:2004ef,Shaw:2009nf,Shaw:2010ea}.
\begin{figure}[t!]
    \centering
    \includegraphics[width=0.5\textwidth]{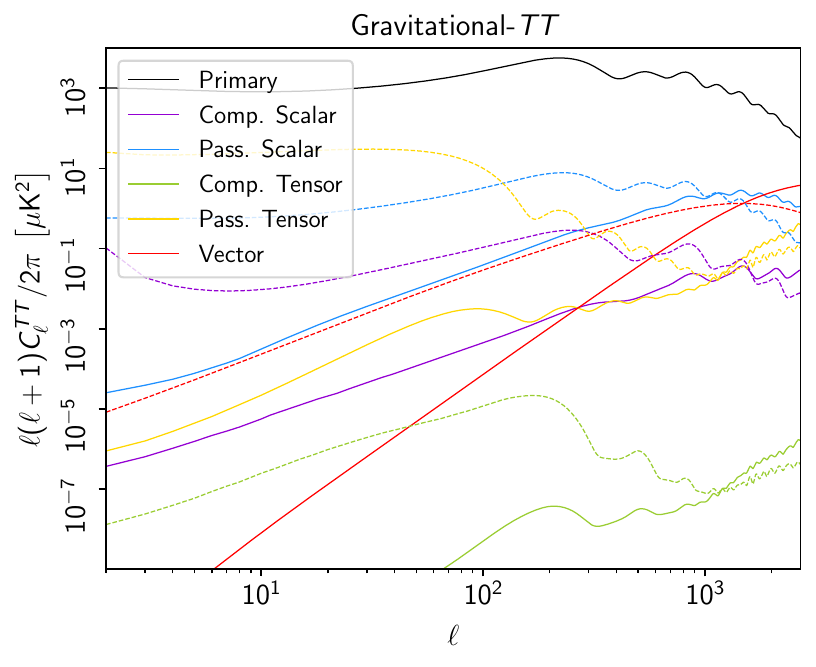}\includegraphics[width=0.5\textwidth]{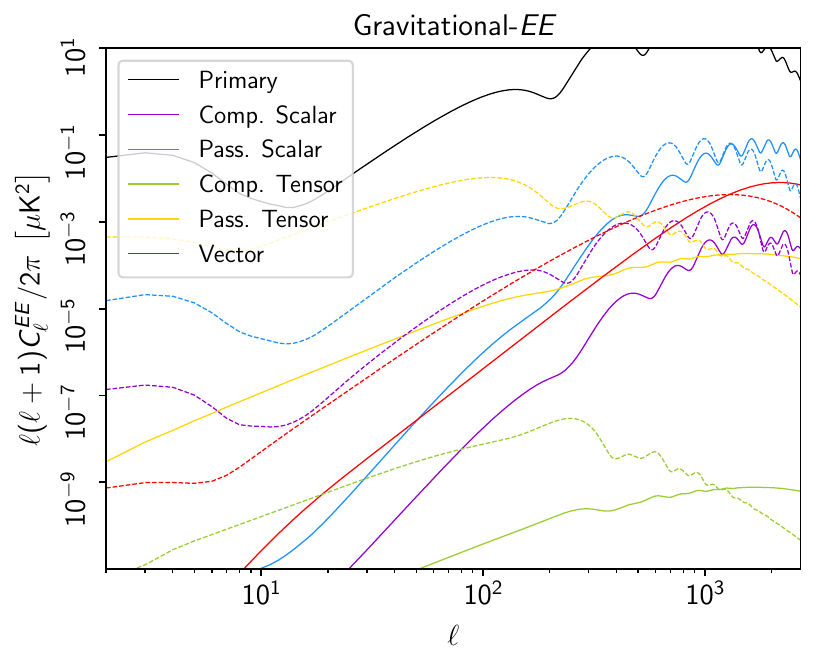}\\
        \includegraphics[width=0.5\textwidth]{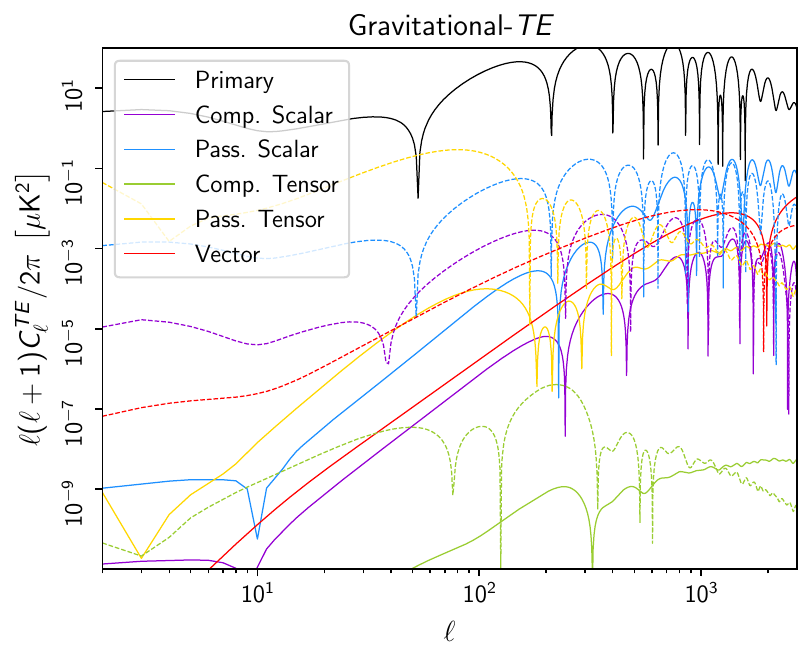}\includegraphics[width=0.5\textwidth]{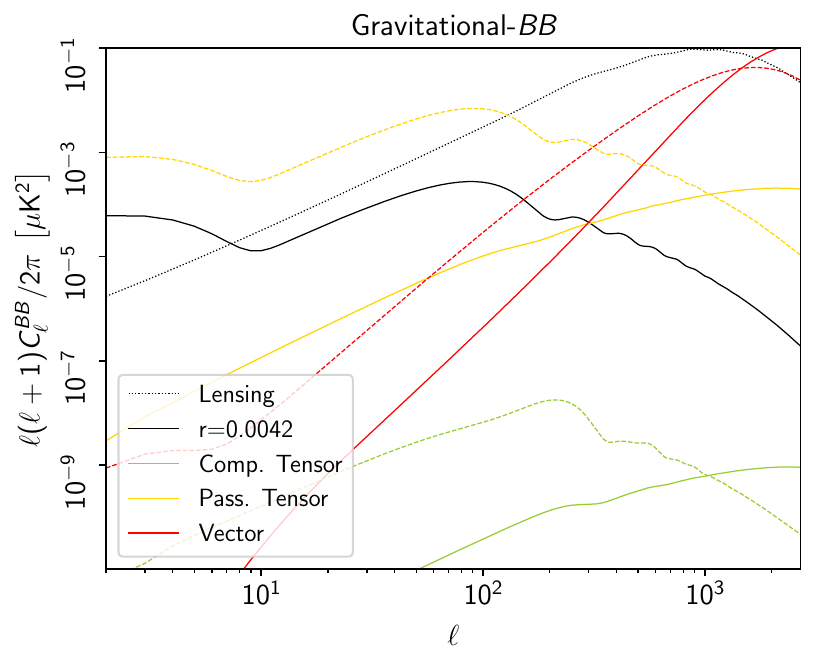}\\
    \caption{Temperature and polarization power spectra of magnetically induced modes, for $\nB=2$ (solid lines) and $\nB=-2.9$ (dashed lines). The amplitude values are compatible with the limits derived from current experiments in Ref. \citep{Paoletti:2019pdi}. In black are shown the standard non-magnetic modes.}
    \label{GRav}
\end{figure}
\noindent
There are initial conditions related to the inflationary generation mechanism. In this case the mode has a shape similar to the passive one, but its amplitude depends on the coupling taking place during inflation \cite{Bonvin:2011dt,Bonvin:2013tba}. We will not consider this initial condition in this work, since we prefer to maintain an agnostic approach with respect to the generation mechanism of the fields.

For the full theoretical treatment of scalar, vector and tensor magnetically induced perturbations we rely on Refs. \citep{Lewis:2004ef,Finelli2008,Paoletti2009,Shaw:2009nf,Paoletti:2010rx,Shaw:2010ea,Paoletti:2012bb,Planck:2015zrl} and use the code developed in Ref. \cite{Paoletti:2019pdi}. In \autoref{GRav}, we compare the magnetically-induced modes and the standard primary perturbations with the same underlying background cosmology. We plot the two extremes of the range of $\nB$ we consider, namely $\nB=2$ and $\nB=-2.9$, with the two different amplitudes compatible with current constraints from real data \citep{Paoletti:2019pdi}. The dominant contribution on large angular scales is given by passive modes and in particular the tensor passive mode. On small scales the dominant contribution is given by the vector modes, which with their distinctive shapes dominate regardless of $\nB$.

\begin{figure}[t!]
    \centering
    \includegraphics[width=0.5\textwidth]{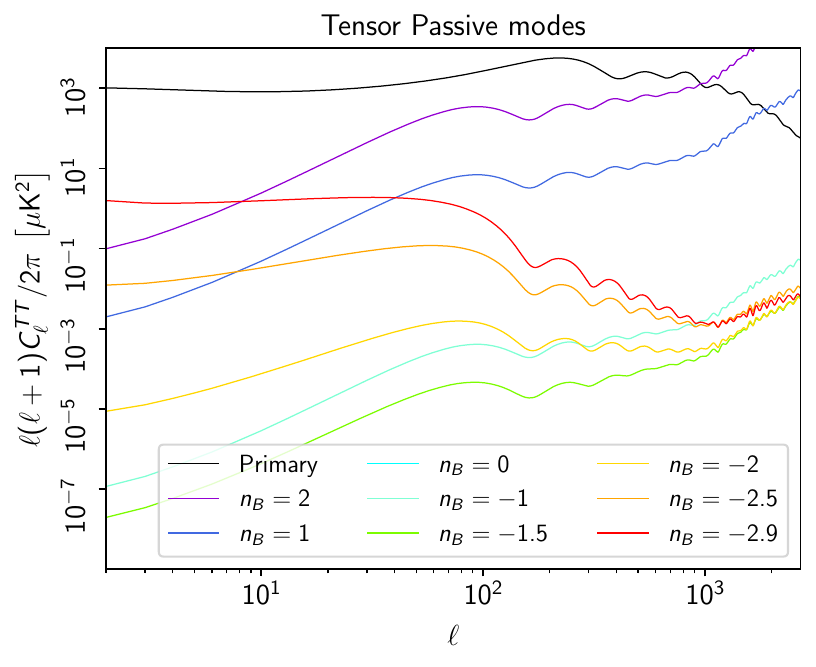}\includegraphics[width=0.5\textwidth]{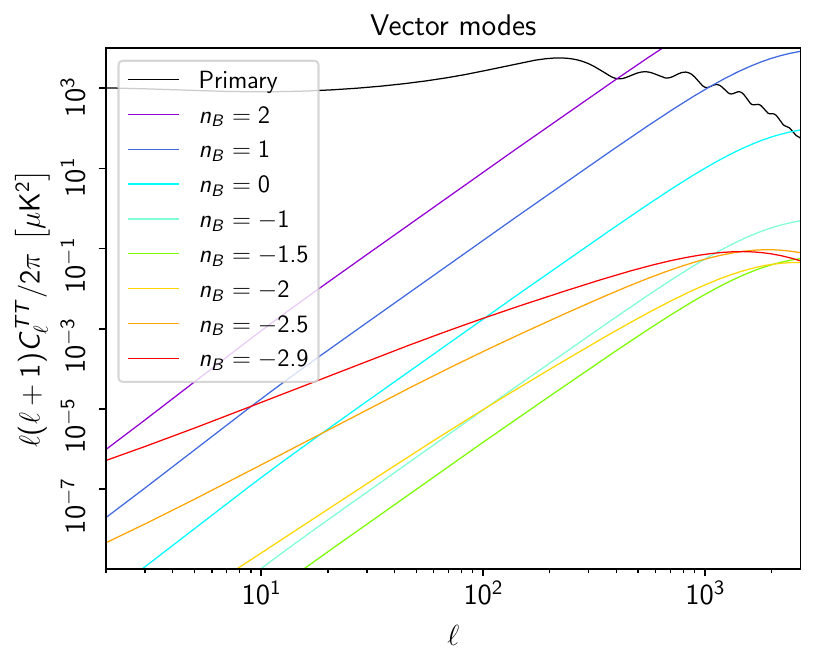}\\
     \caption{Dependence of magnetically-induced mode on $\nB$. We present the main dominant modes, the tensor passive modes (left) and the vector modes (right). Other modes have similar variations. The color code is shown in the legend.}
    \label{GravN}
\end{figure}

In \autoref{GravN}, we show the dependence of the angular power spectra on $\nB$, presenting the results for the temperature channel (although polarization shows a similar behavior). The peculiar spectral dependence is related to the energy-momentum tensor of PMFs, which is dominated by a white noise term for $\nB>-1.5$, whereas for lower $\nB$ it is dominated by the infrared term which goes as $k^{2\nB+3 }$ \citep{Finelli2008,Paoletti2009}. The index $\nB=-1.5$ represents the transition between the two regimes and provides the minimum of the contribution of PMFs to the CMB angular power spectra. This behavior is reflected in the constraints on PMF amplitude which are weaker for this index.
\subsection{Constraints from the gravitational effect}
We will now explore the constraints that \lb can provide on PMF characteristics by means of the gravitational effect. For this effect we will employ all the CMB parity-even channels, $TT$, $TE$, $EE$, $BB$, with different assumptions for the mock data, from the simplest to the most realistic.
\subsubsection{Ideal settings}
We begin with the ideal case: instrumental white noise only for all the channels and the lensing BB signal as an additional noise contribution.
This setting represents the maximum level of constraining power possible were the instrument and data analysis perfect and the sky only made of CMB in both temperature and polarization.
\begin{table}[t!]
\center
\begin{tabular}{|c c c c|}
\hline
Data&\lb-ideal&\lb-baseline&\lb-baseline+\Planck\\
\hline
$\nB$ & $B_{1\,\mathrm{Mpc}} \, [\mathrm{nG}] $&$B_{1\,\mathrm{Mpc}} \,[\mathrm{nG}]$&$B_{1\,\mathrm{Mpc}} \, [\mathrm{nG}] $ \\
\hline
Marginalized &$<2.9$ & $<2.9$ &  $<2.2$\\
\hline
 2&$<0.005$ & $<0.005$&$<0.003$\\
\hline
 1& $<0.06$ &$<0.06$& $<0.031$\\
\hline
 0& $<0.50$ &$<0.51$&$<0.27$\\
\hline
 $-1$& $<2.4$ &$<2.4$& $<1.5$\\
\hline
 $-2$& $<2.5$ &$<2.7$& $<2.3$\\
\hline
 $-2.9$&$<0.6$ & $<0.8$& $<0.8$\\
\hline
\end{tabular}
\caption{\label{tab:GLB}
Constraints on the PMF parameters for \lb and the combination of \lb with \Planck, for the case marginalized over $\nB$ and for each of the $\nB$.}
\end{table}
The results are shown in the second column of \autoref{tab:GLB}. 

\begin{figure}[t!]
    \centering
    \includegraphics[width=\textwidth]{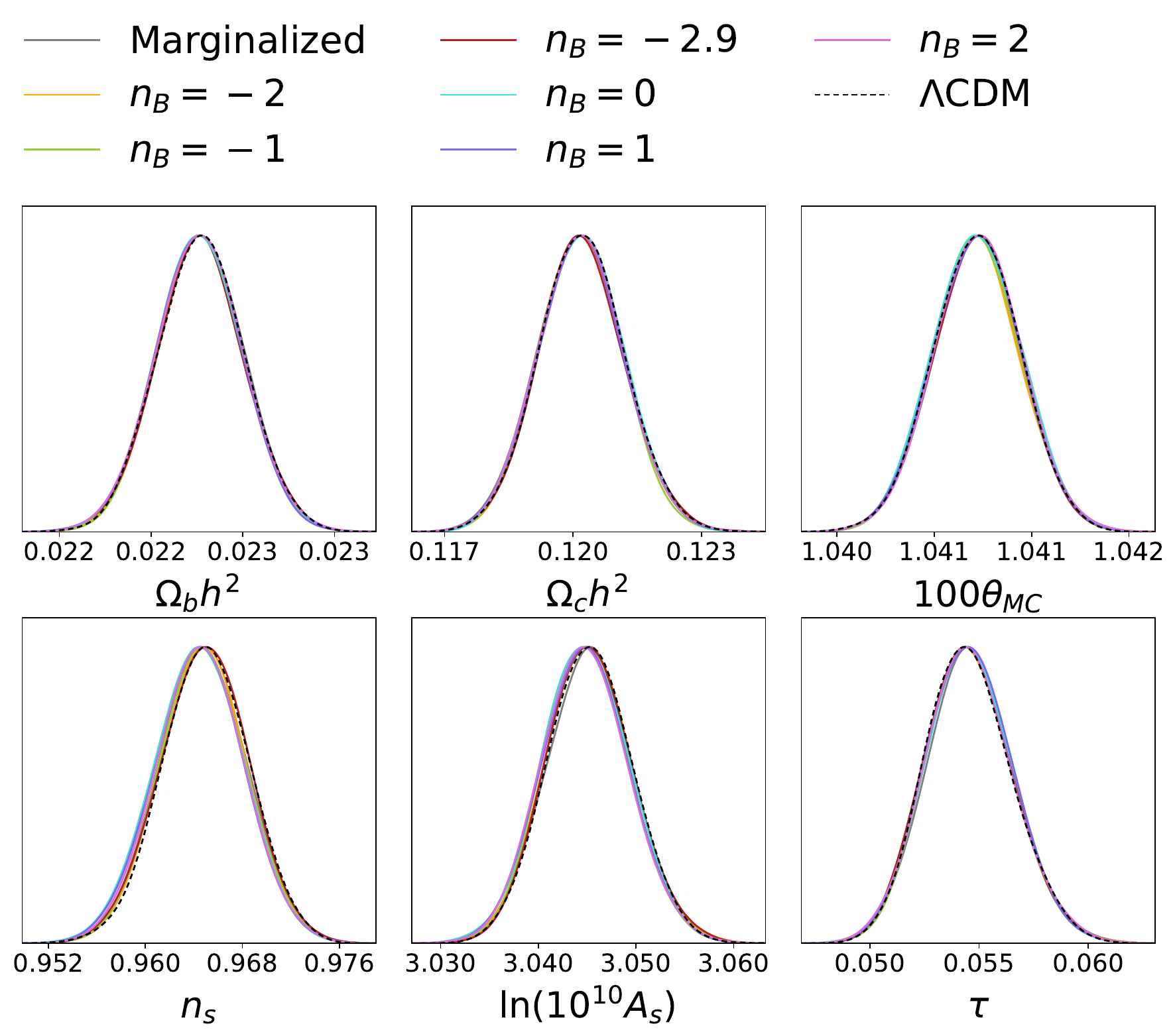}
    \caption{Posterior distributions for the standard 6 $\Lambda$CDM model parameters with and without the PMF contribution.}
    \label{fig:G1D_LB}
\end{figure}

Current constraints on the PMF amplitude from real data, with the same assumptions made here, are provided by the combination \Planck 2018+BICEP/KECK 15 (BK15). The limits for different PMF configurations are: $B_{1\,\mathrm{Mpc}}<3.5$ nG (95\% C.L.) when marginalized over $\nB$; $B_{1\,\mathrm{Mpc}}<0.006$ nG for $\nB=2$; $B_{1\,\mathrm{Mpc}}<2.5$ nG for $\nB=-2.9$ \citep{Paoletti:2019pdi}. \lb alone in this ideal setting is capable of improving the constraints from \Planck and BK15 that used the high multipoles from \Planck in temperature and $E$-modes.
We also investigate possible correlations with standard $\Lambda$CDM parameters. The comparison among the different posterior distributions is shown in \autoref{fig:G1D_LB}. We find that with the \lb sensitivities no significant 
bias is observed in the gravitational effect.
\subsubsection{Baseline case}
\begin{figure}[t!]
    \centering
    \includegraphics[width=0.97\textwidth]{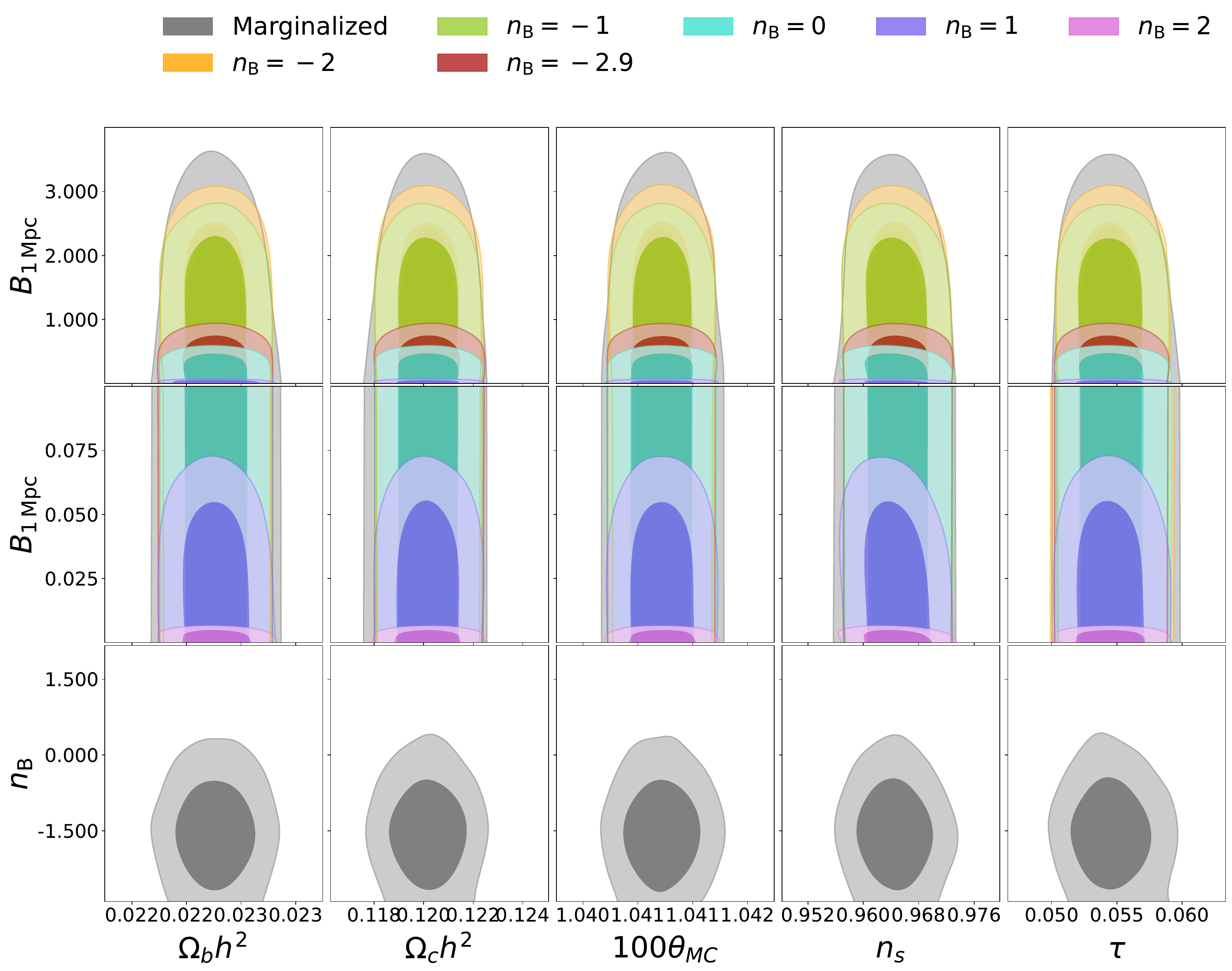}\\
    \caption{Two-dimensional posterior distributions for the cosmological and magnetic parameters for the case of \lb. In the top row we show the amplitude of the fields versus all the other cosmological parameters, while in the second row we show a focus on the higher spectral index constraints. The third row presents the two dimensional distributions for the spectral index.}
    \label{fig:G2D_LB}
\end{figure}

We now move to a non-ideal but optimal case, which we treat as the baseline. We consider the contamination of $B$-mode polarization with foregrounds, but we also assume a perfect cleaning by component separation.
We are left in $B$-mode polarization with a boosted noise and the contribution of statistical foreground residuals up to $\ell=191$ \citepalias{LiteBIRD:2022cnt}. 
The results are shown in the third column of \autoref{tab:GLB}. We find that component separation only minimally affects the results and the effect is limited to smaller $\nB$. This is due mainly to the fact that apart from the lowest multipoles the major contributor to the noise is given by lensing and this reduces the impact of foreground residuals except for very small $\nB$ where the lowest multipoles (due to the shape of the magnetically-induced power spectra) are most relevant. In \autoref{fig:G2D_LB} we show the two-dimensional posterior distributions for the correlation of magnetic and standard $\Lambda$CDM cosmological parameters. We do not see any significant correlations except for a slight impact on the scalar spectral index ($n_{\rm s}$), especially for PMF configurations that have more power on small scales. The weaker constraints belong to the case $\nB=-1.5$ because of the shape of the energy-momentum tensor of PMFs. As shown in Refs. \cite{Finelli2008, Paoletti2009} the magnetic source terms transition from a simple white noise rescaling to an infrared dominated spectrum at $\nB=-1.5$. As a result, their contribution to the CMB angular power spectra in \autoref{GravN} is the smallest for $\nB=-1.5$. This explains a preference for $\nB=-1.5$ in the constraints shown in the bottom panels of \autoref{fig:G2D_LB}, even when no PMF signal is considered in the fiducial model.

We find that \lb improves on current constraints from \Planck when marginalization over $\nB$. An improvement of a factor  of 3 is observed for the almost scale invariant case. This is expected, since the almost scale invariant case is mostly constrained by the $B$-mode polarization signal on large and intermediate scales, where \lb substantially improves things with respect to \Planck. 

\begin{figure}[t!]
    \centering
    \includegraphics[width=0.6\textwidth]{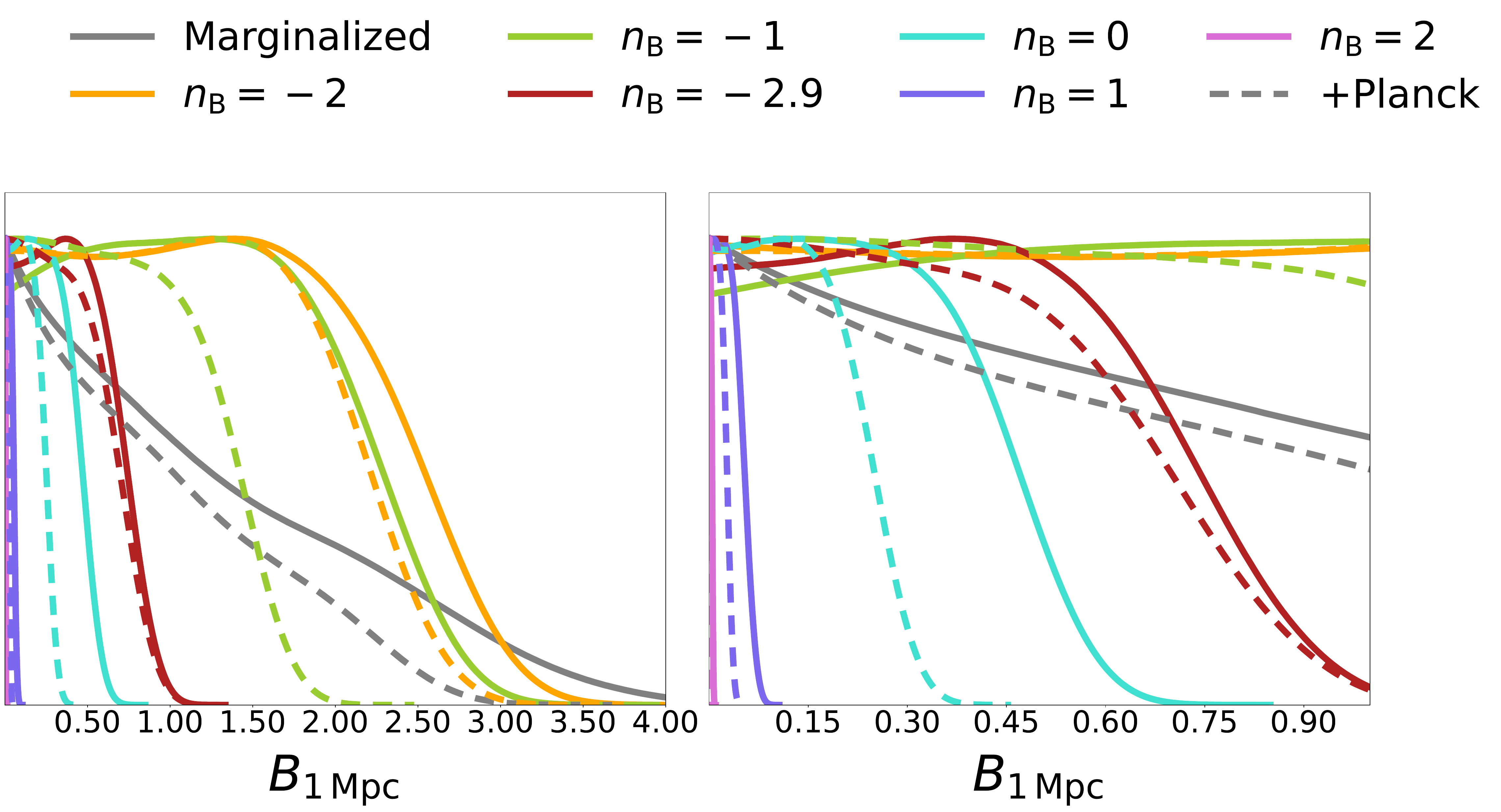}
    \includegraphics[width=0.33\textwidth]{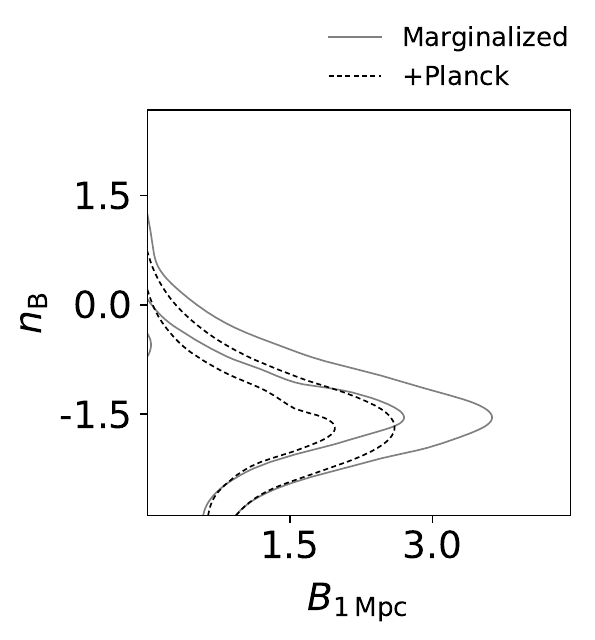}
    \caption{One-dimensional posterior distributions for the magnetic parameters for the case of \lb combined with \Planck (dashed lines) compared with \lb alone (solid) and the two-dimensional contours for the marginalized case (right panel). The middle panel is the focus of the left panel to highlight positive $\nB$.}
    \label{fig:G1D_PLB}
\end{figure}

In order to extend \lb's angular scale range we complement the \lb data set with the simulated \Planck data as described in the previous section.
The constraints are presented in the fourth column of \autoref{tab:GLB} and in \autoref{fig:G1D_PLB}. The addition of \Planck at the higher multipoles demonstrates the complementarity between small and large scales for PMFs. The combination of \lb and \Planck improves the results by a factor of 2 for the blue spectral indices, thanks to the vector modes, whereas there is minimal to no improvement for the infrared indices which are dominant on large scales that are fully probed by \lb. These results offer a very good potential for the combination of \lb and future high resolution ground-based observatories.

\subsubsection{Realistic cases}
The forecasts presented in this subsection are a semi-idealistic representation of what will be the development of the real data pipeline. We have so far assumed that we are capable of perfectly cleaning the foreground residuals and the lensing $BB$ power spectrum, which leave imprints of their presence only as an increased noise power spectrum. In this subsection we use a more realistic data-oriented approach, where we still consider the boosted noise and statistical residuals from component separation at the lowest multipoles, but we now consider the lensing $BB$-signal as a sky component that is varied with the rest of the cosmological model and marginalized over a nuisance amplitude of the lensing $BB$ spectrum. We then increase the complexity by also including the foreground residual bias and the systematic biases, all again varied with a nuisance parameter and marginalized over as in Ref. \citetalias{LiteBIRD:2022cnt}. Within this framework we will consider both non-magnetic fiducials and a fiducial with non-zero primordial tensor modes from inflation.

\begin{table}[t!]
\center
\begin{tabular}{|c| c c |c c |c c|}
\hline
\multicolumn{7}{|c|}{\lb with lensing marginalization}\\
\hline
\multicolumn{1}{|c|}{Lensing} &\multicolumn{2}{|c|}{Full}&\multicolumn{2}{|c|}{43\% delensed}&\multicolumn{2}{|c|}{80\% delensed}\\
\hline
$\nB$ &$B_{1\,\mathrm{Mpc}} \, [\mathrm{nG}] $  & $A_{\mathrm{Lens}}^{BB}$&$B_{1\,\mathrm{Mpc}} $  & $A_{\mathrm{Lens}}^{BB}$ & $B_{1\,\mathrm{Mpc}}$  & $A_{\mathrm{Lens}}^{BB}$ \\
\hline
Marg. &  $<3.5$& $0.985\pm0.020$&  $<3.3$& $0.561 \pm 0.012$& $<3.3$ & $0.196_{-0.006}^{+0.007} $\\
\hline
 2& $<0.007$& $0.985\pm0.020$&  $<0.007$& $0.560\pm 0.013$& $<0.006$ & $0.195 \pm 0.006$\\
\hline
 1& $<0.075$& $0.985_{-0.020}^{+0.021}$&  $<0.072$& $0.560\pm 0.013$& $<0.067$ & $0.195\pm 0.006$\\
\hline
 0& $<0.56$& $0.986_{-0.021}^{+0.020}$  &  $<0.54$& $0.562\pm 0.012$& $<0.51$ & $0.196\pm 0.006$\\
\hline
 
 $-1$& $<2.75$& $0.986_{-0.020}^{+0.021}$&  $<2.67$& $0.561\pm 0.013$& $<2.58$ & $0.195\pm 0.006$\\
\hline
 $-2$& $<3.17$& $0.985\pm0.021$&  $<3.06$& $0.560\pm 0.013$& $<2.90$ & $0.195\pm 0.006$\\
\hline
 $-2.9$& $<0.80$& $0.988\pm0.020$&  $<0.76$& $0.563\pm 0.012$& $<0.73$ & $0.197\pm 0.006$\\
\hline
\end{tabular}
\caption{\label{tab:GLBLBB}
Constraints on the PMF parameters for \lb with marginalization over the $BB$ lensing signal. The left column is the case with the full lensing signal, the middle is with 43\,\% delensing and the right is the more optimistic case with 80\,\% delensing.}
\end{table}

{\textbf{\emph{Lensing signal-}}}
We first consider only the inclusion of the lensing $BB$ signal in the sky and include a marginalization over its amplitude represented by a nuisance parameter $A_{\mathrm{Lens}}^{BB}$ centered on 1 with a flat prior [0-2] and varied with all the other parameters. This marginalization enables the study of possible degeneracies of the lensing $BB$ signal with either PMFs or primordial GW $B$-mode signals.
The mock realization of the $BB$ lensing signal is derived from the same standard-$\Lambda$CDM cosmological parameters since we do not account for any PMF contribution to the lensing.\footnote{A full rendition of the effects of PMFs on the lensing would require a fully nonlinear treatment of the smallest scale perturbations, introducing large theoretical uncertainties.}
\begin{figure}[t!]
    \centering
    \includegraphics[width=\textwidth]{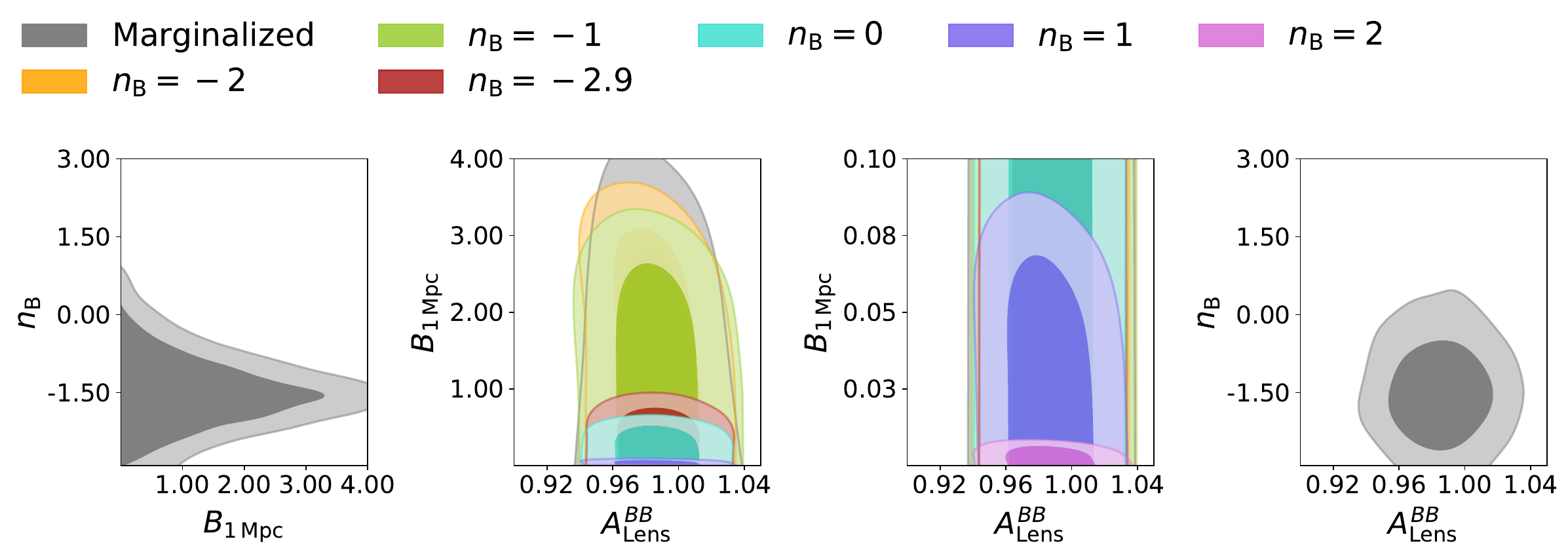}
    \caption{Two-dimensional posterior distributions for the magnetic parameters and lensing $BB$ amplitude. }
    \label{fig:G2D_LBLBB2}
\end{figure}
\noindent
Results for the magnetic and lensing parameters are shown in the second column in \autoref{tab:GLBLBB} and the two-dimensional posteriors are presented in \autoref{fig:G2D_LBLBB2}.
In general we have a good recovery of the nuisance parameter for the lensing $BB$ amplitude and constraints on the PMF amplitude, which are at the level of the ones obtained in the baseline case. The slightly lower value for the $BB$ lensing amplitude might indicate a mild degeneracy, which we see also in the two-dimensional posteriors in  \autoref{fig:G2D_LBLBB2}, where we find a slight tilt of the contours especially for lower and intermediate $\nB$.

{\textbf{\emph{Delensing-}}}
For this case we consider a simplified model of delensing (for the study of realistic \lb delensing, see Ref.\cite{namikawa2023litebird}). It was shown in Ref. \citep{Paoletti:2019pdi} how, in an approach similar to our ideal case, the delensing is not really effective except for very red indices and with an optimistic delensing possibility. We want to investigate this impact in our realistic settings where the lensing $BB$ signal is fitted along with the rest of the sky. We use a simplified delensing approach where we multiply the lensing $BB$ signal in our mock data by a different factor depending on the delensing option considered. In particular, we consider two cases of delensing: a standard case based on current delensing capabilities that rely on cosmic infrared background data, assuming a 43\,\% cleaning of the lensing signal and a more futuristic case, which assumes a more optimistic 80\,\% delensing capability. 
\begin{figure}[t!]
    \centering
    \includegraphics[width=\textwidth]{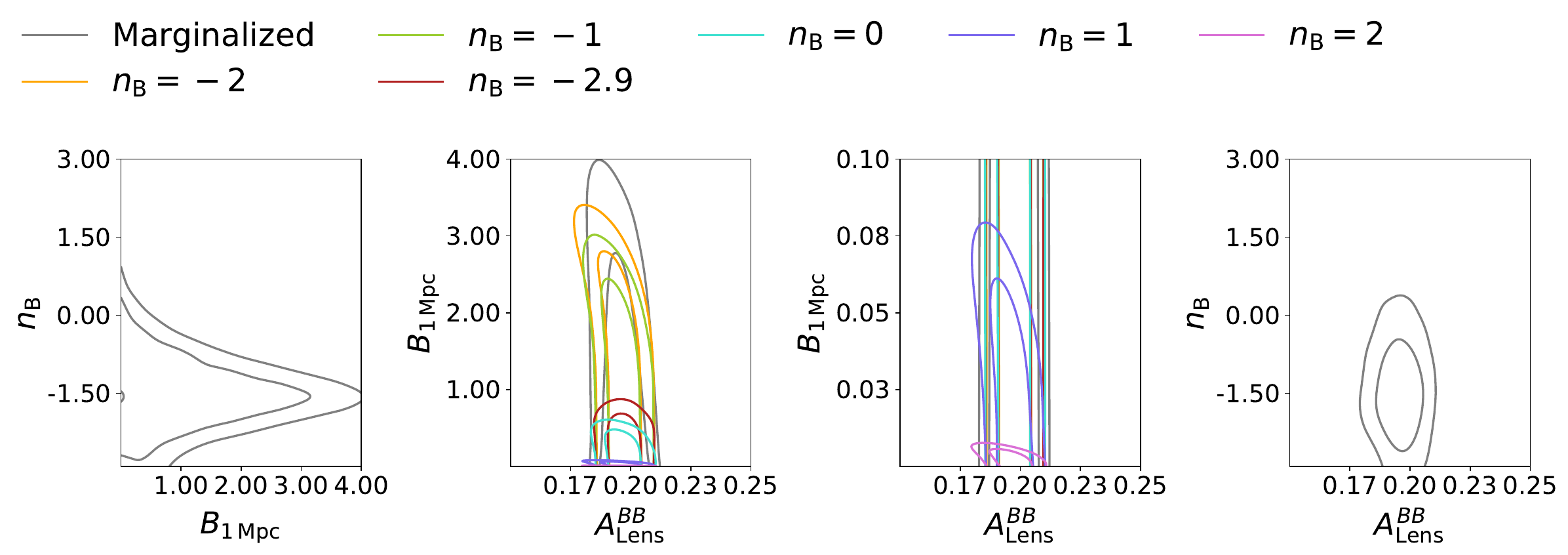}
    \caption{Two-dimensional posterior distributions for the magnetic parameters and lensing $BB$ amplitude when 80\% delensing is used. }
    \label{fig:G2D_LBLBB3}
\end{figure}
Results are shown in the third and fourth columns of \autoref{tab:GLBLBB}. Contrary to what happens for the ideal case \citep{Paoletti:2019pdi}, in this more realistic approach the delensing is capable of improving the constraints on the amplitude of PMFs, especially for redder indices. We can perfectly recover within the error bars the delensed amplitude. An interesting trend is visible in the two-dimensional posteriors of \autoref{fig:G2D_LBLBB3}, where we find that, as expected, lowering the lensing signal increases the degeneracies with the magnetic parameters except for the almost scale invariant and marginalized cases.

{\textbf{\emph{Non-zero inflationary signal-}}} One of the main gravitational effects of PMFs is the creation of  $B$-mode polarization. In particular the signal from passive, almost scale invariant tensor modes resembles what we can have from GWs from inflation. It is crucial to understand the correlation between the tensor-to-scalar ratio $r$ and PMFs, especially for \lb  whose primary target is $B$-mode polarization. 
We therefore investigate what happens when we consider a non-zero primordial gravitational signal from inflation and fit it with both PMFs and primary tensor modes active.
\begin{figure}[t!]
    \centering
    \includegraphics[width=0.5\textwidth]{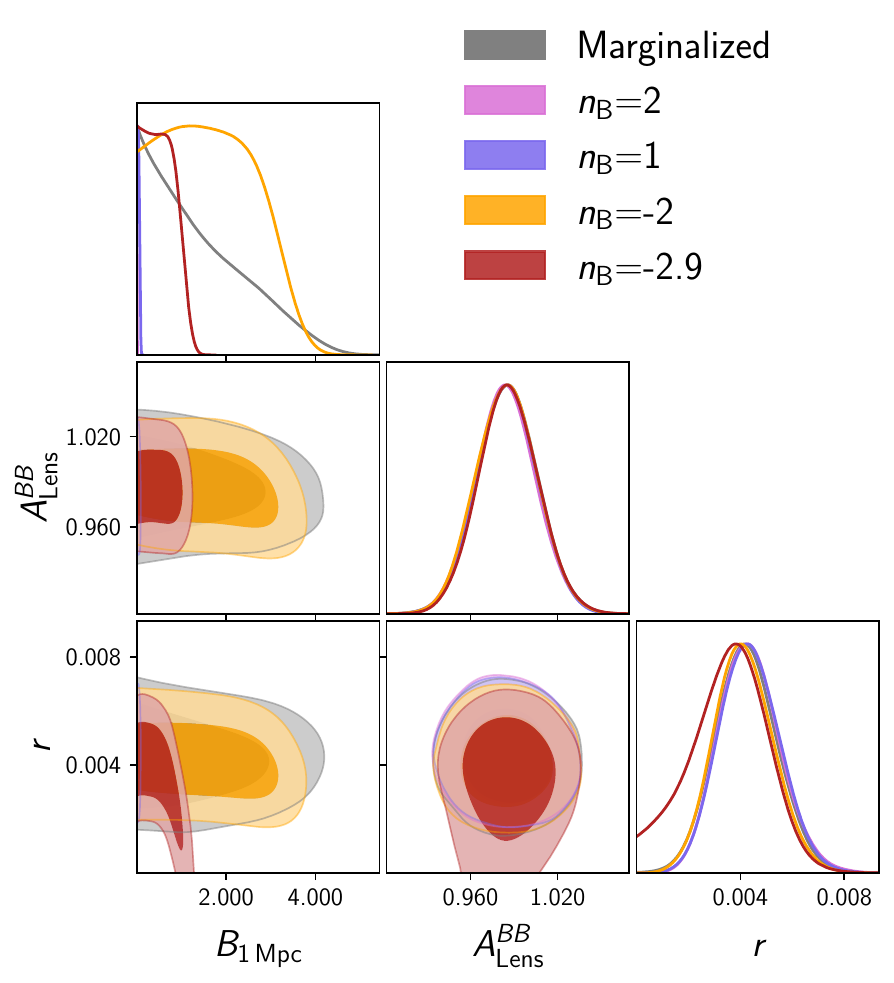}\includegraphics[width=0.5\textwidth]{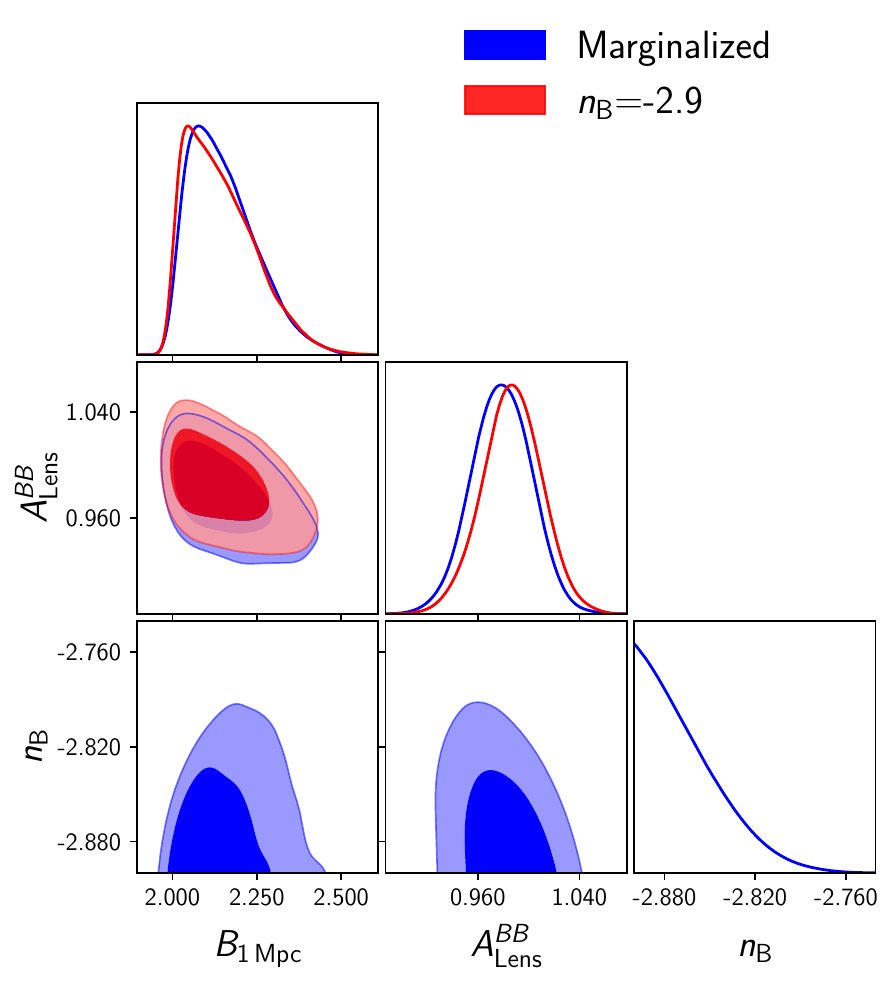}
    \caption{Triangle plot for the magnetic, lensing $BB$ amplitude and tensor-to-scalar ratio, $r$, parameters. (Left) $R^2$-like model of inflation without PMF in the fiducial model. (Right) Non-zero PMF in the fiducial model.}
    \label{fig:alensrandDet}
\end{figure}
 We consider one of the target inflationary models with underlying base cosmology as \Planck 2018, assuming $r=0.0042$ similar to the $R^2$ model of inflation \cite{Starobinsky:1980te}. The results are presented in \autoref{tab:GLBLBBr} and in  the left panel of \autoref{fig:alensrandDet}. 
 As expected, we do not observe a strong correlation with the positive $\nB$ fields or the marginalized case. Infrared-dominated spectra instead tend to correlate with $r$. In particular, we observe a strong degeneracy with the almost scale invariant case. The degeneracy with the lensing amplitude is almost unchanged.
We also investigate the impact of delensing and present the results in the last two blocks of \autoref{tab:GLBLBBr}. We find the same trend as for the $r=0$ case, with delensing being mostly effective for negative $\nB$; for detected primordial GWs the delensing does not worsen the degeneracy with PMFs, although it remains present, especially for the almost-scale invariant case.

\begin{table}[t!]
\center
\begin{tabular}{|c c c c|}
\hline
\multicolumn{4}{|c|}{\lb with lensing Marginalization and non-zero inflationary signal}\\
\hline
\multicolumn{4}{|c|}{Full lensing}\\
\hline
$\nB$ &$B_{1\,\mathrm{Mpc}}$ [nG]  & $A_{\mathrm{Lens}}^{BB}$ & $r$\\
\hline
Marginalized &  $<3.47$& $0.986\pm 0.021$& $0.0042_{-0.0012}^{+0.0011}$\\
\hline
 2& $<0.007$& $0.984\pm 0.0021$& $0.0044_{-0.0013}^{+0.0010}$\\
\hline
 1& $<0.075$&$0.985\pm 0.021$& $0.0044_{-0.0013}^{+0.0010}$\\
\hline
 0& $<0.553$& $0.987\pm 0.020$& $0.0049_{-0.0013}^{+0.0011}$\\
\hline
 $-1$&$<2.71$& $0.987\pm 0.020$&$0.0049_{-0.0013}^{+0.0011}$\\
\hline
 $-2$&$<3.27$& $0.985\pm 0.021$& $0.0041_{-0.0012}^{+0.0010}$\\
\hline
 $-2.9$&$<1.06$& $0.986_{-0.020}^{+0.021}$& $0.0035_{-0.0013}^{+0.0017}$\\
\hline
\hline
\multicolumn{4}{|c|}{43\% Delensed}\\
\hline
$\nB$ &$B_{1\,\mathrm{Mpc}}$ [nG] & $A_{\mathrm{Lens}}^{BB}$ & $r$\\
\hline
Marginalized & $<3.30$&$0.569\pm 0.013$&$ 0.0042\pm 0.0010$\\
\hline
 2& $<0.007$&$0.568\pm 0.013$&$0.0043_{-0.0010}^{+0.0009}$\\
\hline
 1& $<0.071$&$0.568\pm 0.013$&$0.0043_{-0.0011}^{+0.0009}$\\
\hline
 0& $<0.547$&$0.569\pm 0.013$&$0.0043_{-0.0010}^{+0.0009}$\\
\hline
 $-1$& $<2.66$&$0.569\pm 0.013$&$0.0043_{-0.0011}^{+0.0009}$\\
\hline
 $-2$&$<3.16$&$ 0.568\pm 0.013$&$ 0.0041\pm 0.0010$\\
\hline
 $-2.9$&$<1.04$&$0.570\pm 0.013$&$0.0036_{-0.0011}^{+0.0015}$\\
\hline
\hline
\multicolumn{4}{|c|}{80\% Delensed}\\
\hline
$\nB$ &$B_{1\,\mathrm{Mpc}}$ [nG] & $A_{\mathrm{Lens}}^{BB}$ & $r$ \\
\hline
Marginalized &$<3.31$&$0.198_{-0.006}^{+0.007}$&$0.0042\pm 0.0009$\\
\hline
 2& $<0.0062$&$0.197_{-0.006}^{+0.007}$&$0.0043_{-0.0009}^{+0.0008}$\\
\hline
 1& $<0.067$&$0.198\pm 0.006$&$0.0043_{-0.0009}^{+0.0008}$\\
\hline
 0&$<0.514$&$0.199\pm 0.006$&$0.0043_{-0.0009}^{+0.0008}$\\
\hline
 $-1$&$<2.57$&$0.198\pm 0.0063$&$0.0042\pm 0.0008$\\
\hline
 $-2$&$<2.99$&$0.198_{-0.006}^{+0.007}$&$0.0042\pm0.0009$\\
\hline
 $-2.9$&$<1.03$&$0.200\pm 0.006$&$0.0035_{-0.0009}^{+0.0015}$\\
\hline
\end{tabular}
\caption{\label{tab:GLBLBBr}
Constraints on the PMF parameters for \lb, with  marginalization over the $BB$ lensing signal and a non-negligible primary tensor contribution.}
\end{table}

{\textbf{\emph{Non-zero PMF signal-}}} We now investigate the capabilities of \lb to detect PMFs with different characteristics. We will consider both a targeted detection, in which we assume the same PMF model as input in the mock data analysis, and a blind reconstruction where we just leave the PMF configuration free without any a priori assumption. We consider two fiducial cases, the first is a realistic case employing the current limits from the gravitational effect on the almost scale invariant model:
$B_{1\,\mathrm{Mpc}}=2.2\, \mathrm{nG}$ with $\nB =-2.9$ \citep{Paoletti:2019pdi}. By fitting the sky with an almost scale invariant PMF we obtain $B_{1\,\mathrm{Mpc}}=2.14_{-0.14}^{+0.06}$ nG and $A_{\rm lens}^{BB}=0.987\pm 0.023$, showing the capability of detecting the PMF signal when we have knowledge of their characteristics. In the case where we explore the parameter space in a blind way we obtain $-2.9< \nB <-2.86$, $B_{1\,\mathrm{Mpc}}=2.15_{-0.13}^{+0.06}$ nG and $A_{\rm lens}^{BB}=0.979_{-0.024}^{+0.023}$ showing again no issues in the recovery of the input sky. As demonstrated in the right panel of \autoref{fig:alensrandDet}, we have highly non-Gaussian posteriors for the PMF amplitude, but nonetheless we can recover the input sky.

The second case we consider is the causal case, $\nB =2$, where we fix the value of the amplitude to 1\,nG, far higher than allowed by current data, but it is nevertheless useful to investigate the degeneracy with the lensing signal that we expect to be stronger for positive $\nB$.
In this case we assume that we know the PMF configuration in the sky obtaining $B_{1\,\mathrm{Mpc}}=0.999\pm 0.007$ nG and $A_{\rm lens}^{BB}$ remains unconstrained, showing the degeneracy between the two signals. When we blindly try to recover the input sky, we do not manage to reach a convergence even after hundreds of thousands of samples, as we fall into a region of the parameter space that provides very odd theoretical angular power spectra with strong effects.
We conclude that in the case of a detected signal, thanks to \lb sensitivity, we are capable of recovering either PMFs or jointly PMFs and primary tensor modes, although we are still affected by some degeneracies. This illustrates the importance of \lb in putting together different probes that provide constraints that are complementary to each other.

\begin{figure}[t!]
    \centering
    \includegraphics[width=\textwidth]{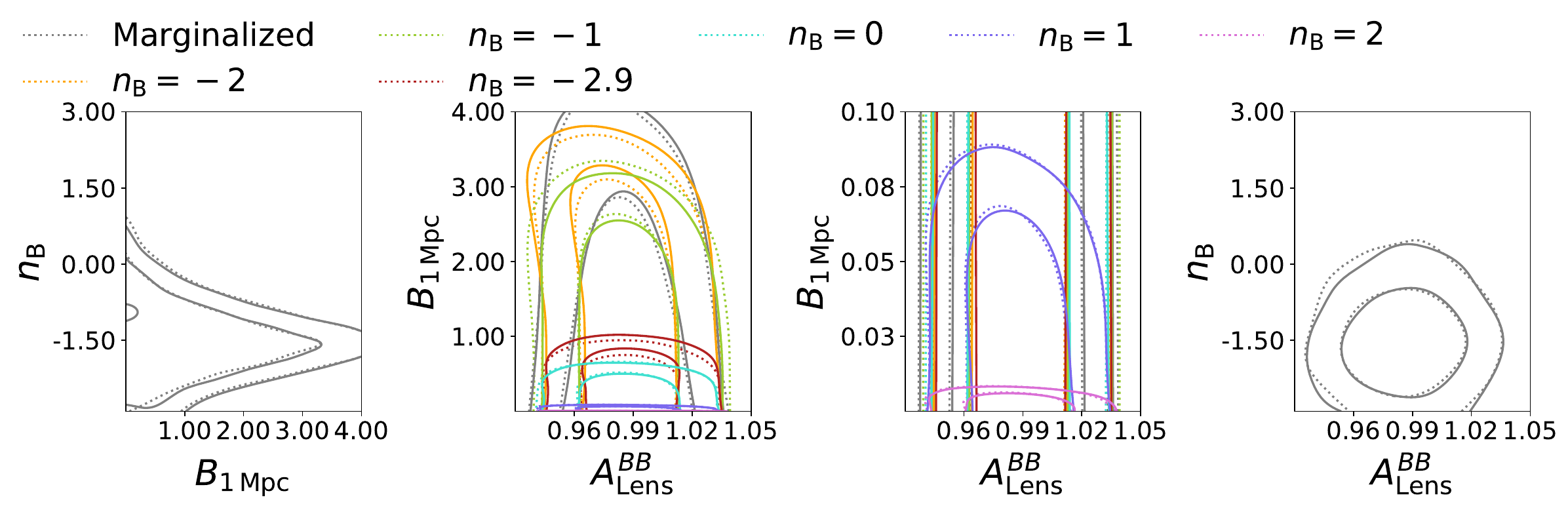}
    \caption{Two-dimensional posterior distributions for the magnetic parameters and lensing $BB$ amplitude when the foreground bias is also included in the signal but not fitted for (solid lines). The dotted lines include only the lensing signal. }
    \label{fig:G2D_LBLBB2FGBIAS}
\end{figure}

{\textbf{\emph{Data complexity: foreground and systematic bias-}}}
We keep increasing the complexity of our simulated data and in this layer we add to the simulated sky signal the residual foreground bias contamination, which mimics the astrophysical residuals coming from the component-separation algorithm as in Ref. \citetalias{LiteBIRD:2022cnt}. Since the signal is weak, in this first test we marginalize only over the lensing amplitude without also considering marginalization over the foreground residuals bias. 
Results are presented in \autoref{fig:G2D_LBLBB2FGBIAS}, compared with the lensing-only signal in dotted lines.
Overall we do not observe a significative degradation of the constraints, with just a minimal effect for red spectral indices such as $-2$ and $-2.9$. Positive $\nB$ are not affected and indeed in some cases show a marginal improvement related to the marginalization on the lensing signal that also includes the bias, absorbing part of the PMF signal. We conclude that with the current sky models the constraints are not affected by the residual foreground contamination from component separation. 

The final layer of complexity of the simulated data involves the systematic biases as presented in Ref. \citetalias{LiteBIRD:2022cnt}. We include the total sum of systematics, which does not include the cosmic rays that are treated in a different manner at the level of the noise. So, finally we have a sky in $B$-mode polarization composed of: CMB; lensing; foreground residuals; and systematic biases.
In the first setup, we only marginalize over the lensing $B$-mode amplitude and consider the foreground and systematics as pure contaminants of the signal (a sort of \textit{testing the unknown unknowns} hypothesis). The results are presented in the second column of \autoref{tab:GLBLBBBias}. Again we find the same trend as with the foreground bias alone. The presence of the additional biases that are not fitted for degrades the constraints on the PMF amplitude for infrared dominated fields, whereas ultraviolet and intermediate indices are left almost unchanged except for some marginal improvement due to the larger recovered lensing signal caused by the presence of the additional biases reducing the signal in the PMFs.
\begin{table}[t!]
\center
\begin{tabular}{|c|c c|c c c|}
\hline
\multicolumn{6}{|c|}{\lb with all biases}\\
\hline
\multicolumn{1}{|c|}{Fiducial} &\multicolumn{2}{|c|}{Only lensing marginalization}&\multicolumn{3}{|c|}{Lensing and biases marginalization}\\
\hline
$\nB$ &$B_{1\,\mathrm{Mpc}} \, [\mathrm{nG}] $  & $A_{\mathrm{Lens}}^{BB}$&$B_{1\,\mathrm{Mpc}} \, [\mathrm{nG}] $  & $A_{\mathrm{Lens}}^{BB}$ & $A_{\mathrm{FG-syst-bias}}^{BB}$\\
\hline
Marginalized &  $<3.38$& $0.987\pm0.020$&  $<3.51$& $0.985 \pm 0.021$& $\dots$\\
\hline
 2& $<0.007$& $0.987 \pm 0.021$&$<0.007$& $0.985\pm 0.020$& $\dots$\\
\hline
 1& $<0.072$& $0.986_{-0.020}^{+0.021}$&$<0.074$&$0.985_{-0.021}^{+0.020}$& $\dots$\\
\hline
 0& $<0.55$& $0.988 \pm 0.021$  &$<0.57$& $0.986\pm 0.021$& $\dots$\\
\hline
 $-1$& $<2.70$& $0.987\pm 0.020$&$<2.79$& $0.985\pm 0.020 $& $\dots$\\
\hline
 $-2$& $<3.30$& $ 0.985\pm 0.022$&$<3.21$& $0.984_{-0.021}^{+0.020}$& $\dots$\\
\hline
 $-2.9$& $<0.89$& $0.988\pm 0.020$&$<0.83$& $0.988\pm 0.021$& $0.946_{-0.806}^{+0.438}$\\
\hline
\end{tabular}
\caption{\label{tab:GLBLBBBias}
Constraints on the PMF parameters including all the biases. In the second column we show the constraints with the marginalization over the $BB$ lensing signal only. In the third column we also marginalize over the foreground and systematic biases.}
\end{table}
In the second setting, which more accurately represents what will be in the real data pipeline, we also marginalize over the foreground and systematic biases. Following Ref. \citetalias{LiteBIRD:2022cnt} we marginalize over the same input signal for the biases, just varying an overall amplitude multiplying both input signals. This approach is optimistic, since it supposes we perfectly know the expected bias signal except for the amplitude. The results are presented in the third column of \autoref{tab:GLBLBBBias}.
We find that the bias amplitude remains  unconstrained, but the marginalization slightly affects the results. We also find a minimal improvement in the constraints of the lower $\nB$, which are more affected by the systematics signals, while at the same time for intermediate and high $\nB$ we have a partial degeneracy with the lensing fitting that leads to some small changes in the constraints on the PMF and lensing amplitudes. The marginalization can therefore improve the most infrared index constraints.
We conclude that the main contamination we expect for the gravitational effect is the contamination by lensing. Foreground residuals and expected systematics do not have a significant impact on the results.

\noindent
\subsubsection{Root mean square parametrization}

\begin{figure}[t!]
    \centering
    \includegraphics[width=0.5\textwidth]{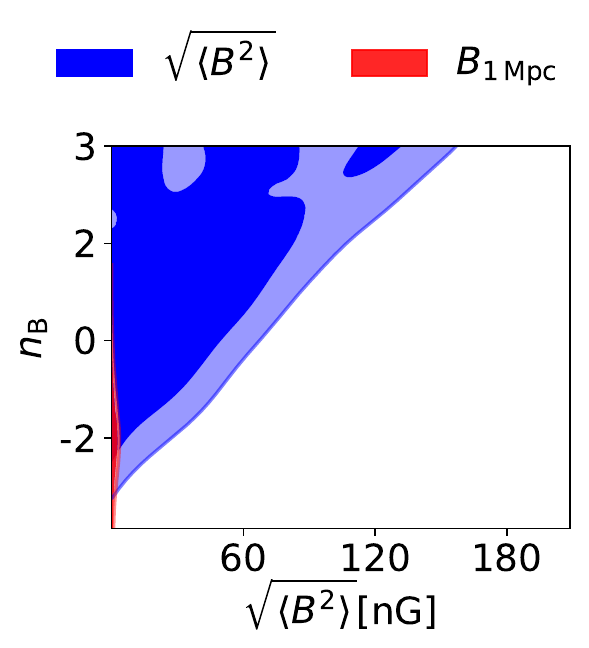}
    \caption{Comparison of the two-dimensional posterior distributions for the magnetic parameters between the rms and the smoothed parametrizations.}
    \label{fig:2DRMS}
\end{figure}

In Section~\ref{sec:formalism} we discussed how the resulting constraints depend on the parametrization used for the PMFs. Along the same lines, here we close the gravitational effect section by investigating 
the constraints with an alternative parametrization, the rms, as shown in \autoref{mean-squared}. This parametrization will be used in the following section for the effect on the ionization and thermal history and it is therefore useful to have a comparison with what we can obtain from the gravitational effect.
The results are shown in \autoref{tab:RMS} and in \autoref{fig:2DRMS}. We observe how the change of parametrization drastically affects the constraints, with a complete inversion of the $\nB$ dependence and much weaker constrains. This is expected from a naive consideration of the two parametrizations: PMFs smoothed on a small scale of 1\,Mpc are strongly affected by the small scale power of ultraviolet indices, whereas in the rms case we are considering an averaged field that erases the small scale power. The almost scale invariant case is nearly independent of the choice of parametrization except for the numerical accuracy\footnote{The perfect equivalence is valid only for exact scale invariance.}. 

\begin{table}[t!]
\centering
 \begin{tabular}{|c c|}
\hline
\multicolumn{2}{|c|}{\lb with rms}\\
\hline
$\nB$ &$\sqrt{\langle B^2\rangle} \, [\mathrm{nG}] $ \\
\hline
Marginalized &  $<122.40$\\
\hline
 2& $<120.43$\\
\hline
 1& $<90.07$\\
\hline
 0& $<66.91$\\
\hline
 $-1$& $<43.45$\\
\hline
 $-2$& $<12.25$\\
\hline
 $-2.9$& $<0.76$\\
\hline
\end{tabular}
      \caption{Constraints on the PMF amplitude using the rms parametrization.}
\label{tab:RMS}
\end{table}


\section{Effect on thermal and ionization history}
\label{sec:thermalhistory}
In this section we describe the effects of PMFs on the thermal and ionization history of the Universe and investigate the constraints on PMFs that \lb can provide through them.
Before recombination, the environment of the primordial plasma is sufficient to maintain the ideal MHD limit, at least at first order and on linear scales, the regime we are interested in for large and intermediate scales of the CMB. In this regime, the magnetic fields are decoupled from the fluid and are flux frozen in the plasma, providing a simple passive dilution of the field amplitude with the Universe expansion, as shown in Section \ref{sec:formalism}.
This situation drastically changes when recombination takes place. The reduction in the ionization fraction and the coupling between the photon and baryon fluids leads to the development of two dissipative effects, namely MHD decaying turbulence and ambipolar diffusion. 
These effects dissipate the PMF energy, heating the plasma, with a strong impact on CMB anisotropies and frequency spectrum. We are interested in the former for \lb.
Differently from the gravitational case, this is a global effect that distorts the primary CMB anisotropy pattern. Therefore, in $\Lambda$CDM the only relevant channels are the temperature, the $E$-mode polarization and their cross-correlation. We will show later the effects on a model that also includes primordial GWs from inflation.

In this section we will use the $\sqrt{\langle B^2\rangle}$ parametrization because of numerical instabilities that we encounter in the 1 Mpc parametrization; positive $\nB$ have a very strong effect which is boosted for that parametrization, making the injection rates too high to be numerically treated. 

We will refer to the treatment and numerical code derived in Refs.\citep{2015MNRAS.451.2244C,Paoletti:2018uic,Paoletti:2022gsn}. We use the extension developed and optimized in Refs. \citep{2015MNRAS.451.2244C,Paoletti:2018uic,Paoletti:2022gsn}, with the same settings as in Ref. \cite{Paoletti:2022gsn},  of the {\tt Recfast++} routine within the {\tt CosmoRec} code \citep{2011MNRAS.412..748C}.
%
%
Our treatment is based on analytical approximations for MHD decaying turbulence and ambipolar diffusion. A full account of the thermal and ionizing effects and of the dissipation of the fields would require numerical simulations of the development of the turbulence (considering also the feedback on the PMF spectral shape and time evolution due to the coupling with the kinetic component of the plasma), and of the development of ambipolar diffusion accounting for the coupling of neutral and ionized plasma components (see, e.g., Refs.\cite{Christensson:2000sp,Christensson:2002xu,Saveliev:2013uva,Kahniashvili:2016bkp,Brandenburg:2016odr} for some turbulence evolution simulations). But our aim is to use CMB data, for this work simulated \lb data, and this requires a likelihood approach. In a likelihood approach hundreds of thousands of samples need to be calculated, making it impossible to use accurate simulations for each step. For this reason we are forced to use some analytical approximation of the effects, which are acceptable on the scales we consider. Specific numerical treatments are already being studied (see for example Ref. \cite{2018MNRAS.481.3401T}) offering a good prospect for a future work employing much more precise analytical approximations based on accurate numerical simulations.

The treatment consists  of  representing the dissipation of PMFs as energy injection rates modifying the matter temperature of the plasma, which can be described as \citep{Sethi2005}
\begin{equation}
\label{eq:dT_dt}
\frac{dT_{\mathrm{e}}}{dt}=- 2 H T_{\mathrm{e}} + \frac{8 \sigma_T\, n_{\mathrm{e}} \,\rho_\gamma}{3\,m_{\mathrm{e}} c N_{\rm tot}} (T_\gamma-T_{\mathrm{e}}) + \frac{\Gamma}{(3/2)k N_{\rm tot}} \,,
\end{equation}
with
\begin{itemize}
    \item $\rho_\gamma=a_{\rm R} T_{\gamma}^4\approx 0.26 \, {\rm eV} (1+z)^4$ the CMB energy density;
    \item $N_{\rm tot} =N_{\rm H}(1+f_{\rm He}+x_{\rm e})$ the number density of all ordinary matter particles sharing thermal energy, where $N_{\rm H}$ is  the number density of H nuclei and $x_{\rm e}$ the free electron fraction;
    \item $f_{\rm He}\approx Y_{\mathrm{p}}/ 4(1-Y_{\mathrm{p}})$ with $Y_{\mathrm{p}}$ the primordial helium mass fraction. 
\end{itemize}
The two effects we consider are then encapsulated at the level of the energy injection rate $\Gamma$, as will be detailed in the following subsections.

\subsection{Magneto-hydrodynamic decaying turbulence}
After recombination the drop in the ionization fraction and the decoupling of photons from the baryonic fluid causes a drop in the fluid viscosity. The reduced viscosity leads to a dominance of the magnetic terms in the fluid dynamics, increasing the Reynolds number and enabling the development of MHD decaying turbulence.
The turbulence moves energy from large to small scales where the energy is dissipated in the plasma. 
The injection rate can be written as \citep{Sethi2009}
\begin{align}
 \Gamma_{\rm turb}=\frac{3 m}{2}\, 
 \frac{\left[\ln \left(1+\frac{t_i}{t_{\rm d}}\right)\right]^m}{\left[\ln \left(1+\frac{t_i}{t_{\rm d}}\right)
 + \frac{3}{2} \ln \left( \frac{1+z_i}{1+z}\right)\right]^{m+1}} H(z)\,\rho_{\rm B}(z),
 \label{Eqn:rate}
\end{align}
with $t_i/t_\mathrm{d}\approx 14.8 (\langle  B^2 \rangle^{1/2} / \mathrm{nG})^{-1}(k_\mathrm{D}/ \Mpc^{-1})^{-1}$ the ratio between the initial time of the decay and the tur\-bu\-lence time\-scale, $m=2(\nB+3)/(\nB+5)$ and $\rho_\mathrm{B}(z)=\langle  B^2 \rangle (1+z)^4/ (8\pi) \approx {9.5}\times 10^{-8} (\langle  B^2 \rangle/ \mathrm{nG}^{2})\,\rho_\gamma(z)$.

\begin{figure}[t!]
    \centering
    \includegraphics[width=0.5\textwidth]{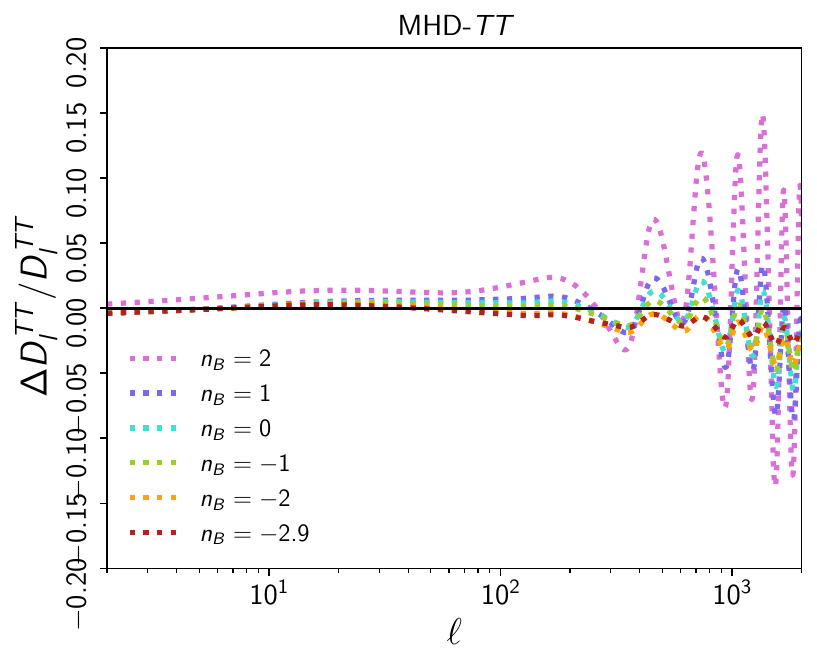}\includegraphics[width=0.5\textwidth]{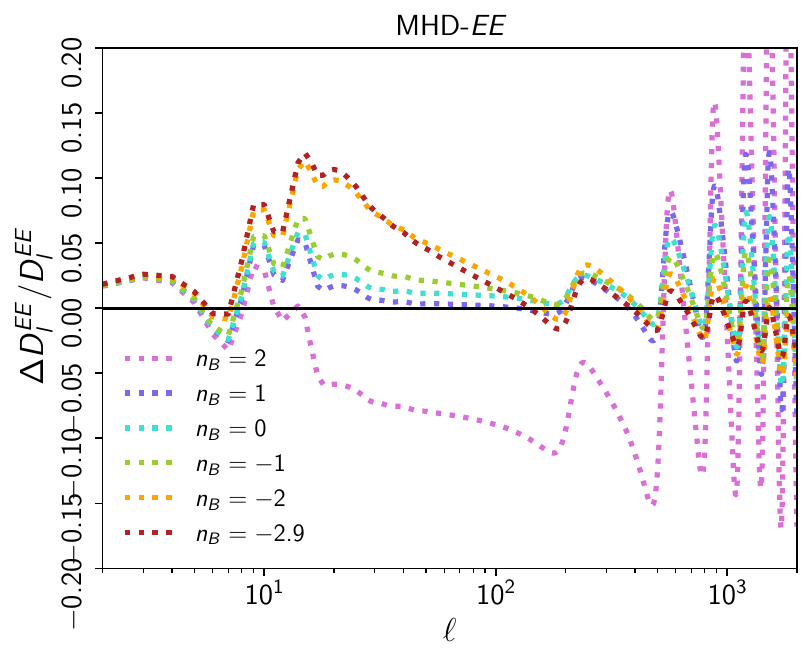}
    \caption{Effect of the MHD decaying turbulence on the CMB anisotropy angular power spectra in temperature (left) and polarization (right) for different spectral indices $\nB$. We display the relative differences compared with the corresponding $\Lambda$CDM model.}
    \label{MHDCl}
\end{figure}

In \autoref{MHDCl} we present the effect on the CMB anisotropy angular power spectra in temperature and $E$-mode polarization. The effect on temperature anisotropies is limited to the smallest scale part, which shows oscillations. The effect in polarization shows the same features on small scales as temperature, but in addition it also shows effects on intermediate and large angular scales.
\subsection{Ambipolar diffusion}
\begin{figure}[t!]
    \centering
    \includegraphics[width=0.5\textwidth]{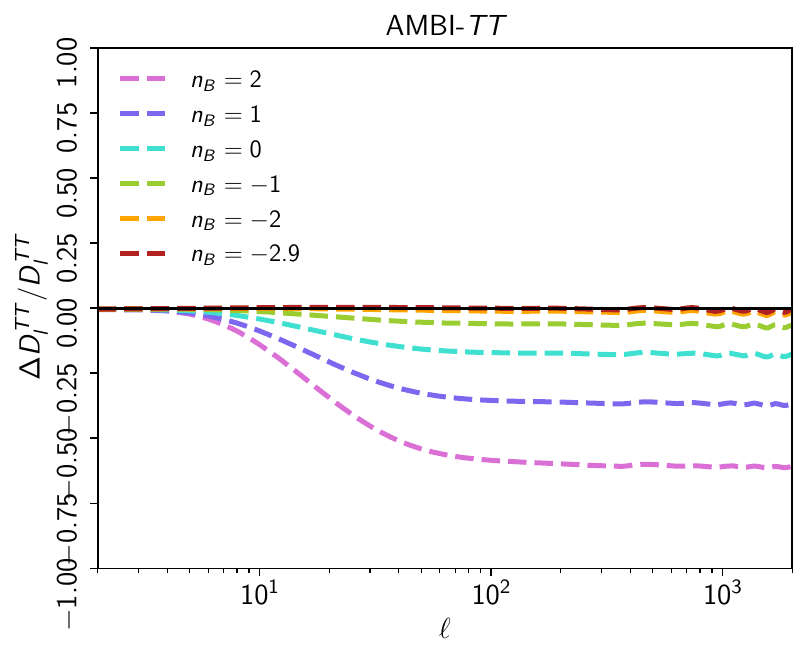}\includegraphics[width=0.5\textwidth]{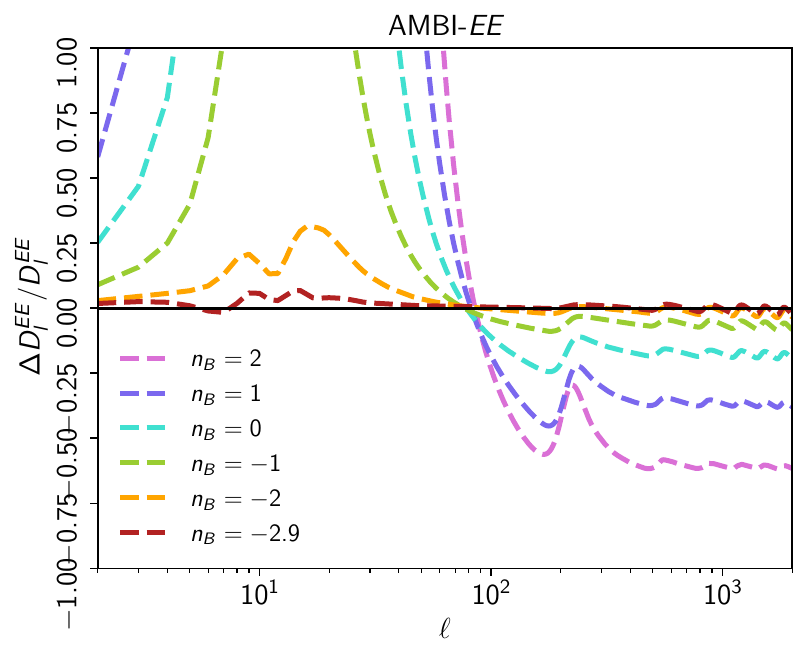}
    \caption{Same as \autoref{MHDCl} but for the ambipolar diffusion effect.}
    \label{AMBICl}
\end{figure}

We now describe the second effect taking place after recombination, ambipolar diffusion. Dissipation is caused by the dropping of the ionization fraction, leading to a large neutral component in the plasma, which, due to the presence of PMFs, has a different velocity with respect to the residual ionized part. The thermalization due to the transfer of energy to the neutral component dissipates magnetic energy, heating the plasma.
This can be described by the energy injection rate \citep{Sethi2005, Schleicher2008b}
\begin{align}
\Gamma_{\rm am}\approx \frac{(1-X_{\rm p})}{\gamma X_{\rm p}\, \rho_{\rm b}^2} \left< {\bf L}^2\right>,
\end{align}
where $\left< {\bf L}^2\right>$, $\rho_{\rm b}$ and  $X_{\rm p}$ are the Lorentz force, the baryon density and the coupling between the two components, neutral and ionized.\footnote{For the Lorentz force we use the same treatment as in \citep{Paoletti:2022gsn,Paoletti:2018uic}, which is the one relative to the sharp cut off damping model for the fields.}
In \autoref{AMBICl} we show the effect of ambipolar diffusion on the CMB anisotropy angular power spectra in temperature and polarization. We find the completely different imprint of this effect compared to the MHD turbulence. We find an overall depletion of power on intermediate and small angular scales in temperature and an important effect on both large and small scales in polarization. In particular we have a drastic change in the reionization bump in EE. Ambipolar diffusion also shows a remarkable dependence on $\nB$ compared to MHD turbulence, with positive $\nB$ providing the strongest effect.

\subsection{Joint effect}
\begin{figure}[t!]
    \centering
    \includegraphics[width=0.5\textwidth]{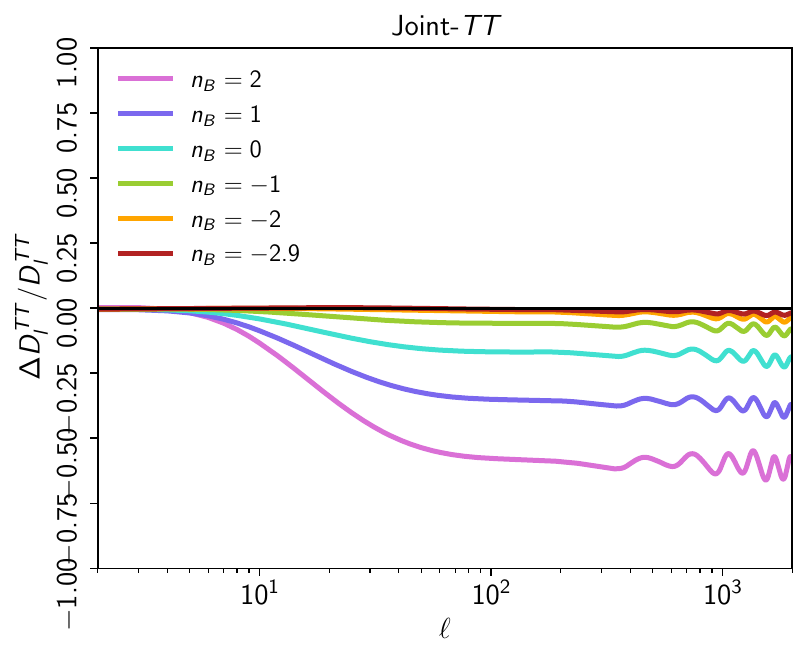}\includegraphics[width=0.5\textwidth]{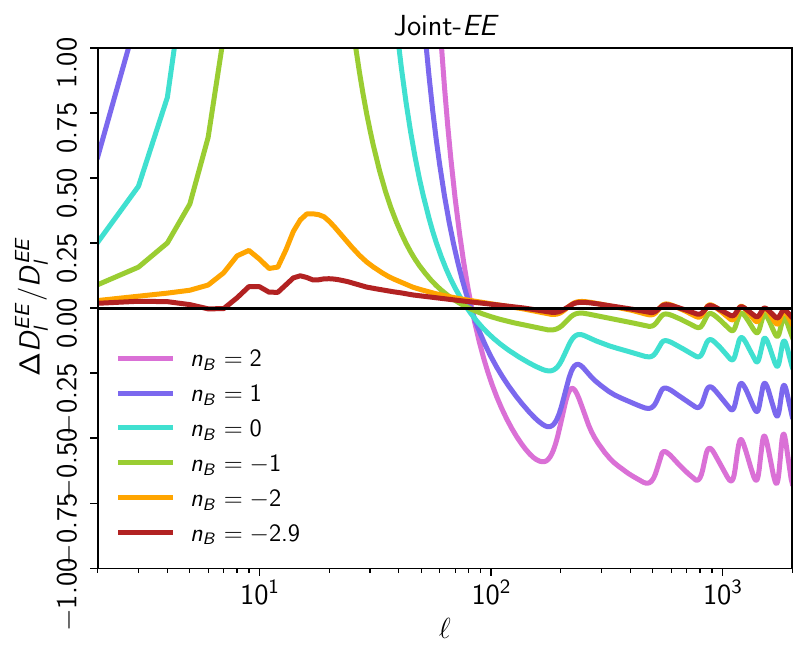}
    \caption{Same as \autoref{MHDCl} but for the combined effect of ambipolar diffusion and MHD decaying turbulence.}
    \label{MHDAMBICl}
\end{figure}

The combination of ambipolar diffusion and MHD decaying turbulence has the power to massively affect both large and small scales in temperature and polarization. There is strong dependence on $\nB$ in ambipolar diffusion, enabling the constraints on blue indices whereas the MHD turbulence dominates the constraints on infrared indices.
In \autoref{MHDAMBICl} we present the joint effect on the angular power spectra. The combination shows imprints on both intermediate and small angular scales in temperature and a strong effect for all $\nB$ in polarization.

\subsection{Results}
We will now go through the results of the MCMC parameter exploration. As for the gravitational effect in Section \ref{sec:gravitational}, we vary the amplitude in rms of the PMFs (\autoref{mean-squared}) together with cosmological parameters from the $\Lambda$CDM model. In the marginalized case we also vary $\nB$. The impact on the thermal and ionization history is strong especially in $E$-mode polarization. For this reason  we will focus on $TT$-$TE$-$EE$ in this section. We consider the ideal case, since the post-component-separation contamination in the $E$-mode polarization is expected to be very low considering the sensitivity and frequency range of \lb \citepalias{LiteBIRD:2022cnt}. We will also consider the combination with the \Planck simulated dataset.
We will first investigate the two effects separately with their specific characteristics and regions of interest, and then proceed with their combination.
\subsubsection{Magneto-hydrodynamics decaying turbulence}
\begin{figure}[t!]
    \centering
     \includegraphics[width=\textwidth]{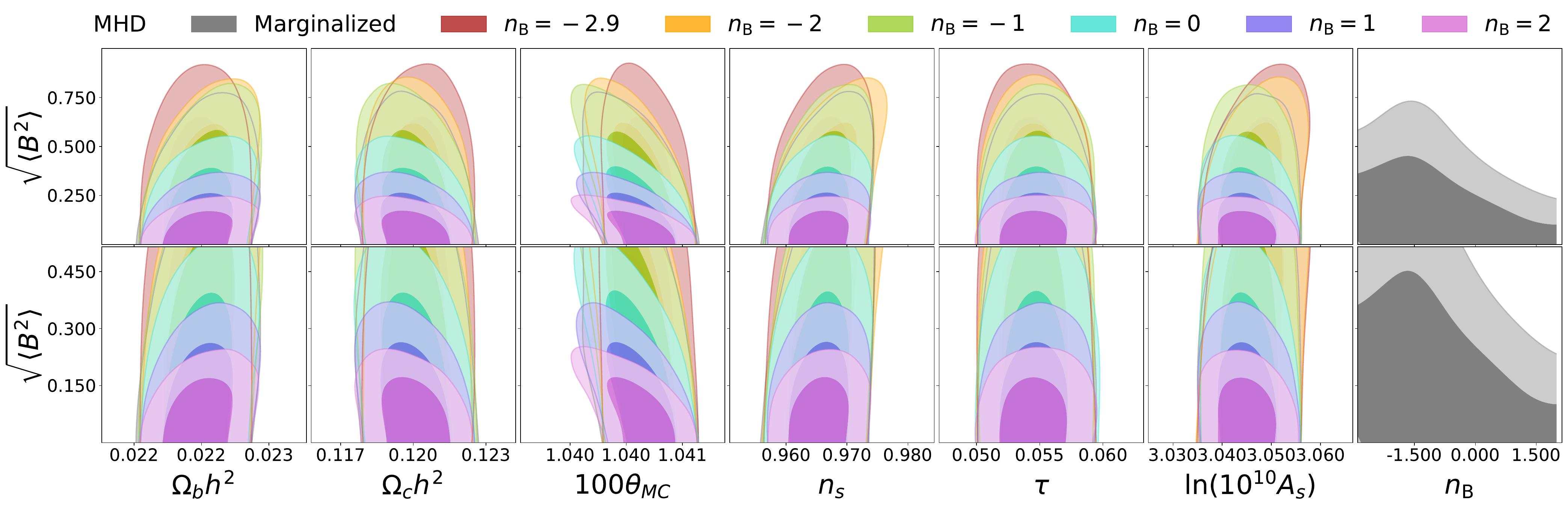}
    \caption{Two-dimensional posterior distributions for the magnetic amplitude versus the other cosmological parameters for the MHD decaying turbulence. We show the fixed spectral index $\nB$ and the marginalized case (in grey). The lower panels focus on the higher $\nB$.}
    \label{fig:MHD2D}
\end{figure}

We start from the constraints derived with the MHD decaying turbulence effect alone. In this case we have seen that the effect is mainly focused on the intermediate and small scales in polarization and there is a mild dependence on $\nB$. This is reflected in the constraints we obtain and shown in \autoref{tab:HMHD} and in two-dimensional contours in \autoref{fig:MHD2D}.
\begin{table}[t!]
\center
\begin{tabular}{|c|c c|}
\hline
\multicolumn{3}{|c|}{MHD decaying turbulence}\\
\hline
$\nB$ &\multicolumn{2}{|c|} {$\sqrt{\langle  B^2 \rangle}$ [nG] }  \\
\hline
Dataset & \lb (\Planck 2018) & \lb + \Planck \\
\hline
Marginalized &  $<0.60$ ($<0.68$)&$<0.58$  \\
\hline
$2$ & $<0.20$ ($<0.18$) &$<0.15$\\
\hline
$1$ & $<0.30$ ($<0.27$)&$<0.22$\\
\hline
$0$ & $<0.46$ ($<0.41$)&$<0.34$\\
\hline
$-1$ & $<0.67$ ($<0.63$)&$<0.54$\\
\hline
$-2$ & $<0.70$ ($<0.79$) &$<0.70$\\
\hline
$-2.9$ &  $<0.76$ ($<1.05$)& $<0.72$\\
\hline
\end{tabular}
\caption{\label{tab:HMHD}
Constraints on the PMF amplitude both for a fixed spectral index and the marginalized case over $\nB$ by using only the MHD decaying turbulence effect. Constraints are at 95\,\% C.L.}
\end{table}
\begin{figure}[t!]
    \centering
     \includegraphics[width=1.0\textwidth]{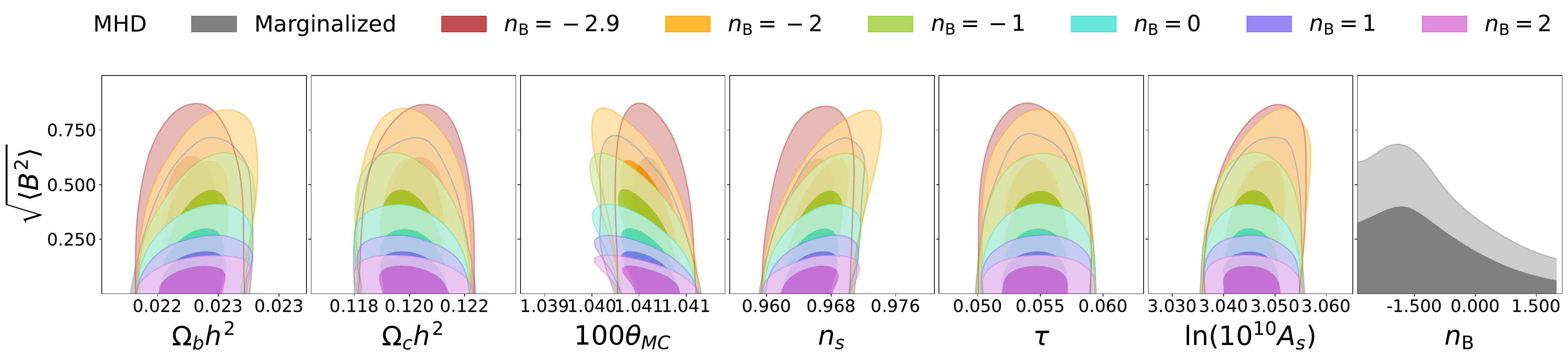}
       \caption{Same as \autoref{fig:MHD2D} but for \lb combined with \Planck.}
    \label{fig:MHD2DP}
\end{figure}
With respect to current constraints from \Planck data (in parentheses in the table) \citep{Paoletti:2022gsn}, we find that with \lb  sensitivities we are capable of improving the constraints for negative $\nB$. This is due to the shape of the MHD decaying turbulence effect. MHD strongly affects the region of the acoustic peaks in the angular power spectra and  only to a lesser extent the intermediate and large angular scales (where \lb has the most constraining power) and for this reason the improvement is limited. 
The two-dimensional posteriors also show how with \lb sensitivities (which basically means cosmic variance limited temperature and $E$ modes) we observe the degeneracies found in Ref. \cite{Paoletti:2022gsn} with the standard cosmological parameters, especially with the angular diameter distance and the scalar fluctuations amplitude, although those are much reduced compared with the ones of the current data.
The addition of \Planck, shown in \autoref{fig:MHD2DP} enables the improvement of all the constraints. Thanks to data at high multipoles and the high sensitivity provided by \lb on intermediate and small multipoles, constraints are improved over the whole range of $\nB$. At the same time, we find how the addition of the high multipoles worsens the degeneracies with cosmological parameters, especially for the scalar spectral index due to the enhanced sensitivity provided by the small-scale addition. 

\subsubsection{Ambipolar diffusion}
We proceed with the ambipolar diffusion effect. In this case we have shown how the effect strongly depends on $\nB$ due to the contribution of the Lorentz force, and how it is really strong on the lowest multipoles in $E$-mode polarization.
\begin{table}[t!]
\center
\begin{tabular}{|c|c c|}
\hline
 \multicolumn{3}{|c|}{Ambipolar diffusion}\\
\hline
$\nB$ &\multicolumn{2}{|c|} {$\sqrt{\langle  B^2 \rangle} \, [\mathrm{nG}] $}   \\
\hline
 &\lb(\Planck)&\lb+\Planck  \\
\hline
Marginalized &  $<2.05$ ($<3.40$)&$<1.95$ \\
\hline
$2$ & $<0.018$ ($<0.058$) &$<0.018$\\
\hline
$1$ & $<0.037$ ($<0.12$) & $<0.036$\\
\hline
$0$ & $<0.080$ ($<0.26$) & $<0.078$\\
\hline
$-1$ & $<0.19$ ($<0.62$) & $<0.19$\\
\hline
$-2$ & $<0.57$ ($<1.84$) &$<0.58$\\
\hline
$-2.9$ &  $<3.6$ ($\dots$)&$<3.6$\\
\hline
\end{tabular}
\caption{\label{tab:HAMBI}
Same as \autoref{tab:HMHD} but for the ambipolar diffusion effect.}
\end{table}
\begin{figure}[t!]
    \centering
    \includegraphics[width=\textwidth]{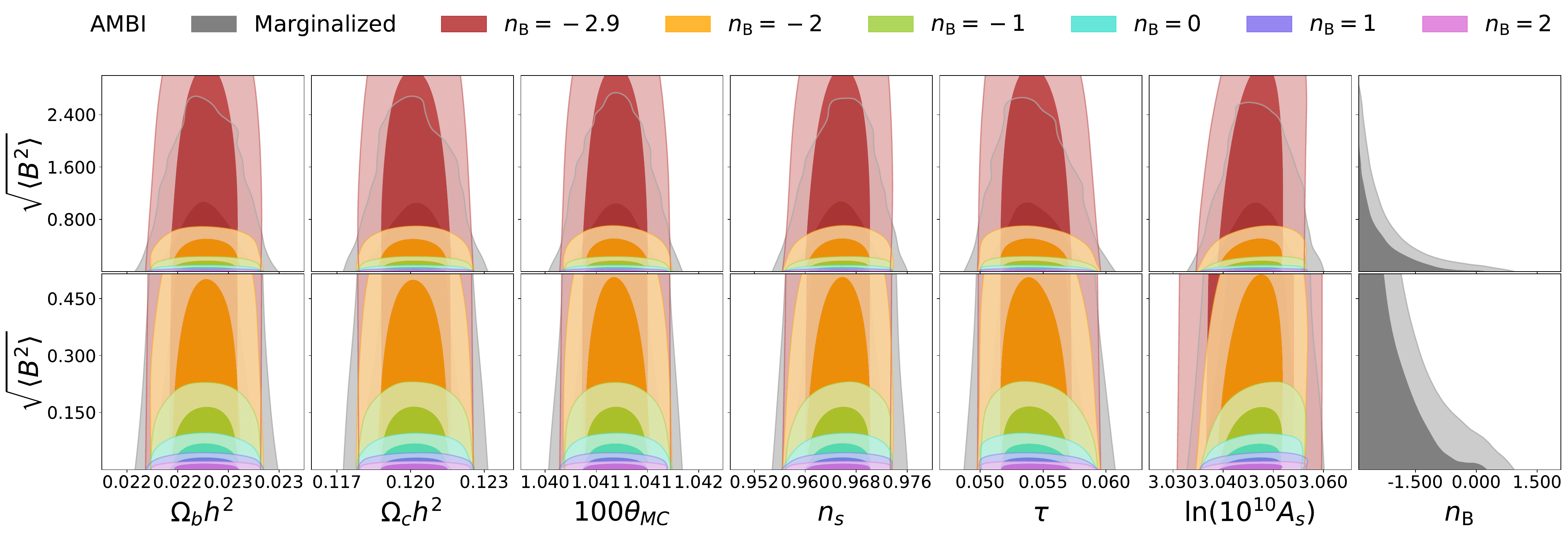}\\
    \caption{Same as \autoref{fig:MHD2D} but for the ambipolar diffusion effect.}
    \label{fig:AMBI2D}
\end{figure}
These characteristics of ambipolar diffusion make the related CMB signal an excellent target for \lb polarization, as reflected in the constraints in \autoref{tab:HAMBI}. \lb is capable of improving the constraints for all $\nB$ with respect to the current results using \Planck. \lb is furthermore able to constrain the almost scale invariant case, which is currently unconstrained. The improvement for positive $\nB$ reaches up to a factor of 3. The two-dimensional contours are shown in \autoref{fig:AMBI2D}; we find how the powerful data of \lb in the $E$-mode polarization are capable of almost completely removing the degeneracies observed with current data \citep{Paoletti:2022gsn}, especially in the optical depth and the amplitude of scalar fluctuations. This is again a demonstration of the impressive gains that can be reached at the sensitivity levels of \lb.
Because the effect is stronger on the lowest and intermediate multipoles, as expected, the addition of \Planck only leads to modest improvements for intermediate $\nB$ (close to 1 and 0). 

\subsubsection{Combined effect}
We now consider both effects together. 
We perform the analysis for both \lb and its combination with \Planck.
\begin{table}[t!]
\center
\begin{tabular}{|c| c c|}
\hline
 \multicolumn{3}{|c|}{Combined effect}\\
\hline
$\nB$ & \multicolumn{2}{|c|}{$\sqrt{\langle  B^2 \rangle} \, (\mathrm{nG}) $}   \\
\hline
 &\lb (\Planck)&\lb+\Planck   \\
\hline
Marginalized&$<0.50$ ($<0.69$)&$<0.48$ \\
\hline
$2$ & $<0.018$ ($<0.06$) &$<0.018$\\
\hline
$1$ & $<0.037$ ($<0.12$) &$<0.037$\\
\hline
$0$ & $<0.080$ ($<0.26$) &$<0.079$\\
\hline
$-1$ & $<0.20$ ($<0.56$) &$<0.19$\\
\hline
$-2$ & $<0.48$ ($<0.79$) &$<0.49$\\
\hline
$-2.9$ &$<0.73$ ($<1.06$)  &$<0.69$\\
\hline
\end{tabular}
\caption{\label{tab:HCOMBO}
Same as \autoref{tab:HMHD} but for the combination of ambipolar diffusion and MHD decaying turbulence.}
\end{table}
\begin{figure}[t!]
    \centering
    \includegraphics[width=\textwidth]{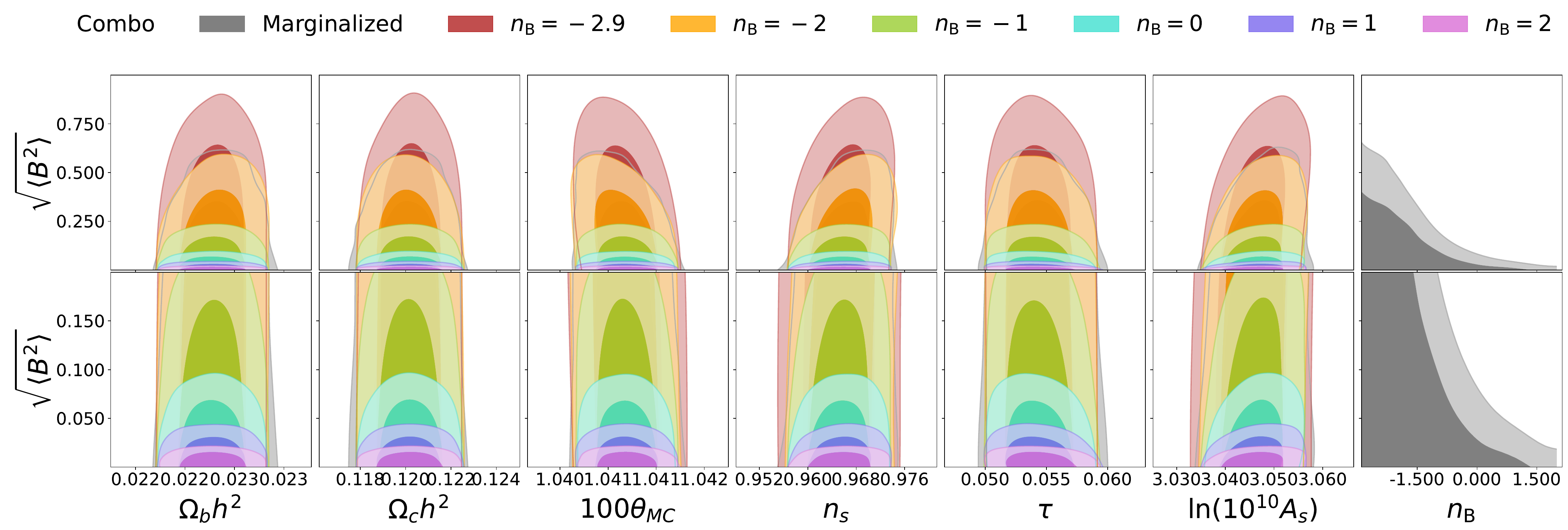}\\
    \caption{Same as \autoref{fig:MHD2D} but for the combination of ambipolar diffusion and MHD decaying turbulence.}
    \label{fig:MHDAMBI2D}
\end{figure}
The combined results are presented in \autoref{tab:HCOMBO} and represent the perfect combination of the two effects, with ambipolar diffusion and MHD turbulence constraining the ultraviolet and infrared spectral indices, respectively. The combination of both effects is therefore capable of strongly constraining the whole range of $\nB$ that we considered, thanks to the complementarity of the two effects.  This is even more reinforced by the addition of \Planck, which improves on the infrared indices. 
In \autoref{fig:MHDAMBI2D} we present the two dimensional contours which simply reflect the status of the constraints. We again find the reduction of the degeneracy of the MHD turbulence effect with respect to current data and the almost complete disappearance of the ones coming from ambipolar diffusion.

\subsection{Effect of the \texorpdfstring{$\boldsymbol{B}$}{B}-mode signal}
\begin{figure}[t!]
    \includegraphics[width=0.49\textwidth]{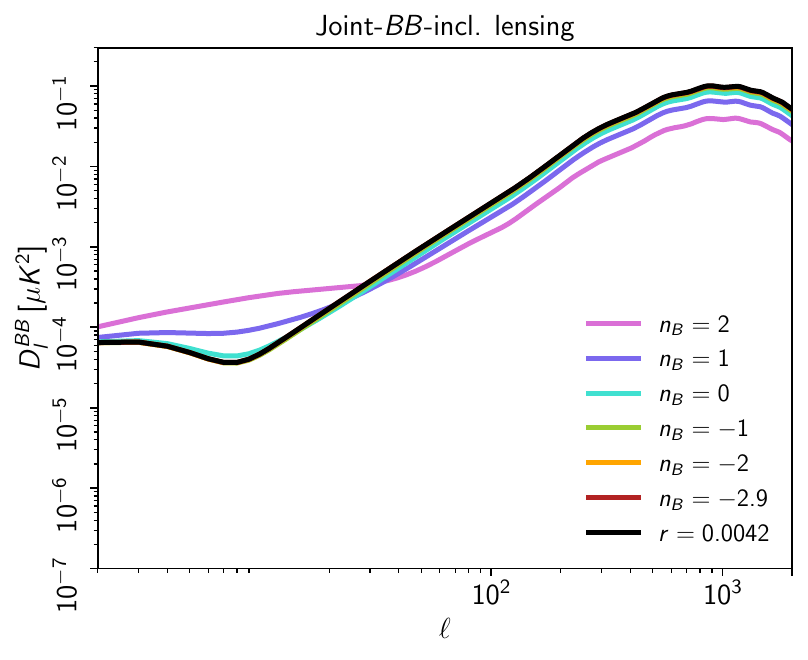}
    \includegraphics[width=0.49\textwidth]{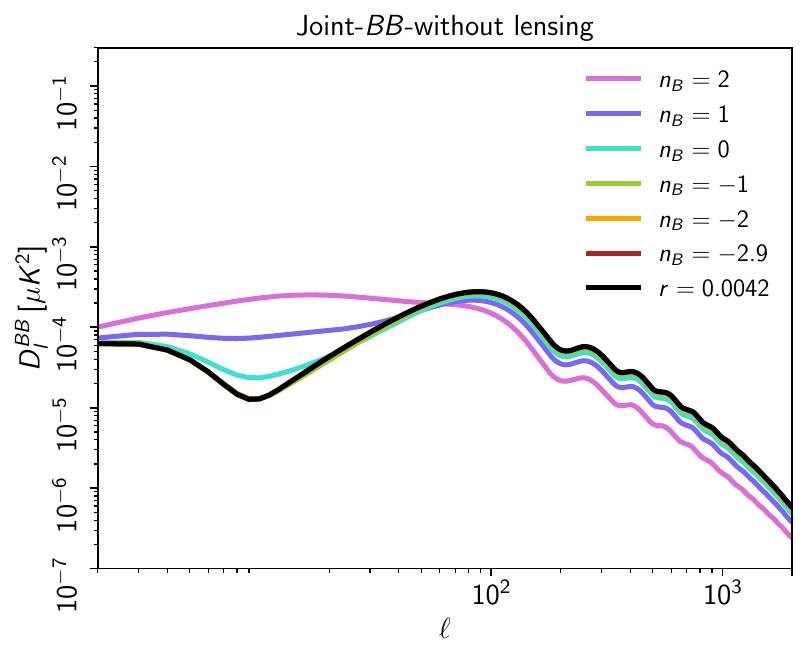}
    \caption{Effect of the combination of MHD turbulence and ambipolar diffusion on the $BB$ angular power spectrum for a primordial tensor mode with $r=0.0042$ and 1 nG PMFs. On the left we show the total signal in $BB$ including also the lensing contribution, whereas in the right panel we present the effect on the primordial tensor mode alone.}
    \label{fig:SpectraBB}
\end{figure}

The effect on the thermal and reionization history impacts the primary CMB anisotropies, modifying the angular power spectra\footnote{Contrary to the gravitational effect, which changes initial conditions and generates additional fluctuations.}. This implies that the presence of PMFs affects all the channels of CMB anisotropies in temperature and polarization, with the only condition that the channel has a non-negligible primary contribution, including the primordial tensor $B$-mode signal. For this reason we now test the impact of the heating effect on primordial GWs from inflation. In \autoref{fig:SpectraBB} we show how the combined MHD turbulence and ambipolar diffusion effects impact primordial tensor modes. We assume (as in the gravitational case) the same $R^2$-like model of inflation with $r=0.0042$. We find a strong impact at the level of the reionization bump especially for the ultraviolet indices caused by ambipolar diffusion. The combination of the two effects instead modifies the oscillation region imprinting an overall damping on the high multipole tail.
In light of this effect we derive the forecasts including the $B$-mode channel in our mock data. We consider the same setup as the baseline case in the gravitational effect. As a reminder, it consists of lensing considered as additional noise, the contribution of statistical foreground residuals, and the increased noise from component separation.

\begin{table*}[t!]
\centering
\begin{tabular}{|c|c c|c c|c c|}
\hline
 \multicolumn{7}{|c|}{\lb TEB}\\
\hline
& \multicolumn{2}{|c|} {AMBI} & \multicolumn{2}{|c|} {MHD} & \multicolumn{2}{|c|} {Combined} \\
\hline
$\nB$ & \multicolumn{1}{|c|}{$\sqrt{\langle  B^2 \rangle} \, (\mathrm{nG}) $}& $r$ [95\,\%] & \multicolumn{1}{|c|}{$\sqrt{\langle  B^2 \rangle} \, (\mathrm{nG}) $}& $r$ [95\,\%]& \multicolumn{1}{|c|}{$\sqrt{\langle  B^2 \rangle} \, (\mathrm{nG}) $}& $r$ [95\,\%]  \\
\hline
Marg. &$<1.82$& $<0.0018$&$<0.64$& $<0.0018$ &$<0.49$& $<0.0018$\\
\hline
$2$ &$<0.018$ & $<0.0018$&$<0.20$ & $<0.0018$ & $<0.018$ & $<0.0019$ \\
\hline
$1$ &$<0.037$ & $<0.0018$&$<0.30$ & $<0.0019$ & $<0.037$& $<0.0018$\\
\hline
$0$  &$<0.080$& $<0.0017$&$<0.46$ & $<0.0018$ & $<0.080$& $<0.0018$\\
\hline
$-1$ &$<0.19$& $<0.0018$&$<0.65$& $<0.0018$ & $<0.19$& $<0.0018$\\
\hline
$-2$ &$<0.58$ & $<0.0017$&$<0.70$ & $<0.0018$ & $<0.50$& $<0.0018$\\
\hline
$-2.9$ &$<3.4$& $<0.0018$&$<0.75$& $<0.0019$ &$<0.75$& $<0.0018$\\
\hline
\end{tabular}
\caption{\label{tab:HCOMBOteb}
Constraints on the PMF amplitude and the tensor-to-scalar ratio $r$, both for a fixed $\nB$ and the marginalized case over $\nB$, by using the combination of ambipolar diffusion and MHD decaying turbulence. Constraints are at 95\,\% C.L, assuming $r=0$ in the input.}
\end{table*}
\begin{figure*}[htbp]
    \centering
    \includegraphics[width=0.5\textwidth]{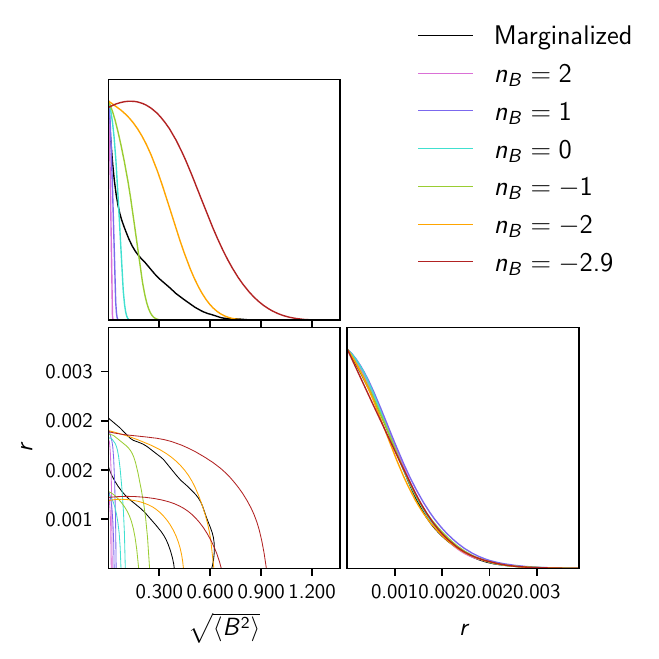}\includegraphics[width=0.5\textwidth]{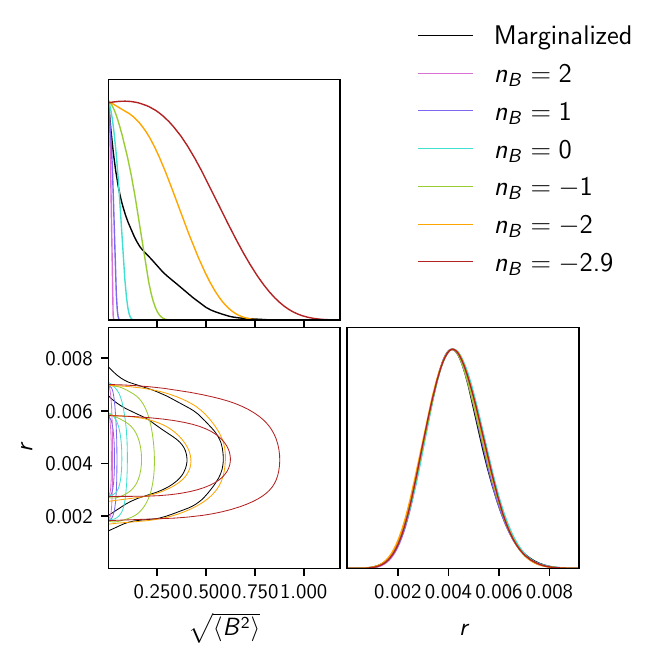}
    \caption{Triangle plot for the magnetic and primordial GW parameters for the combined effect. (Left) $r=0$ in the input. (Right) $r=0.0042$ in the input.}
    \label{fig:triHeatTEB}
\end{figure*}
The results for the case where we assume a zero contribution from primordial GW  from inflation ($r=0$) in the fiducial are shown in \autoref{tab:HCOMBOteb} for the two separate effects and their combination. We find that the addition of the $B$-mode channel and the sampling of $r$ does not change the constraints very much; minimal variations are due to either better fitting thanks to the additional channel or otherwise the minimal degeneracy with the tensor contribution from inflation. The two-dimensional contours are shown in the left panel of \autoref{fig:triHeatTEB}, showing no strong degeneracies between the PMF signal and the inflationary one.
\begin{table}[t!]
\center
\begin{tabular}{|c c c|}
\hline
\multicolumn{3}{|c|}{Combined heating and inflation}\\
\hline
$\nB$ &$\sqrt{\langle  B^2 \rangle} \, (\mathrm{nG}) $  & $r$ [68\,\%]\\
\hline
Marginalized &  $<0.47$& $0.0043_{-0.0012}^{+0.0010}$\\
\hline
 2& $<0.018$ & $0.0043_{-0.0012}^{+0.0010}$\\
\hline
 1&  $<0.037$& $ 0.0043_{-0.0012}^{+0.0010}$\\
\hline
 0& $<0.081$& $0.0043_{-0.0012}^{+0.0011}$ \\
\hline
 $-1$&  $<0.20$& $0.0043_{-0.0012}^{+0.0010}$\\
\hline
 $-2$&  $<0.50$& $ 0.0043_{-0.0012}^{+0.0011}$\\
\hline
 $-2.9$&  $<0.74$& $0.0044_{-0.0012}^{+0.0010}$\\
\hline
\end{tabular}
\caption{\label{tab:AMBIMHDr42}
Constraints on the PMF amplitude and $r$ for the combined heating effect and $r=0.0042$ in the input. Constraints on PMFs and $r$ are at 95 and 68\,\% C.L., respectively.}
\end{table}
For the combined effect we consider also the case with a non-negligible fiducial for primordial GWs. Again we consider the $r=0.0042$ case. The results are shown in \autoref{tab:AMBIMHDr42}. As in $r=0$, we do not observe strong degeneracies, as shown also in the right panel of \autoref{fig:triHeatTEB}.

\subsection{Combination with the gravitational effect for the almost scale invariant case}
To combine the gravitational and the heating effects, we must rely on the same parametrization and setting for the PMF characteristics. The results of the gravitational effect in the rms parametrization demonstrate that the only relevant case to combine them is the one where the two parametrizations almost coincide, the almost scale invariant case. 
The resulting constraint on the amplitude of PMFs is $\sqrt{\langle  B^2 \rangle}<0.64$\,nG. For this particular PMF configuration the combination of heating and gravitational effects improves the constraints. For other configurations we have shown that the gravitational effect using this kind of parametrization provides much looser constraints, seemingly favouring the heating effect. We stress the importance of the complementarity of these two constraints. While the gravitational effect mainly relies on $B$-mode polarization, the heating effect relies mostly on $E$-mode polarization. While the gravitational effect is based on a well established semi-analytical treatment, the heating effect still relies on some approximations. Finally, while the gravitational effect may be degenerate with $r$, the heating effect is mostly degenerate with other parameters. Therefore the two effects are both relevant and crucial in the determination of PMF characteristics, especially in the case of a possible detection, for example in $B$-mode polarization.

\section{Faraday rotation}
\label{sec:faraday}

The existence of PMFs at the last scattering surface and later epochs induces a Faraday rotation (hereafter FR) signal in the CMB polarization anisotropies \citep{1996ApJ...469....1K}. The effect is proportional to the Faraday depth, i.e., the integral along the line of sight of the product of the parallel magnetic field component and the electron density. The expected rotation angle in a given direction has an amplitude that scales with a characteristic frequency behaviour of $\lambda^2 (\propto \nu^{-2})$, which can in principle be used to separate the FR signal from other ones such as inflationary GWs or cosmic birefringence \cite{Komatsu:2022nvu}.

The detailed modifications of the Boltzmann equations for the Stokes parameters in the presence of PMFs have been derived in several works, both for homogeneous 
\citep{2004PhRvD..70f3003S} and for stochastic PMF distributions \citep{Kosowsky:2004zh}. In this paper, we will focus on a stochastic PMF distribution, using the same notation and formalism presented in Section \ref{sec:formalism}. The FR effect converts some $E$ modes to $B$ modes. Here we forecast the constraints on PMFs using the $BB$ power spectrum induced by FR. 

\subsection{Inputs}
\label{sec:fr_inputs}
In order to quantify the potential impact of foreground residuals on single frequency CMB spectra needed for the FR analysis, we generated synthetic foreground residual spectra with the assumption that the component-separation algorithms will be able to clean foregrounds at the 1\,\% level in each \lb frequency channel.
The expected levels of contamination are shown in Ref. \cite{Krachmalnicoff:2018imw} and the current \lb simulations show a reduction compatible with that assumed in Ref. \citetalias{LiteBIRD:2022cnt}. 

\begin{figure}[t!]
    \centering
\includegraphics[width=0.49\columnwidth]{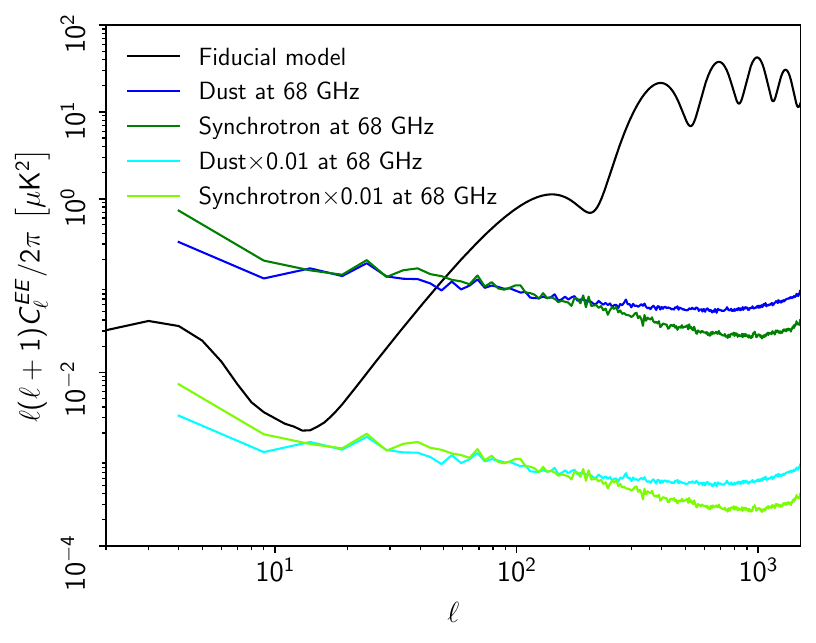}    \includegraphics[width=0.49\columnwidth]{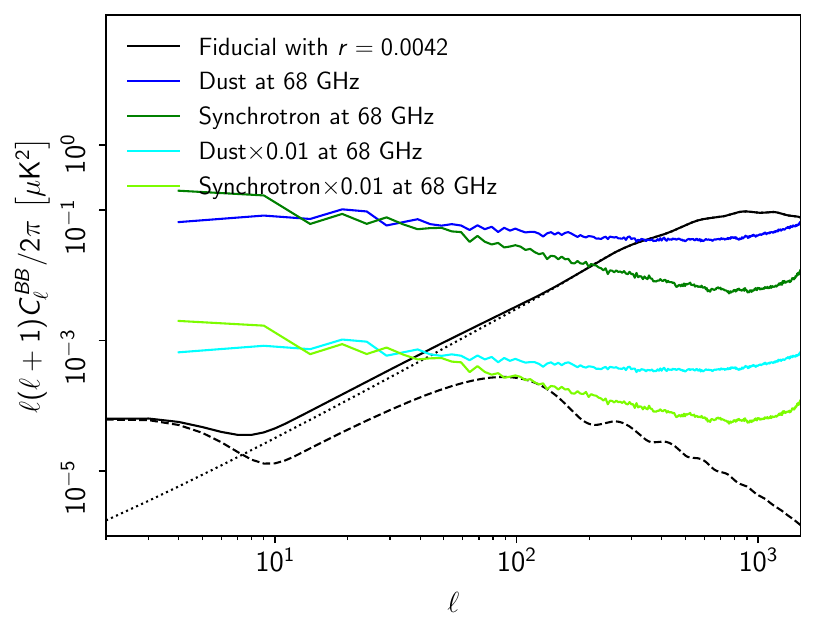}
    \caption{Synchrotron (green) and dust (blue) $EE$ (left) and $BB$ (right) binned spectra estimated at 68\,GHz and the corresponding expected amplitude of residual synchrotron (light green) and dust (cyan) contamination after the component-separation cleaning procedures. }
    \label{fig:fr_inputs}
\end{figure}

We started from a set of synchrotron and dust polarization maps generated at 100\,GHz using the \texttt{PySM code} \citep{Thorne:2016ifb,Zonca:2021row} matching the model used in Ref. \citetalias{LiteBIRD:2022cnt}. The maps are generated at $N_{\rm side}=512$ with a 5 arcmin Gaussian beam. We extracted the angular power spectra using \texttt{cROMAster}, a pseudo-C$_\ell$ algorithm implementing a geometrical correction for the loss of orthonormality of the spherical harmonic functions in the cut sky \citep{Hivon:2001jp,Polenta:2004qs}, using the \Planck likelihood polarization mask at 100\,GHz \citep{Planck:2019nip} leaving 79\,\% of the sky pixels available for the analysis. The amplitudes of the synchrotron and dust spectra at 100\,GHz are then renormalized to account for the frequency dependence of the two foreground signals to obtain the amplitudes in all the \lb channels. Following the \texttt{PySM} templates for the synchrotron signal we considered a power-law behaviour with spectral index equal to $-3$. For the dust signal we followed a modified black body behaviour with a spectral index of $1.54$ and a temperature of 20\,K.
As an example, we show the resulting $EE$ and $BB$ spectra of synchrotron and dust at 68\,GHz in \autoref{fig:fr_inputs}.

All the resulting spectra at each \lb frequency channel are then divided by a factor of 100 to mimic the presence of a residual foreground signal from the component-separation cleaning procedure. In \autoref{fig:fr_inputs} we also show the expected residuals at 68\,GHz. These residuals are used as input to the FR analysis described in the following subsections.

\subsection{Methodology}
\begin{figure}[t!]
    \centering
    \includegraphics[width=0.5\columnwidth]{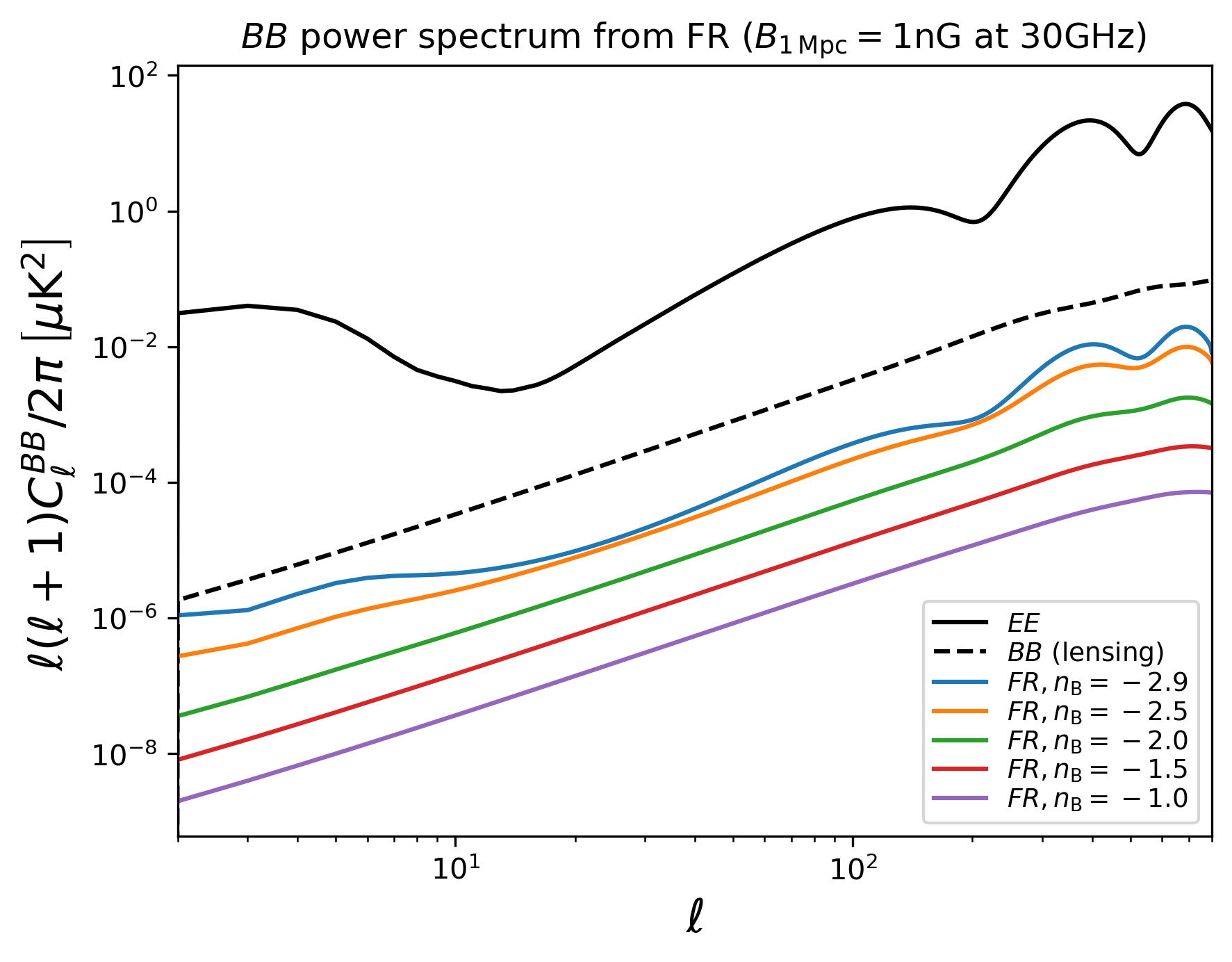}%
     \includegraphics[width=0.5\columnwidth]{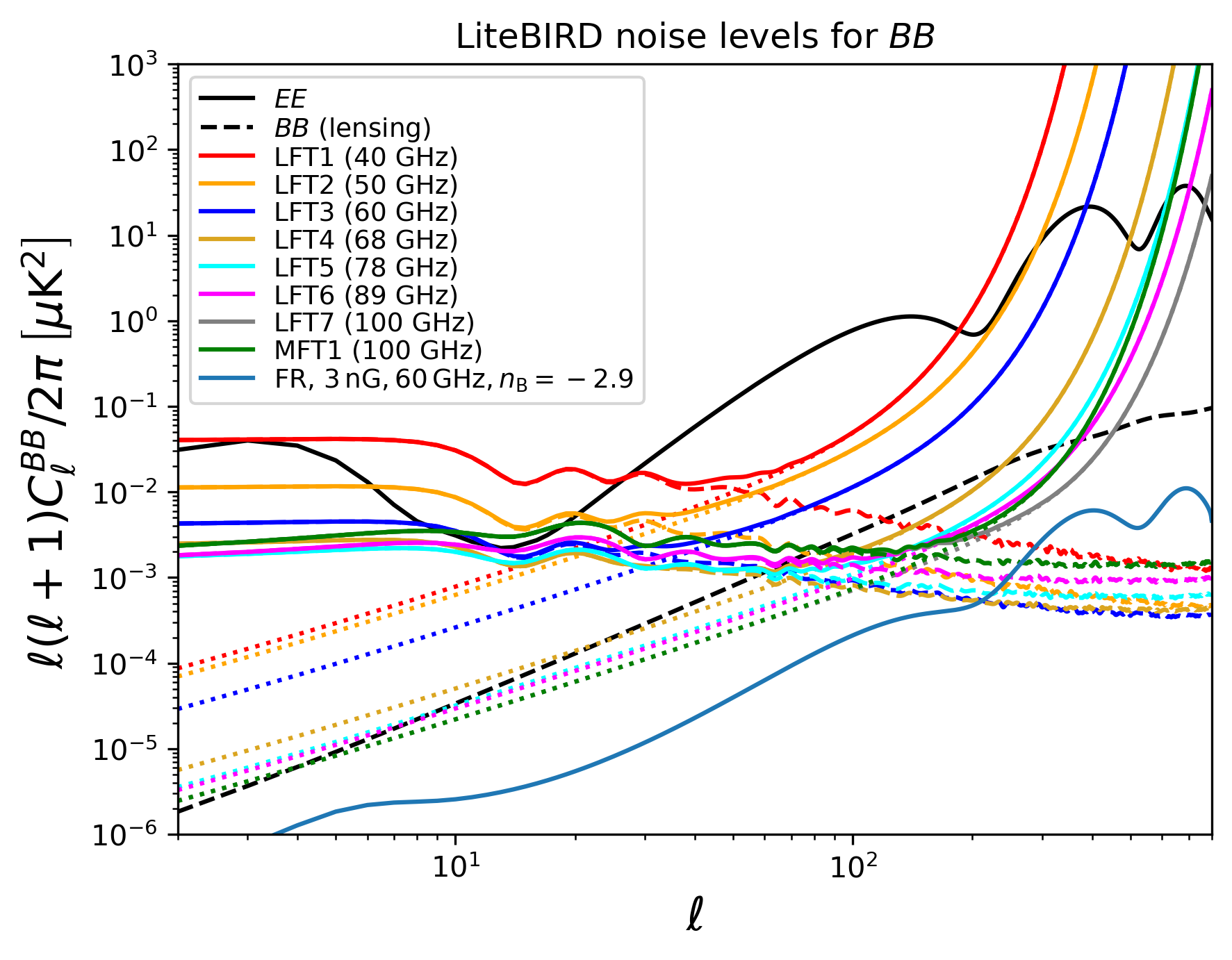}
    \caption{(Left) $B$-mode signal  generated by Faraday rotation with $B_{1\,{\rm Mpc}}=1$\,nG, $\nu_0=30$\,GHz, and different values of $\nB$. For comparison purposes, we show the level of the $EE$ and lensing $BB$ spectra in our fiducial model (see Section \ref{sec:forecast}). (Right) Noise levels on the BB spectrum for the detection of FR signals using different \lb channels. The noise levels have two contributions, instrumental noise and foreground residuals (see text for details). For comparison, the FR signal for a 3\,nG field with $\nB=-2.9$ at 60\,GHz is also shown.  }
    \label{fig:aps_fr}
\end{figure}

In this work, we constrain the angular power spectrum of the $B$-mode polarization arising from small FR due to stochastic PMFs, following the formalism presented in Refs. \citep{Kosowsky:2004zh, 2009PhRvD..80b3009K}, which was also used in Ref. \citep{Planck:2015zrl}.
The stochastic PMF distribution is described here by the power law given in \autoref{gaussian}. Note that any helical part of the field does not contribute to the FR signal \citep{Campanelli:2004pm}. The power spectrum of the generated $B$ modes coming from FR of $E$ modes is given by \cite{Kosowsky:2004zh}
\begin{equation}
C_\ell^{\rm BB} = N_\ell^2 \sum_{\ell_1, \ell_2} \frac{(2\ell_1+1)(2\ell_2+1)}{4\pi(2\ell+1)} N_{\ell_2}^2 K(\ell,\ell_1,\ell_2)^2 C_{\ell_2}^{\rm EE} C_{\ell_1}^{\rm \alpha} (C_{\ell_1 0 \ell_2 0}^{\ell 0})^2\,,
\end{equation}
where $C_{\ell_1 0 \ell_2 0}^{\ell 0}$ are Clebsch-Gordan coefficients, $N_\ell = (2(\ell-2)!/(\ell+2)!)^{1/2}$ is a normalization factor, and $K(\ell,\ell_1,\ell_2)\equiv-1/2(L^2+L_1^2+L_2^2 -2L_1 L_2 -2L_1 L + 2L_1 -2L_2 -2L)$ with $L\equiv\ell(\ell+1)$, $L_1\equiv\ell_1(\ell_1+1)$ and $L_2\equiv\ell_2(\ell_2+1)$. Finally, the power spectrum of the rotation angle, $C_{\ell}^{\rm \alpha}$, is given by
\begin{equation}
C_{\ell}^{\rm \alpha} = \frac{9 \ell(\ell+1)}{(4\pi)^3 e^2} \lambda_0^4 \frac{B_\lambda^2}{\Gamma(\nB +3/2)} \Big( \frac{\lambda}{\eta_0}\Big)^{\nB+3} \int_{0}^{x_{\rm D}} dx x^{\nB} j_\ell^2(x).
\label{eq:c_l_alpha}
\end{equation}
Here $x_{\rm D} =k_{\rm D} \eta_0$, where $\eta_0$ is the conformal time today, and $k_{\rm D}$ is the magnetic field cutoff given by the Alfvén-wave damping scale as defined in \autoref{kd_def}.
In \autoref{eq:c_l_alpha} we explicitly see the frequency dependence of the signal via the $\lambda_0^4$ factor, $\lambda_0$ being the observing wavelength. 
\autoref{fig:aps_fr} shows examples of $B$-mode signals generated by FR of PMFs for different values of $\nB$. 
Note that our numerical implementation of this computation has been tested and compared with other independent approaches; see e.g. Refs. \citep{2022PhRvD.106f3505G, 2022PhRvD.105f3536C}, yielding consistent results within the numerical precision.  

\subsection{Results}
Here we provide forecasts for future capabilities of \lb to constrain PMFs using FR. Following Refs. \citep{Kahniashvili:2008hx,2011PhRvD..84d3530P, Planck:2015zrl}, our analysis is based on the $BB$ spectrum alone. The theoretical $C_\ell^{\rm BB}$ spectra are compared with the expected \lb measurements using a Gaussian likelihood. Instrumental noise power spectra for each individual channel are computed as described in Section \ref{sec:forecast} and \autoref{eq:noise}. The variance for each individual measurement of $C_\ell^{BB}$ is computed by adding the two contributions. The first one accounts for the cosmic variance and instrumental noise terms (i.e., $2 (2\ell+1)^{-1} f_{\rm sky}^{-1}$ times the square of the sum of the signal and noise power spectra), with $f_{\rm sky}=0.7$. The second contribution accounts for a residual foreground contribution, which is not subtracted after the component-separation process at each individual frequency. For this second term, we add in quadrature to the error a term corresponding to $0.01$ times the total foreground contribution computed above in Section \ref{sec:fr_inputs}.
The right panel in \autoref{fig:aps_fr} shows the total noise contribution for all the \lb channels with effective frequencies equal to or below 100\,GHz, where the FR signal is larger due to $1/\nu_0^4$. For completeness, the noise contribution is separated into the two components of noise (dotted lines) and foreground residuals (dashed lines). 

\begin{figure}[t!]
    \centering    \includegraphics[width=0.7\columnwidth]{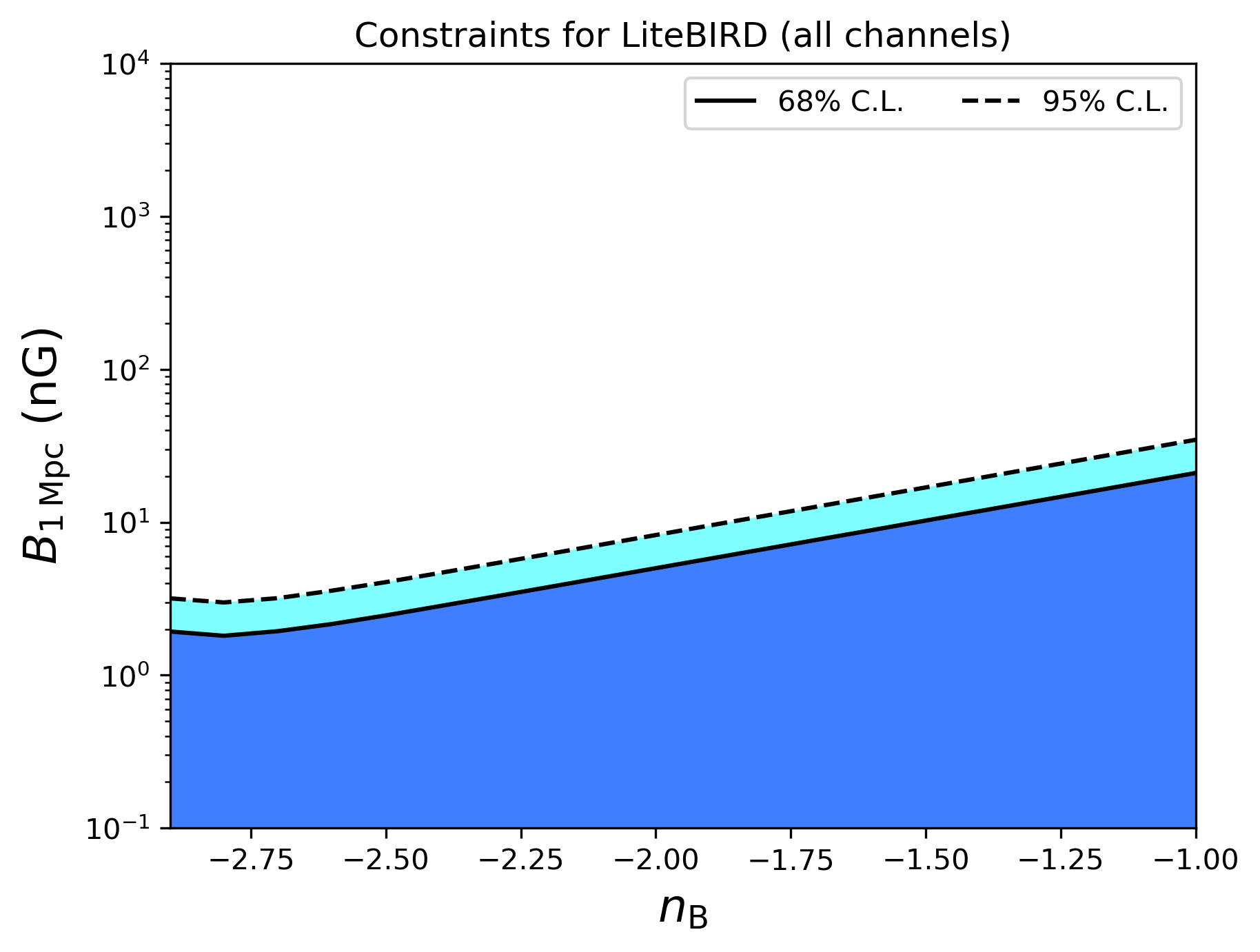}
    \caption{Constraints from Faraday Rotation combining all \lb channels. }
    \label{fig:fr}
\end{figure}

The results of our analysis are presented in \autoref{tab:fr} and \autoref{fig:fr}. 
For each individual \lb frequency channel, we constrain the PMF amplitude as a function of $\nB$. We also derive the overall constraints for all channels by combining the noise levels with inverse variance weighting. However, in this case, we have to account for the fact that the FR signal scales as $1/\nu_0^4$, so the noise has to be scaled as $(\nu_0/60\,{\rm GHz})^4$, if we use for example the 60\,GHz channel as the reference frequency. 
\autoref{tab:fr} shows the result for 68\,GHz (second column), and the combination of all channels (third). For the case of $\nB=-2.9$, the 95\,\% C.L. is expected to be at the level of 3\,nG. 

\begin{table}[t!]
    \centering
    \begin{tabular}{|c c c c|}
    \hline
    $\nB$ & $B_{1\,{\rm Mpc}}$ [nG] & $B_{1\,{\rm Mpc}}$ [nG] & $B_{1\,{\rm Mpc}}$ [nG]\\
    & ($68$\,GHz) & (all) & (no foreg.)\\
    \hline
  $-2.9$ & $<    4.5$ & $<    3.2$ & $<    3.0$ \\  
  $-2.7$ & $<    4.5$ & $<    3.2$ & $<    3.0$ \\  
  $-2.5$ & $<    5.7$ & $<    4.1$ & $<    3.8$ \\  
  $-2.3$ & $<    7.4$ & $<    5.4$ & $<    5.0$ \\  
  $-2.1$ & $<    9.8$ & $<    7.2$ & $<    6.6$ \\  
  $-1.9$ & $<   12.9$ & $<    9.5$ & $<    8.7$ \\  
  $-1.7$ & $<   17.1$ & $<   12.7$ & $<   11.6$ \\  
  $-1.5$ & $<   22.7$ & $<   16.9$ & $<   15.4$ \\  
  $-1.3$ & $<   30.2$ & $<   22.5$ & $<   20.5$ \\  
  $-1.1$ & $<   40.1$ & $<   30.0$ & $<   27.3$ \\  
  \hline
\end{tabular}
    \caption{Constraints from Faraday rotation for \lb, using the 68 GHz channel only (second column), all channels (third) and all channels without foreground residuals (fourth). Constraints are at 95 \% C.L.. }
\label{tab:fr}
\end{table}

Finally, as can be seen in the right panel of \autoref{fig:aps_fr}, most of the constraining power of this analysis comes from the high multipole region around $\ell \gtrsim 300$ where the overall noise contribution in the $BB$ spectrum is dominated by instrumental noise rather than foreground residuals. Indeed, in \autoref{tab:fr} we show that the constraints are almost identical without foreground residuals in the noise variance. This result relies on the fact that the foreground residuals have been decreased to the 0.01 level in the power spectrum for each individual frequency, as might be expected from the overall performance of the component-separation algorithms, as shown in Ref. \citetalias{LiteBIRD:2022cnt}. 

\section{Non-Gaussianities}
\label{sec:nongaussian}
As described in Section \ref{sec:gravitational}, PMFs contribute to cosmological perturbations through their energy-momentum tensor, with its different projections sourcing magnetically-induced perturbations in scalar, vector, and tensor modes.
In the PMF energy-momentum tensor, the magnetic contribution to the source terms of cosmological perturbations is quadratic in the field amplitude. But the square of a random Gaussian field is a $\chi^2$-distribution, meaning that the contribution of PMFs to cosmological perturbations is highly non-Gaussian. This leads to non-vanishing higher order statistical moments in CMB anisotropies such as the bispectrum \cite{Brown:2006wv,Seshadri:2009sy,Caprini:2009vk,Shiraishi:2010yk,Trivedi:2010gi,Shiraishi:2011xvp,Shiraishi:2011fi,Shiraishi:2011dh,Shiraishi:2012sn,Shiraishi:2012rm,Shiraishi:2013wua,Ade:2015cva} and the trispectrum \cite{Trivedi:2011vt,Trivedi:2013wqa}.

Here we are interested in the analyses that can benefit from \lb measurements of CMB polarization. Hence we will investigate the case of the passive tensor mode bispectrum, which generates a  specific signature in $B$-mode polarization.
PMFs can generate tensor modes via the gravitational effect, i.e., GWs, sourced by the magnetic anisotropic stress fluctuations. This GW production is maintained from the birth of the PMFs ($\tau_{\mathrm{B}}$) until the neutrino decoupling epoch ($\tau_\mu$) when we have the excitation of the compensated mode and the generation of the passive one as a remaining offset from the matching of initial conditions at neutrino decoupling. The super horizon mode of the resultant GWs takes the form
\begin{align}
  h_{ij}({\bf k}) \approx -1.8 \frac{\ln(\tau_\nu / \tau_{\mathrm{B}})}{4 \pi \rho_{\gamma, 0}} \sum_{s = \pm 2} e_{ij}^{(s)}(\hat{k}) e_{kl}^{(s) *}(\hat{k}) \int \frac{d^3 p}{(2\pi)^3} B_k({\bf p}) B_l({\bf k} - {\bf p}),  
\end{align}
where $\rho_{\gamma, 0}$ is the present photon energy density and $e_{ij}^{(s)}$ is the spin-2 transverse-traceless polarization tensor normalized as $e_{ij}^{(s)}(\hat{k}) e_{ij}^{(s')*}(\hat{k}) = 2 \delta_{s,s'}$. The tensor passive mode is linearly transformed to the CMB $B$-mode field, as in the standard adiabatic GWs; 
it becomes a $\chi^2$ (i.e, highly non-Gaussian) field because of the assumption of the Gaussianity of $B_i$. 
Since an induced $B$-mode bispectrum scales as $\langle a_{\ell_1 m_1}^B a_{\ell_2 m_2}^B a_{\ell_3 m_3}^B \rangle \propto \langle h_{i_1 j_1}({\bf k}_1) h_{i_2 j_2}({\bf k}_2)h_{i_3 j_3}({\bf k}_3) \rangle$,
its magnitude simply depends on the following parameter including the sixth power of PMF amplitude:
\begin{align}
  A_{\rm bis} \equiv \left(\frac{B_{1 \, \rm Mpc}}{1 \, \rm nG} \right)^6 \left(\frac{\ln (\tau_\nu/\tau_{\mathrm{B}})}{\ln(10^{17})}\right)^3 ,
  \end{align}
and $\nB$ determines its scale dependence. For nearly scale invariant $\nB \approx -3$, the signal at the squeezed configurations: $\ell_1 \ll \ell_2 \approx \ell_3 $, $\ell_2 \ll \ell_3 \approx \ell_1$ and $\ell_3 \ll \ell_1 \approx \ell_2$, dominates \cite{Shiraishi:2011dh,Shiraishi:2012rm,Shiraishi:2013vha,Shiraishi:2019yux}.%
\footnote{The compensated vector and tensor modes can also create a $B$-mode bispectrum; however, the dominant signal is located at such high $\ell$ \cite{Shiraishi:2010yk} that \lb would be insensitive to it. For this reason, we do not consider these cases, but only the passive tensor mode.}

We examine the detectability of $A_{\rm bis}$ for $\nB = -2.9$, computing the Fisher matrix for $A_{\rm bis}$:

\begin{align}
  F = f_{\rm sky} \sum_{\ell_1 \ell_2 \ell_3}  \frac{|B_{\ell_1 \ell_2 \ell_3}|^2}{6 C_{\ell_1} C_{\ell_2} C_{\ell_3}}\,,
\end{align}

where $B_{\ell_1 \ell_2 \ell_3}$ is the angle-averaged $B$-mode bispectrum for $A_{\rm bis} = 1$ that does not vanish only for $\ell_1 + \ell_2 + \ell_3 = \rm odd$ and hence takes pure imaginary numbers, and $f_{\rm sky}$ is the fraction of the sky coverage. Here we assume a Gaussian covariance; thus, the denominator is simply given by the triple of the observed $B$-mode power spectrum $C_\ell$.  This is given by the sum of the contributions of primordial GW, lensing, and the noise bias determined by all experimental features and foreground residuals, i.e., $C_\ell = C_\ell^{\rm prim} + C_\ell^{\rm lens} + N_\ell$. For the foreground residuals we are assuming the baseline case. Moreover, we assume that $C_\ell^{\rm prim}$ has a scale-invariant shape and is proportional to $r$. In the computation of $C_\ell^{\rm lens}$, we do not take delensing into account here. Furthermore, in this analysis, any other non-Gaussian source than PMFs is not taken into account.

\begin{figure}[t!]
    \centering
    \includegraphics[width=0.5\textwidth]{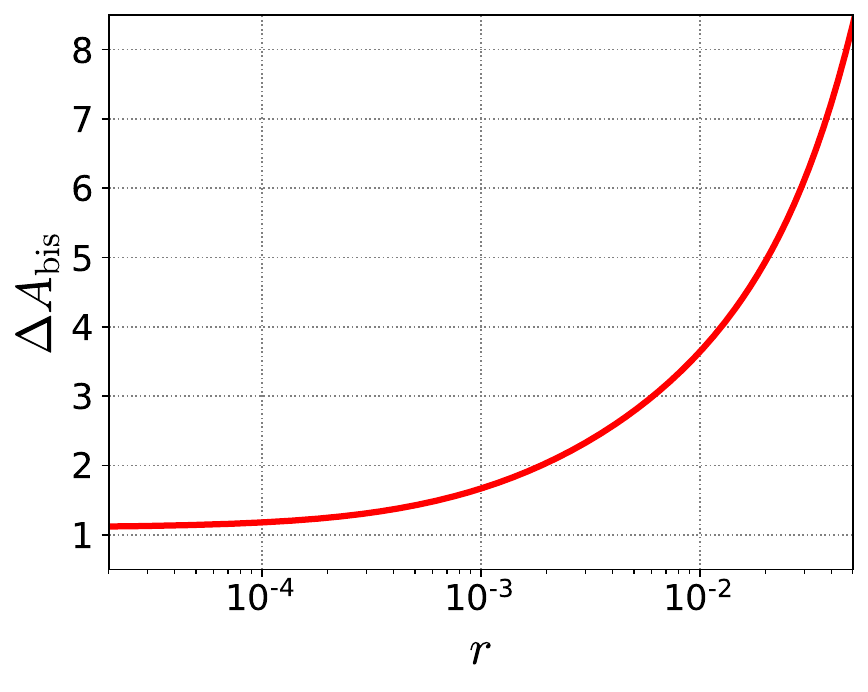}
    \caption{Expected $1\sigma$ error on the size of the $B$-mode bispectrum $A_{\rm bis}$ as a function of $r$.}
    \label{fig:error_Abis_vs_r}
\end{figure}

\autoref{fig:error_Abis_vs_r} shows the expected $1\sigma$ error, $\Delta A_{\rm bis} = 1/\sqrt{F}$, for various $r$. As $r$ becomes smaller, $C_\ell$ decreases, resulting in decreasing $\Delta A_{\rm bis}$. However, for $r \lesssim 10^{-4}$, $\Delta A_{\rm bis}$ becomes constant in $r$, since $C_\ell^{\rm prim}$ is subdominant enough compared with $C_\ell^{\rm lens}$ and $N_\ell$, and $C_\ell$ becomes independent of $r$.

From this figure, we can find that $A_{\rm bis} = {\cal O}(1)$ would be measurable independently of a value of $r$. In other words, the \lb $B$-mode data could improve $\Delta A_{\rm bis}$ by three orders of magnitude in comparison with the \Planck  \cite{Planck:2015zrl} and WMAP \cite{Shiraishi:2013wua} results.%
\footnote{
Note that $A_{\rm bis}$ and $A_{\rm bis}^{\rm MAG}$ values in Ref. \cite{Planck:2015zrl} are different by a factor of $3^6$.
}
This is consistent with the prediction in Ref. \cite{Shiraishi:2019yux}. 
This is a huge improvement in terms of the amplitude of the bispectrum, possible thanks to the sensitivity in polarization of \lb. Although it is such an improvement in terms of the bispectrum amplitude, the improvement in terms of the PMF amplitude is much reduced by the sixth power dependence of the bispectrum amplitude on the PMF amplitude. 
Nevertheless, the estimated improvement on the bispectrum detection leads to the breaking of the $B_{1 \, \rm Mpc} \simeq 1 \, \rm nG$ threshold also for non-Gaussianities, and hence such values could be captured for a wide range of $\tau_{\mathrm{B}}$. 

\section{Conclusions}
\label{sec:conclusions}
The \lb satellite, with its groundbreaking sensitivity in polarization, will open a new era in cosmology, allowing us to probe a wide range of fundamental physics.
PMFs represent an unconventional observational window on fundamental physics in the early Universe.
We have provided a series of forecasts for \lb's capabilities of constraining (or eventually detecting) PMFs. 
\lb offers a variety of probes for primordial magnetism, all within the same experiment. As for \Planck, with \lb we can potentially detect magnetically induced perturbations, post-recombination dissipative effects on the thermal and ionization history, Faraday rotation and non-Gaussianities, but unlike for \Planck, with \lb the improvement is much higher, thanks to polarization. 
We have focused on these effects of PMFs on the CMB anisotropies where polarization is particularly relevant, but for each effect in a different way, hence the multiplicity of constraints.

Magnetically-induced perturbations, i.e. the gravitational effect, mainly rely on $B$-mode polarization from \lb. We have presented a whole series of different mock data conditions, from the ideal case to the realistic case in terms of $B$-mode data contamination. We have shown that the data contamination even in the realistic case does not change our conclusions, highlighting the robustness of \lb. The major improvement is for the inflationary and in particular the almost scale invariant configuration, which breaks the nG threshold, improving current constraints by a factor of 3. The degeneracy with the primordial tensor mode from inflation still remains for infrared power-law indices  and will require the combination of all \lb probes. Overall \lb will be able to improve on all the PMF configurations providing stringent constraints; it will be also able to detect PMFs with a configuration such as the one currently constrained by \Planck data for an almost scale invariant input sky, either with a blind or informed reconstruction.

The post-recombination dissipative effects that damp PMFs heat the plasma and modify the ionization history (differently from the gravitational effect) rely mainly on temperature, which is already cosmic variance limited by \Planck and $E$-mode polarization which would likely be cosmic variance limited in \lb. In this case \lb is capable of strongly improving all the constraints, but complementary to the gravitational effect, the most constrained configurations are the ultraviolet indices, with the causal-limit fields constrained to a factor of 3 better than current results.
For the first time we have also investigated the impact of this effect on the primordial $B$-mode polarization from inflation. Contrary to the gravitational case, where PMFs create new additional signals (magnetically induced modes), the heating effect modifies the primary CMB signal. As a result, the $B$-mode power spectrum is modified in a way similar to what happens for the $E$-mode polarization. We have shown how constraints on the PMF amplitude remain mostly unaffected by the presence of primordial GWs, with only modest differences for both PMFs and $r$. This is thanks to \lb's accuracy in $E$-mode polarization, which carries enough information to disentangle the PMF effects and therefore does not affect the $B$-mode measurement.

Faraday rotation has always been a crucial independent CMB probe for PMFs. Although its strong frequency dependence makes it a hard target, dominating mainly at lower frequencies, we have shown how \lb will open a scenario in which FR is one of the main players. Predictions of constraints from \lb are orders of magnitude better than the current ones, reaching a few nG, at the same level as all the other probes. This additional constraining power enabled by \lb's sensitivity and frequency coverage is a game changer for constraints from FR, which will be crucial to determine the characteristics of PMFs especially in the case of a detection.

Finally also in non-Gaussianities we have demonstrated how, thanks to the first measurement of the $B$-mode bispectrum, \lb will improve by orders of magnitude the current bispectrum amplitude measurements, leading to the breaking of the nG threshold in the non-Gaussianities sector.

We have derived a range of different complementary constraints on PMFs with \lb simulated data. Although we considered different levels of complexities of the data and rely mostly on the detailed results from Ref. \citetalias{LiteBIRD:2022cnt}, we acknowledge that these are realistic forecasts, but with some optimistic assumptions. The full account of all possible complex systematics in all the channels is still in development. One example is the possible contamination caused by multipole-couplings induced by our motion relative to the CMB rest frame \citep{Challinor2002, Amendola2011, Jeong2014, Dai2014}, which should be modelled carefully. Nevertheless, even if on the optimistic side, the stability of the constraints across the different mock data sets is an indication of the robustness of \lb, and a good indication of its capability to extract the signals at the best possible level.

Overall \lb's capabilities to constrain PMFs from different perspectives, employing different data characterization and products, will allow us to improve all the current constraints at least by a factor of 3, 
if not orders of magnitude for some probes. But the most important point is the multiplicity of probes within the same experiment, not only to improve the constraints, but in the case of a detection only the joint evidence of varied complementary probes would lead to a clear understanding and interpretation of the data. We have shown how this is something only an experiment such as \lb can achieve.

\begin{acknowledgments}
%
This work is supported in Japan by ISAS/JAXA for Pre-Phase A2 studies, by the acceleration program of JAXA research and development directorate, by the World Premier International Research Center Initiative (WPI) of MEXT, by the JSPS Core-to-Core Program of A. Advanced Research Networks, and by JSPS KAKENHI Grant Numbers JP15H05891, JP17H01115, and JP17H01125.
The Canadian contribution is supported by the Canadian Space Agency.
The French \textit{LiteBIRD} phase A contribution is supported by the Centre National d’Etudes Spatiale (CNES), by the Centre National de la Recherche Scientifique (CNRS), and by the Commissariat à l’Energie Atomique (CEA).
The German participation in \textit{LiteBIRD} is supported in part by the Excellence Cluster ORIGINS, which is funded by the Deutsche Forschungsgemeinschaft (DFG, German Research Foundation) under Germany’s Excellence Strategy (Grant No. EXC-2094 - 390783311).
The Italian \textit{LiteBIRD} phase A contribution is supported by the Italian Space Agency (ASI Grants No. 2020-9-HH.0 and 2016-24-H.1-2018), the National Institute for Nuclear Physics (INFN) and the National Institute for Astrophysics (INAF).
Norwegian participation in \textit{LiteBIRD} is supported by the Research Council of Norway (Grant No. 263011) and has received funding from the European Research Council (ERC) under the Horizon 2020 Research and Innovation Programme (Grant agreement No. 772253 and 819478).
The Spanish \textit{LiteBIRD} phase A contribution is supported by the Spanish Agencia Estatal de Investigación (AEI), project refs. PID2019-110610RB-C21,  PID2020-120514GB-I00, ProID2020010108 and ICTP20210008.
Funds that support contributions from Sweden come from the Swedish National Space Agency (SNSA/Rymdstyrelsen) and the Swedish Research Council (Reg. no. 2019-03959).
The US contribution is supported by NASA grant no. 80NSSC18K0132.
%
\\
We thank Franco Vazza for the useful discussions, and Y. Guan for the numerical comparison of our FR code with his code. 
We acknowledge the use of the computing centre  of Cineca and INAF, under the coordination of the ``Accordo Quadro MoU per lo svolgimento di attività congiunta di ricerca Nuove frontiere in Astrofisica: HPC e Data Exploration di nuova generazione", for the availability of computing resources and support with the project INA23\_C9A05. 
We acknowledge financial support from the Spanish Ministry of Science and Innovation (MICINN) under the project PID2020-120514GB-I00.
 We acknowledge financial support from the ERC Consolidator Grant {\it CMBSPEC} (No.~725456) and as a Royal Society University Research Fellow at the University of Manchester, UK (No.~URF/R/191023).
We acknowledge financial support from the JSPS KAKENHI Grant Nos.~JP19K14718, JP20H05859 and JP23K03390. We acknowledge the Center for Computational Astrophysics, National Astronomical Observatory of Japan, for providing the computing resources of the Cray XC50.

\end{acknowledgments}

\bibliographystyle{JHEP}
\bibliography{Lit}

\end{document}